\documentclass[aps,prd,reprint,nofootinbib,superscriptaddress]{revtex4-1}
\usepackage{amsmath}
\usepackage{amssymb}
\usepackage{amsthm}
\usepackage{MnSymbol}
\usepackage{nicefrac}
\usepackage{graphicx}
\usepackage{amsfonts}
\usepackage{dsfont}
\usepackage[hidelinks]{hyperref}  
\usepackage{xspace}
\usepackage{comment}
\usepackage[normalem]{ulem}

\usepackage{slashed}
\usepackage{xcolor}

\definecolor{lightgray}{gray}{0.95}

\def \and{\textmd{and}}

\def\Nf{N_\text{f}}
\def\Ns{N_\text{s}}
\def\Nv{N_\text{v}}

\newcommand{\fg}{\ensuremath{f_g}}
\newcommand{\fgtil}{\ensuremath{\tilde{f}_g}}
\newcommand{\fgcrit}{\ensuremath{f_{g\,,\mathrm{crit}}}}

\newcommand{\fgtilLit}{\ensuremath{\tilde{f}_{g,\,\mathrm{Lit.}}}}
\newcommand{\fgtilExp}{\ensuremath{\tilde{f}_{g,\,\mathrm{Exp.}}}}

\newcommand\ie{\mbox{\textit{i.\,e.}}\xspace}
\newcommand\cf{\mbox{c.\,f.}\xspace}
\newcommand\eg{\mbox{e.\,g.}\xspace}

\newcommand{\Regk}{\ensuremath{\mathcal{R}_k}}

\newcommand{\hypC}{\ensuremath{g_{\scriptstyle{Y}}}}
\newcommand{\hypCFP}{\ensuremath{g_{\scriptstyle{Y},\,*}}}

\newcommand{\hypCFPInt}{\ensuremath{g_{\scriptstyle{Y},\,*\,,\mathrm{int}}}}

\newcommand{\GN}{\ensuremath{G}}
\newcommand{\GNcrit}{\ensuremath{G}_{\mathrm{crit}}}

\newcommand{\CC}{\ensuremath{\lambda}}

\newcommand{\Int}{\int\!\!\!\mathrm{d}^4x\sqrt{g}\,} 
\newcommand{\Intb}{\int\!\!\!\mathrm{d}^4x\sqrt{\bar{g}}\,} 

\newcommand{\GFalpha}{\ensuremath{\alpha_{h}}}
\newcommand{\GFbeta}{\ensuremath{\beta_{h}}}

\newcommand{\RegParLit}{\ensuremath{a_{\mathrm{L}}}}
\newcommand{\RegParExp}{\ensuremath{a_{\mathrm{E}}}}

\newcommand{\RegParLitPMS}{\ensuremath{a_{\mathrm{L,\,PMS}}}}

\newcommand{\Regh}{\ensuremath{\mathrm{Reg}_h}}
\newcommand{\RegA}{\ensuremath{\mathrm{Reg}_A}}

\newcommand{\FTT}{\ensuremath{\mathrm{P}_{\mathrm{TT}}}}
\newcommand{\FZ}{\ensuremath{\mathrm{P}_{\mathrm{0}}}}
\newcommand{\FA}{\ensuremath{\mathrm{P}_{\mathrm{A}}}}

\begin{document}
	\title{Regulator and gauge dependence of the Abelian gauge coupling in asymptotically safe quantum gravity}

\author{Maksym Riabokon}
\email{maksym.riabokon@partner.kit.edu}
\affiliation{Institut f{\"u}r Theoretische Teilchenphysik (TTP),
Karlsruher Institut f{\"u}r Technologie (KIT), 76131 Karlsruhe, Germany.}

\author{Marc Schiffer}
\email{marc.schiffer@ru.nl}
\affiliation{High Energy Physics Department, Institute for Mathematics, Astrophysics, and Particle Physics, Radboud University, Nijmegen, The Netherlands}

\author{Fabian Wagner}
\email{f.wagner@thphys.uni-heidelberg.de}
\affiliation{Institute for Theoretical Physics, Heidelberg University, Philosophenweg 16, 69120 Heidelberg, Germany}

\begin{abstract}
    Both General Relativity and the Standard Model of particle physics are not UV complete. General Relativity is perturbatively non-renormalizable, while the Standard Model features Landau poles, where couplings are predicted to diverge at finite energies, \eg, in the Abelian gauge sector. Asymptotically safe quantum gravity may resolve both of these issues at the same time. 
    In this paper, we assess the systematic uncertainties associated with this scenario, in particular with the gravitationally induced UV-completion of the Abelian gauge sector. Specifically, we study the dependence of this qualitative feature, namely the existence of a UV-complete gauge sector, on unphysical choices like the gauge, and the regulator function. Intriguingly, in some scenarios, we find simultaneous points of minimal sensitivity relative to both the regulator and gauge parameters, which allow for a UV completion. This provides further indications that the simultaneous UV-completion of quantum gravity and matter via an asymptotically safe fixed point is a robust physical feature, and that physical quantities, like scaling exponents, can become independent of unphysical choices.
\end{abstract}

\maketitle

\section{Motivation}

The quantum nature of gravity remains one of the big unresolved mysteries of theoretical physics. Despite over a century of research, no consensus has emerged on a prevailing theory. Numerous candidate approaches exist, but the absence of experimental input makes it difficult to evaluate their respective merits.

Instead, theoretical arguments, \ie, arguments from consistency have to make do. For instance, a na\"ive quantisation of General Relativity (GR) along the lines of the Standard Model of particle physics (SM) leads to a perturbatively non-renormalizable theory, rendering it unpredictive. 
Existing approaches to quantum gravity propose different mechanisms for UV completions of GR.
 
Even without considering gravity, the SM by itself is UV-incomplete: although perturbatively renormalizable, some couplings, including the Abelian gauge coupling, diverge at finite energy scales, at so-called Landau poles \cite{Landau:1954nau,Landau:1955ip,Landau:1956zr,Yndurain:1991ez}. Avoiding these divergences requires removing the associated interactions at all energies, rendering the theory trivial \cite{Pomeranchuk:1956zz}, and phenomenologically inconsistent (see \cite{Wilson:1973jj} for a review). 

The existence of the Landau poles has been proven in dimensions $d>4$ \cite{Aizenman:1982ze}. In $d=4$ it has been confirmed by non-perturbative computations using lattice  \cite{Frohlich:1982tw,Luscher:1987ay,Gockeler:1997dn} and functional methods \cite{Gies:2004hy}. Thus, the triviality problem is not an artefact of perturbative techniques, but a true inconsistency of the theory.
 
The unification of the SM with gravity could also address the triviality problem. Estimates suggest that the Landau pole occurs at energies beyond the Planck scale ($\Lambda_L \sim 10^{34}$ GeV in the SM, reduced to $\Lambda_L \sim 10^{17}$ GeV in some supersymmetric extensions \cite{Yndurain:1991ez,Gockeler:1997dn}). This has two key implications: first, gravity cannot be ignored when assessing the consistency of the theory. Second, the triviality problem is expected in the deeply quantum-gravitational regime, necessitating a quantum theory of gravity valid at those scales to resolve it.

From another perspective, the absence of a Landau pole in the electroweak sector can serve as a nontrivial test for quantum-gravity models: any phenomenologically viable theory of quantum gravity has to resolve it and allow for finite Abelian gauge interactions to arise in the IR. It therefore poses an important phenomenological consistency test for any theory of quantum gravity.

There are different ways, in which theories of quantum gravity could resolve the triviality problem:
\begin{itemize}
	\item The gauge groups of the SM (together with gravity) may be unified in a larger framework, \eg a grand unified theory, whose gauge couplings do not have a Landau pole. This unification would be expected, for example, in string theory \cite{Polchinski:1998rq,Frasca:2021iip,Abel:2024twz}.
	\item If spacetime exhibits a minimal length or area, there may be a physical cut-off in momentum space, which makes it impossible to reach the Landau pole in a physical process. Fundamental discreteness (or uncertainty of length/area measurements) is central, \eg, to causal-set theory \cite{Bombelli:1987aa, deBrito:2023nie} or spinfoam models \cite{Perez:2012wv}.
	\item The graviational contribution to the running of gauge couplings may render the theory asymptotically free or asymptotically safe \cite{Daum:2009dn, Harst:2011zx, Folkerts:2011jz, Christiansen:2017gtg, Eichhorn:2017lry, Christiansen:2017cxa}, see \cite{Eichhorn:2019yzm} for a discussion of seeming discrepancies with perturbative computations. Similar to the minimal-length idea, this behavior leads to the absence of a fundamental scale in the deeply quantum regime, a hallmark of asymptotically safe quantum gravity (ASQG).
\end{itemize}

In this paper we concentrate on the last point. ASQG is based on an enhanced symmetry in the UV, namely quantum scale symmetry. In other words, the theory is UV-completed by an interacting fixed point in the gravitational sector, the Reuter fixed point \cite{Reuter:1996cp,Souma:1999at}. Unlike a perturbative quantization of GR, ASQG is expected to have only a finite number of relevant directions at the fixed point, requiring a finite number of free parameters in the IR and thus providing a predictive framework.

Evidence is mounting that such a fixed point indeed exists and that the number of free parameters in the gravitational sector is three \cite{Lauscher:2001ya, Reuter:2001ag, Lauscher:2002sq, Litim:2003vp, Codello:2006in, Machado:2007ea, Codello:2008vh, Benedetti:2009rx, Eichhorn:2009ah, Manrique:2010am, Eichhorn:2010tb, Groh:2010ta, Dietz:2012ic, Christiansen:2012rx, Rechenberger:2012pm, Falls:2013bv, Ohta:2013uca, Eichhorn:2013xr, Falls:2014tra, Codello:2013fpa, Christiansen:2014raa, Demmel:2015oqa, Gies:2015tca, Christiansen:2015rva, Ohta:2015fcu, Ohta:2015efa, Falls:2015qga, Eichhorn:2015bna, Gies:2016con, Denz:2016qks, Biemans:2016rvp, Falls:2016msz, Falls:2016wsa, deAlwis:2017ysy, Christiansen:2017bsy, Falls:2017lst, Houthoff:2017oam, Falls:2017cze, Becker:2017tcx, Knorr:2017fus, Knorr:2017mhu, DeBrito:2018hur, Eichhorn:2018ydy, Falls:2018ylp, Bosma:2019aiu, Knorr:2019atm, Falls:2020qhj, Kluth:2020bdv, Knorr:2021slg, Bonanno:2021squ, Baldazzi:2021orb, Sen:2021ffc, Mitchell:2021qjr, Knorr:2021iwv, Baldazzi:2021fye, Fehre:2021eob, Baldazzi:2023pep, Saueressig:2023tfy, Kawai:2023rgy}\footnote{The remaining free parameters are the cosmological constant, the gravitational constant and a combination of the couplings of operators quadratic in curvature.}, see \cite{Eichhorn:2017egq, Percacci:2017fkn, Reuter:2019byg, Pereira:2019dbn, Reichert:2020mja, Pawlowski:2020qer, Eichhorn:2022jqj, Eichhorn:2022gku, Saueressig:2023irs, Pawlowski:2023gym} for reviews. 
Long standing criticism of this scenario, such as the extension of studies in Lorentzian signature \cite{Fehre:2021eob, DAngelo:2023wje, Saueressig:2025ypi}, and unitarity \cite{Knorr:2019atm, Platania:2020knd, Knorr:2021niv,Fehre:2021eob, Platania:2022gtt,Pastor-Gutierrez:2024sbt}, are being addressed now, see also \cite{Donoghue:2019clr, Bonanno:2020bil}. 
There is also significant evidence for a fixed point when the SM matter content is coupled to gravity  \cite{Dona:2013qba, Meibohm:2015twa, Biemans:2017zca, Alkofer:2018fxj, Wetterich:2019zdo, Korver:2024sam}.

Intriguingly, the asymptotically safe fixed point appears to be near-perturbative in nature, see \cite{Falls:2013bv, Falls:2014tra, Falls:2017lst, Falls:2018ylp, Eichhorn:2018akn, Eichhorn:2018ydy, Eichhorn:2018nda, Kluth:2020bdv}. This important property suggests that calculations which study operators based on their canonical mass dimension can already produce reliable results.

There is also mounting evidence that ASQG can induce a UV completion of the SM \cite{Shaposhnikov:2009pv, Harst:2011zx, Eichhorn:2017ylw, Eichhorn:2017lry, Eichhorn:2018whv, Alkofer:2020vtb, Kowalska:2022ypk, Pastor-Gutierrez:2022nki}, and even provide predictions for SM parameters \cite{Shaposhnikov:2009pv, Harst:2011zx, Eichhorn:2017ylw, Eichhorn:2017lry, Eichhorn:2018whv}.

The main goal of this paper is to test the mechanisms for possible UV completions, focussing on the Abelian gauge sector. Here, quantum fluctuations of the metric add an anti-screening contribution to the scale-dependence of the Abelian hypercharge. This contribution dominates over the pure-matter screening contribution at small couplings, and therefore induces asymptotic freedom for the Abelian gauge coupling \cite{Daum:2009dn, Harst:2011zx, Folkerts:2011jz, Christiansen:2017gtg, Eichhorn:2017lry, Christiansen:2017cxa, Eichhorn:2019yzm}.

To assess the robustness of this possibility, we introduce free parameters which parameterize unphysical choices necessary in practical computations and, systematically vary these parameters. Understanding how robust the main qualitative feature, \ie, the possibility of a gravity-induced UV-completion of the Abelian hypercharge is from these unphysical choices, gives us crucial insights into whether ASQG can indeed resolve the Landau-pole problem in the Abelian gauge sector of the SM. 

Furthermore, we apply the principle of minimal sensitivity (PMS) \cite{Canet:2002gs, Canet:2003qd, Gies:2015tca, Duclut:2016jct, Balog:2019rrg, DePolsi:2020pjk, DePolsi:2022wyb, Baldazzi:2023pep}\footnote{See \cite{Stevenson:2022gcv} for a monograph in the context of perturbation theory.} to identify parameter values that minimize systematic uncertainties and ensure that results are least sensitive to regulator and gauge choices. A complementary approach to minimize unphysical dependences of the regularization procedure is to optimize the shape of the regulator, see \eg, \cite{Litim:2000ci, Litim:2001up, Pawlowski:2003hq}.

We indeed find such PMS-points when when evaluating them on their respective (parameter dependent) fixed-point values for a minimal matter content, \ie, one scalar, one fermion and one vector. These points suggest that a viable resolution of the triviality problem may be within reach. We then study how these PMS-points change upon addition of more matter fields, and find that generically scalars and gauge fields stabilize the presence of PMS-points, while fermions remove them.

This paper is structured as follows: In Sec. \ref{sec:HyperchargeASQG} we describe the mechanism by which the triviality problem may be resolved in ASQG. In Sec.~\ref{sec:Setup} we introduce the tools we use to extract the scale-dependence of the Abelian gauge coupling under the impact of gravitational fluctuations, in particular, the functional renormalization group (FRG). In Sec.~\ref{sec:RegDepWithCC} we study the regulator and gauge dependence of the critical exponent of the gauge coupling, which is closely linked to the resolution to the triviality problem. In Sec.~\ref{sec:Discussion} we summarize and interpret our findings.

 \section{The Abelian hypercharge in asymptotically safe quantum gravity\label{sec:HyperchargeASQG}}
There is evidence that ASQG  may induce a UV-completion of the Abelian gauge sector of the SM. This UV completion is induced by an anti-screening contribution of gravitational fluctuations to the scale dependence of the Abelian hypercharge $\hypC$ as
 \begin{equation}
 	\label{eq: betagyschem}
 	\beta_{\hypC}=-\fg\,\hypC+ \beta_{\mathrm{SM}}\,\hypC^3 + \mathcal{O}(\hypC^5) \,,
 \end{equation}
where $\beta_{\mathrm{SM}}$ is the one-loop contribution from charged matter, which is screening, \ie, $\beta_{\mathrm{SM}}>0$, and where $\fg$ is the gravitational contribution, parameterized by the Newton coupling and the cosmological constant.
In ASQG this contribution is constant at high energies \ie $\fg\approx \mathrm{const}$ for $k^2>M_{\mathrm{Planck}}$, while $\fg\approx 0$ for $k^2<M_{\mathrm{Planck}}$, due to the scale-dependence of the gravitational couplings. 

As Eq.~\eqref{eq: betagyschem} reveals, different signs of $\fg$ separate different scenarios: if $\fg<0$, the gravitational contribution adds an additional screening term, which worsens the Landau-pole problem by shifting the scale of divergence to lower energies. If $\fg=0$, the gravitational contribution simply vanishes, such that the Landau-pole problem of the SM remains unchanged. Hence, if $\fg\leq0$, ASQG alone cannot provide a UV completion of the SM. If however, $\fg>0$ holds, then gravitational fluctuations add an anti-screening contribution to the scale dependence of  $\hypC$, which dominates for small enough $\hypC$. Hence, the Gaussian fixed point $\hypCFP=0$ becomes IR repulsive, such that finite values of $\hypC$ can be reached at low scales, while $\hypCFP=0$ is realized in the UV. Note that for the SM, this is not necessarily a sufficient requirement, as $\fg$ might have to be larger than a critical value to satisfy observational consistency with the Abelian hypercharge sector, see below for details.

The stability properties of fixed points are encoded in critical exponents, defined as
\begin{equation}
\Theta_i=-\mathrm{Eig}\left(\frac{\partial \beta_{g_i}}{\partial g_j}\right)\bigg|_{g_i=g_{i,\,*}}\,,
\end{equation}
which determine the directions in which a fixed point is attractive or repulsive under flows towards the IR. In this convention, $\Theta_i>0$ corresponds to IR repulsive, so-called relevant directions. If the fixed point UV-completes the theory, relevant directions come with free parameters that need to be fixed by experiment. Conversely, $\Theta_i<0$ corresponds to IR-attractive directions. Those directions do not come with a free parameter and their IR-value is a prediction of the UV-completion of the theory. In a simple approximation, the gravitational contribution $\fg$ \textit{is} the critical exponent of the Abelian gauge coupling $\hypC$ at the Gaussian fixed point $\hypCFP=0$. Therefore, $\fg>0$ indicates asymptotic freedom and a free parameter, while $\fg<0$ provides a prediction of the theory, which, in the case of the Gaussian fixed point $\hypCFP=0$ corresponds to $\hypC=0$ at all scales.

\begin{figure}
\centering
	\includegraphics[width=.9\linewidth]{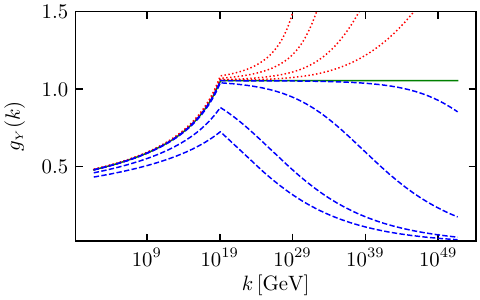}
	\caption{The Abelian hypercharge $\hypC(k)$ in ASQG with anti-screening gravitational contribution $\fg$, see \eqref{eq: betagyschem}. Any blue (dashed) trajectories emanate from the asymptotically free fixed point. Any trajectory emanating from that fixed point is bounded from above by the green (solid) trajectory emanating from the interacting fixed point. Any trajectory above the green one (red, dotted) is not UV-complete.}
	\label{fig:gytraj}
\end{figure}

Besides rendering the Abelian hypercharge asymptotically free, an anti-screening gravitational contribution $\fg>0$ can give rise to an additional fixed point $\hypCFPInt>0$, which is IR attractive. This fixed point, if realized, gives rise to one unique trajectory towards the IR, along which the value $\hypC(k)$ is a prediction of the underlying UV completion at all scales $k$ \cite{Harst:2011zx, Eichhorn:2017lry}, see \autoref{fig:gytraj} for an illustration.

The predicted value of $\hypC$ from the asymptotically safe fixed point is compatible with the observed one, within the estimated systematic uncertainty in determining $\fg$ \cite{Eichhorn:2017lry}.

Besides generating  a predictive trajectory, the interacting fixed point $\hypCFPInt$ also shields trajectories emanating from $\hypCFP=0$ from becoming too large, due to its IR-attractive nature. Hence, $\hypCFPInt$ and the trajectory emanating from it act as an upper bound for any $\hypC(k)$ embedded in a UV-complete quantum field theory, see \autoref{fig:gytraj} for an illustration.

For ASQG to be consistent with low-energy observations, $\hypCFPInt$ has to be large enough to accommodate the observed low-energy value. In the following, we will focus on the size of $\fg$, which in our approximation determines $\hypCFPInt$. In particular, observational consistency requires that
 \begin{equation}
 	\fg\geq\fgcrit=  \frac{0.096}{\pi^2}\,,
 \end{equation}
 see \cite{Eichhorn:2017lry}.

Let us briefly comment on the scheme-dependence of the gravitational contribution $\fg$: While $\fg$ is the critical exponent of $\hypC$ at the Gaussian fixed point\footnote{This holds at leading order, \ie, when no induced interactions are taken into account, see also \cite{YukToApp} for a similar study in the Yukawa sector beyond leading order.}, and should therefore be universal, this is not necessarily the case  in small truncations, and when not evaluated at the gravitational fixed point. In particular, even in perturbative studies both $\fg=0$ and $\fg>0$ are possible, depending on the regularization scheme \cite{Robinson:2005fj, Pietrykowski:2006xy, Toms:2007sk, Ebert:2007gf, Toms:2010vy, Anber:2010uj}. Similarly, studies using the FRG indicate that $\fg\geq0$ holds \cite{Daum:2009dn, Harst:2011zx, Folkerts:2011jz, Christiansen:2017gtg, Eichhorn:2017lry, Christiansen:2017cxa, Eichhorn:2019yzm}, yet regulators that yield $\fg=0$ can be constructed, too \cite{Folkerts:2011jz}, see also \cite{Pastor-Gutierrez:2022nki} for an explicit implementation. On the other hand, \cite{deBrito:2022vbr} argues that $\fg>0$ arises at the gravity-matter fixed point in a limit where the FRG-regulator vanishes. Since $\fg\propto\GN$,  $\fg>0$ arises from carefully considering both the gravitational contributions to $\beta_{\hypC}$, but evaluating it at the asymptotically safe fixed point for $\GN$, which is also regulator-dependent.

 In this paper we will further study the dependence of $\fg$ on both regulator -- by computing it using several one-parameter families of regulators -- and gauge fixing. In particular, we will employ a PMS on $\fg$ to determine the regulator which minimizes residual unphysical dependencies for each of the families. The dependence of $\fg$ on the individual parameters, as well as the dependence of $\fg$ on the different regulator families evaluated at their PMS-point yields insights into the regulator-dependence, and ultimately, on the question of observational consistency of  and systematic uncertainties in ASQG. 

 \section{Technical setup\label{sec:Setup}}
Our goal is to extract the scale dependence of $\hypC$, and in particular the gravitational contribution $\fg$ employing the FRG \cite{Wetterich:1992yh}. The FRG is based on the scale-dependent effective action $\Gamma_k$ and implements the Wilsonian idea of integrating out modes according to their momentum shell. In this sense, $\Gamma_k$ contains all quantum fluctuations with momenta $p^2 \gtrsim k^2$, and lowering $k\to k-\delta k$ integrates out modes with $p^2\approx k^2$. Therefore, $\Gamma_k$ interpolates between a quantity akin to a classical action\footnote{The UV fixed-point action amounts to the classical action up to an operator determinant, yielding the reconstruction problem \cite{Manrique:2009tj,Morris:2015oca,Fraaije:2022uhg}.} for $k\to\infty$, when no quantum fluctuations are integrated out and the full quantum 1-PI effective action $\Gamma_{k\to0}=\Gamma_{1\mathrm{PI}}$ at $k\to0$, when all quantum fluctuations are integrated out. 

This interpolation is implemented via a regulator function $\Regk(p^2)$, which enters as a mass-like term (\ie, quadratically in the fields) in the path integral. The regulator has to satisfy
 \begin{alignat}{1}
 	\Regk(p^2)\,\,&
 		\begin{cases}
 			= 0 \quad & \mathrm{for}\,\,p^2\gtrsim k^2\,,\\
 			> 0 &  \mathrm{for}\,\,p^2\lesssim k^2\,,\\
 		\end{cases}
 		\\
 	 \mathrm{and}\notag\\
 		\Regk(p^2)&\to
 		\begin{cases}
 			0 \quad & \mathrm{for}\,\,k^2\to0\,,\\
 			\infty &  \mathrm{for}\,\,k^2\to \Lambda \to \infty\,,\\
 		\end{cases}
 \end{alignat} 
 where $\Lambda$ is a UV-cutoff which can be removed at an RG fixed point. Due to these properties, the regulator adds a positive mass for modes with $p^2<k^2$, and hence suppresses low-momentum modes in the path integral, while modes with $p^2>k^2$ are not suppressed and hence integrated out.
 
 The FRG provides a differential equation for the scale-dependent effective action $\Gamma_k$, which reads \cite{Wetterich:1992yh, Morris:1993qb, Ellwanger:1993mw, Reuter:1996cp}
 \begin{equation}
 	\label{eq: floweq}
 	\partial_t\,\Gamma_k=\frac{1}{2}\mathrm{STr}\left[\partial_t \Regk \left(\Gamma_k^{(2)}+\Regk\right)^{-1} \right]\,,
 	\end{equation}
 where $\partial_t=k\partial_k$ is the dimensionless scale derivative, $\Gamma_k^{(2)}$ refers to the second functional derivative of $\Gamma_k$ with respect to all fields, and the $\mathrm{STr}$ refers to a trace over all discrete indices, and an integration over the loop momentum, while adding additional signs for Grassmann-valued fields. Beta functions can be extracted from the flow equation by projecting onto the desired field monomial. For a detailed review on the FRG and its applications, see, \eg, \cite{Dupuis:2020fhh}.
 
 The flow equation \eqref{eq: floweq} is in principle exact, but practical computations require to limit the set of operators included in $\Gamma_k$ to a subset of the full theory. Upon these truncations of the theory, the flow equation is not closed any more, and even physical quantities, like critical exponents depend on unphysical choices, like the truncation, gauge-fixing, and the chosen regulator function $\Regk$. Systematic extensions of the truncation are required to assess the robustness of the results in a given truncation.
 
 In this work we are interested in the impact of gravitational fluctuations on the Abelian gauge sector. To include the impact of gravitational fluctuations, we employ the background-field method, where we decompose the full metric into a background metric and fluctuations, \ie,
 \begin{equation}
     g_{\mu\nu}=\bar{g}_{\mu\nu}+h_{\mu\nu}\,.
 \end{equation}
 In principle the background metric does not have to be specified, but specific choices are convenient for practical computations.
Our ansatz for $\Gamma_k$ reads
 \begin{equation}
 	\Gamma_{k}=\Gamma_{k}^{\mathrm{grav}}+\Gamma_{k}^{U(1)}\,,
 \end{equation}
 and we will approximate the dynamics of the gravitational sector by the Einstein-Hilbert action, \ie,
 \begin{equation}
 	\Gamma_{k}^{\mathrm{grav}}=\frac{1}{16\pi\, \GN\,k^{-2}}\Int \left[-R+2\CC\,k^{2}\right]+\Gamma_k^{\mathrm{gf},\,h}\,,
 \end{equation}%
 where $\GN$ and $\CC$ are  the scale-dependent but dimensionless versions of the Newton coupling and the cosmological constant, respectively. Furthermore, $\Gamma_k^{\mathrm{gf},\,h}$ is the gauge-fixing action given by
\begin{equation}
	\label{eq: gf}
	\Gamma_k^{\mathrm{gf},\,h}=\frac{1}{32\pi\,\GN\,\GFalpha}\Intb\bar{g}^{\mu\nu}\mathcal{F}_{\mu}\,\mathcal{F}_{\nu}\,,
\end{equation}
with gauge condition
\begin{equation}
	\mathcal{F}_{\mu}=\left(\delta_{\mu}^{(\alpha}\bar{D}^{\beta)}-\frac{1+\GFbeta}{4}\,\bar{g}^{\alpha\beta}\bar{D}_{\mu}\right)h_{\alpha\beta}\,.
\end{equation}
Here $\bar{D}$ denotes the covariant derivative with respect to the background $\bar{g}$, and $\GFalpha$ and $\GFbeta$ are gauge parameters. In this work, we choose $\GFalpha\to0$, which corresponds to Landau gauge, throughout the paper, as it is a fixed-point for both gauge parameters \cite{Litim:2002ce, Knorr:2017fus}, but leave $\GFbeta$ general.

In the following, we will expand gravitational fluctuations around a flat background, \ie, $\bar{g}=\delta$, and parameterize metric fluctuations via a linear split
\begin{equation}
	g_{\mu\nu}=\delta_{\mu\nu}+\sqrt{16\pi k^{-2}\GN\,Z_{h}}\,h_{\mu\nu}\,,
\end{equation}
where $Z_h$ is the graviton-wavefunction renormalization.
The gauge fixing \eqref{eq: gf} also gives rise to Fadeev-Popov ghosts, which only contribute directly to the running of gravitational couplings.

We approximate the dynamics of the Abelian gauge field as
\begin{equation}
	\label{eq:gammakmat}
	\Gamma_{k}^{U(1)}=\frac{Z_A}{4} \Int 	g^{\mu\nu}g^{\rho\sigma}F_{\mu\rho}F_{\nu\sigma} +\Gamma_k^{\mathrm{gf},\,A}\,,
\end{equation} 
where $F_{\mu\nu}=D_{\mu}A_{\nu}-D_{\nu}A_{\mu}$ is the field strength tensor of the Abelian gauge field $A_{\mu}$, and where $\Gamma_{\mathrm{gf},\,A}$ is the gauge-fixing action of the Abelian gauge field, which reads
\begin{equation}
	\Gamma_k^{\mathrm{gf},\,A}=\frac{1}{2\,\alpha_A} \Intb\left(\bar{D}^{\nu}A_{\nu}\right)\left(\bar{D}^{\mu} A_{\mu}\right)\,,\quad \alpha_A\to 0\,,\label{eq:GFEffAction}
\end{equation}
where again we choose Laundau gauge for the associated gauge parameter $\alpha_A$. 
The minimal coupling between the Abelian gauge field and gravity arises naturally via the inverse metric and $\sqrt{g}$ in Eq. \eqref{eq:gammakmat}. This minimal coupling ultimately gives rise to the gravitational contribution $\fg$ in \eqref{eq: betagyschem}. To extract the scale dependence of $\hypC$, we employ Ward identities which state \cite{Ward:1950xp,Takahashi:1957xn}\footnote{The pure-matter contribution to $\beta_{\hypC}$ is universal at one-loop and can therefore also be read-off from gauge-scalar and gauge-fermion vertices. The gravitational contribution $\fg$ does not need to be universal, see \cite{Eichhorn:2017lry,Pastor-Gutierrez:2022nki, GustavoInPrep} for a comparison.}
\begin{equation}
	\beta_{\hypC}=\frac{\eta_{A}}{2}\,\hypC\quad\Rightarrow\fg=-\frac{\eta_{A}\big|_{\mathrm{grav}}}{2}\,,
\end{equation}
where $\eta_A$ is the anomalous dimension of the gauge field,
\begin{equation}
	\eta_i=-\frac{\partial_t\,Z_i}{Z_i}\,,
	\end{equation}
and where $\eta_{A}\big|_{\mathrm{grav}}$ refers to the gravitational contribution only, \ie, omitting pure-matter diagrams.

Finally, for the regulator $\mathcal{R}$ we use
\begin{equation}%
	\mathcal{R}_k(p^2)=\frac{k^2}{p^2} \,\,r_k\!\left(\frac{p^2}{k^2}\right)\,\Gamma_k^{(2)}\bigg|_{\CC=0,h_{\mu\nu}=0,A_{\mu}=0}\,.
\end{equation}\\%
Here, $r_k$ denotes the shape function. For the purpose of the present paper, we work with the Litim-type \cite{Litim:2000ci, Litim:2001up} and exponential shape functions
\begin{alignat}{1}
	\label{eq: Litim}
\mathrm{Litim}\quad	r_k(z)&=\RegParLit\,(1-z)\Theta(1-z)\,,\\
	\label{eq: Exp}
\mathrm{Exponential}\quad	r_k(z)&=\RegParExp\,\frac{z}{e^z-1}\,,
\end{alignat}
with the free parameters $\RegParLit$, and $\RegParExp$, see \cite{Balog:2019rrg}.  These free parameters govern  which momentum shell is integrated out when lowering $k\to k-\delta k$. The regulator-induced effective mass, \ie, the IR cut-off, increases monotonically with increasing $\RegParLit,$ $\RegParExp.$

Extracting $\eta_A$ in this setup leads to \cite{Daum:2009dn, Harst:2011zx, Folkerts:2011jz, Christiansen:2017gtg, Eichhorn:2017lry, Christiansen:2017cxa, Eichhorn:2019yzm}
\begin{alignat}{1}
	\eta_A=\frac{\GN}{3\pi}\int\!\!&\mathrm{d}q^2\,q^2 \,\notag\\
	\times\,\bigg[& \Regh \left(-5 \FTT^2+\frac{2\,\GFbeta^2}{(\GFbeta-3)^2}\FZ^2\right)\notag\\
	&+\RegA\,\FA^2\left(5\FTT-  \frac{2\,\GFbeta^2}{(\GFbeta-3)^2}\FZ\right)\notag\\
	&+\Regh\,\FA\left(5\FTT^2-\frac{2\,\GFbeta^2}{(\GFbeta-3)^2}\FZ^2\right)\bigg]\,,
	\label{eq:etaAnonint}
\end{alignat}
where we have summarized the regulator contributions in the denominator coming from the $\partial_t \Regk$ insertion in \eqref{eq: floweq} as 
\begin{align}
	\label{eq:regfct}
	\mathrm{Reg}_{\Phi}=(-2+\eta_{\Phi})\,\,r_k\left(\frac{q^2}{k^2}\right) +2 \frac{q^2}{k^2}\, r'_k\left(\frac{q^2}{k^2}\right)\,,&&\Phi=h,A,
\end{align}
and where we have introduced the three propagator structures as
\begin{alignat}{1}
	\FA=&\frac{1}{q^2+k^2\,r_k\left(\frac{q^2}{k^2}\right) }\,,\\
	\label{eq:propTT}
	\FTT=&\frac{1}{q^2+k^2\,r_k\left(\frac{q^2}{k^2}\right) -2\CC\,k^2}\,,\\
	\FZ=&-\frac{1}{k^2\,r_k\left(\frac{q^2}{k^2}\right)+4\CC\,k^2\frac{(\GFbeta^2-3)}{(\GFbeta-3)^2}}\,,
	\label{eq:propz0}
\end{alignat}
which correspond to the regularized propagator of the gauge field, the transverse-traceless mode of metric fluctuations, and the scalar mode of metric fluctuations, respectively, see \cite{Knorr:2021niv}. Here, the loop integration over $q^2$ is performed explicitly after specifying the regulator function $r_k$.

In the following, we will employ a perturbative approximation, where we neglect the anomalous dimensions coming from the regulator insertion, \ie, we set $\eta_{\Phi}=0$ in \eqref{eq:regfct}. This approximation is valid as long as all anomalous dimensions remain small enough, which holds in our present analysis, as $\fg$ itself remains small (in particular $\fg<1$), as we will see.
 
 \section{Regulator dependence of the Abelian gauge-gravity system\label{sec:RegDepWithCC}}
We now study the regulator dependence of $\fg$ at the asymptotically safe fixed point. The beta functions of the gravity-gauge system including regulator and gauge dependence up until now have only been published for $\CC=0$ \cite{deBrito:2022vbr}, and for minimal matter content. Here, we add the scale-dependence of the cosmological constant, as well as the general dependence on gauge parameters and matter content.\footnote{We thank the authors of \cite{deBrito:2022vbr} for providing us with the more general form of the beta-function, and in particular G. de Brito for the beta-functions at $\CC\neq0$, which have not been published yet.}
With these beta functions, we analyse the critical exponents of the system evaluated at the Reuter fixed point, and identify PMS-points. We start with a simple matter system with $\Nf = \Ns = \Nv = 1$, and then study changes to the results upon adding additional matter fields.

\subsection{Fixed-point study for a minimal matter content}
\label{sec:minmatt}

 \begin{figure}
	\includegraphics[width=\linewidth]{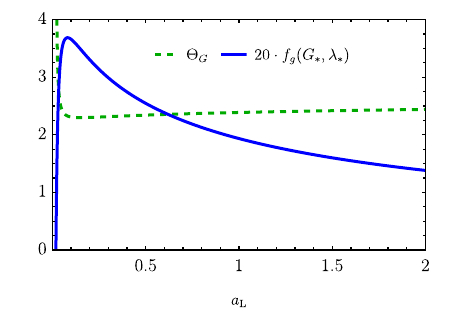}
	\caption{We show the gravitational critical exponent $\Theta_G$ and the gauge critical exponent $\fg(g^*,\lambda^*)$ at the Reuter fixed point for the Litim-type shape function as a function of the regulator parameter $\RegParLit$ for $\GFbeta=1$.}
	\label{fig:LitimCritExpSB0}
\end{figure}

Structurally, in the current approximation $\beta_{\GN}$ and $\beta_{\CC}$ are independent of $\hypC$. Therefore, the eigendirections of the stability matrix do not mix the gravitational couplings with the gauge coupling. Accordingly, one eigendirection, which corresponds to $\fg$, is fully aligned with $\hypC$, while the other two eigendirections mix between the two gravitational couplings.
In particular, their critical exponents form a complex pair, such that we are left with two independent, regulator-dependent critical exponents, $\Theta_{\GN}$ and $\fg$.\footnote{Strictly speaking, the critical exponents will only form a complex pair for $a_i\lesssim2$, after which they split up. However, in the regime beyond the split of the complex pair, there are no additional PMS-points, at least for the gauge- and regulator choices we explicitly checked.} 

In systems with various critical exponents, one generally must decide to which of the critical exponents the PMS should be applied. One could argue that the most relevant (\ie the largest positive) critical exponent is the one that has most impact on the physics of the system and hence minimizing its regulator dependence has the largest effect. Following this argument, we would apply the PMS to $\Theta_{\GN}$. Conversely, in systems with multiple types of degrees of freedom — such as the gauge–gravity system at hand — it is less clear which ones dominate the physical behavior. In particular, in our system, changes to $\fg$ can change qualitative aspects of the fixed point, such as its overall predictivity, and the availability of a solution to the triviality problem, while changes in $\Theta_{\GN}$ will mostly have a quantitative impact on the system. 

In the following we remain agnostic  on the choice of PMS-points, and apply the PMS to both $\fg$ and $\Theta_{\GN}$ to investigate their respective properties, such as gauge dependence. We study the Litim-type shape function and the exponential shape function separately. \\

 \begin{figure}
	\includegraphics[width=\linewidth]{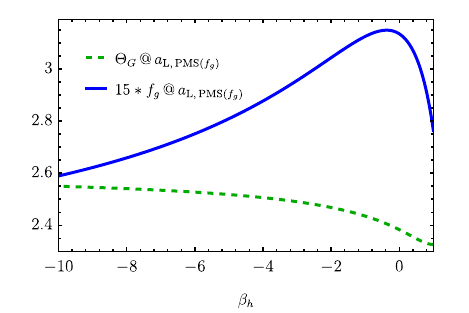}
	\includegraphics[width=\linewidth]{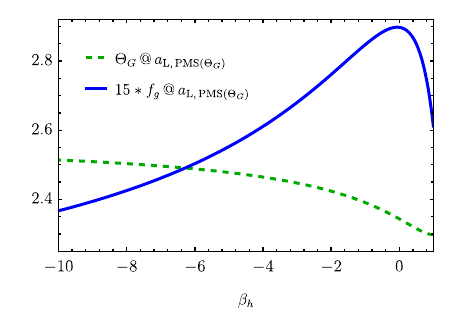}
	\caption{We show the critical exponents at the PMS-point for $\fg$ (upper panel) and $\Theta_G$ (lower panel) with Litim-type shape function.}
	\label{fig:LitPMSfgbetdepCE}
\end{figure}

\paragraph{Litim-type shape function}\textcolor{white}{.}\\[3pt]
In the gauge-gravity system for the Litim-type shape function, we find the Reuter fixed point for all values of $\RegParLit$. In \autoref{fig:LitimCritExpSB0} we display the real part of the critical exponents of the Newton coupling ($\Theta_G$) and the gauge coupling (\fg) as functions of $\RegParLit,$ and for $\GFbeta=1$. Both critical exponents exhibit individual PMS-points, which albeit close, are not at coincident values of $\RegParLit$. While for $\Theta_G$ the PMS-point is a global minimum of $\Theta_G(\RegParLit)$, the PMS-point for $\fg$ is a global maximum.

As discussed above, in models with multiple PMS-points, those points have to be compared by other means than the PMS-criterion. 
As we gather from \autoref{fig:LitimCritExpSB0}, while the gravitational critical exponent appears stable along a rather wide range of values for $\RegParLit,$ $\fg$ is not. 

\begin{figure}
 	\includegraphics[width=\linewidth]{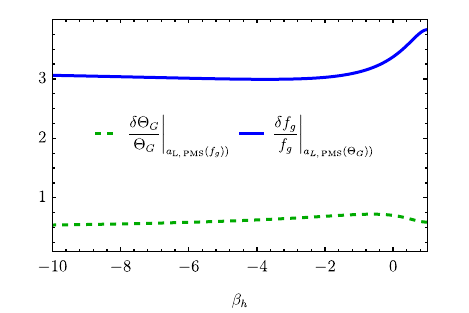}
 	\caption{We show the relative deviation of the critical exponents $\Theta_G$ and $\fg$, when evaluated at the PMS-point of $\RegParLit$ of the respective other critical exponent.}
 	\label{fig:LitPMSbetdepDCEvsCE}
 \end{figure}

To further investigate the gauge dependence of our result, we plot the two critical exponents $\fg$ and $\Theta_G$ evaluated at the two PMS-points as functions of the gauge parameter $\GFbeta$ in \autoref{fig:LitPMSfgbetdepCE}. We find, that $\fg$ evaluated at the PMS-points for both $\fg$ itself and for $\Theta_G$ have additional non-coincident PMS-points for the gauge dependence. The local maxima of $\fg|_{\RegParLitPMS}$ are located at $\GFbeta\approx-0.37$, and $\GFbeta\approx-0.06$, respectively, see \autoref{fig:LitPMSfgbetdepCE}. This is close to the local maximum of $\fg$, when treating $\GN$ and $\CC$ as input parameters, see appendix \ref{app:GravCouplAsPar}. Thus, apart from minimizing the regulator dependence of either $\Theta_G$ or $\fg$, we can further minimize the gauge dependence of $\fg$. We also find that there is no value $\GFbeta$ for which the PMS-points with respect to $\RegParLit$ for $\Theta_G$ or $\fg$ coincide.
To see whether qualitative changes are expected from applying the PMS on one critical exponent over the other, we compare the relative changes of $\Theta_G$ and $\fg$ at the PMS-points of their respective counterparts, \ie $\mathrm{d}\log\Theta_i/\mathrm{d}\RegParLit(\RegParLitPMS(\Theta_j))$ with $\Theta_{i,j}\in(\Theta_G,\fg)$,
as functions of the gauge parameter $\GFbeta$, see \autoref{fig:LitPMSbetdepDCEvsCE}. While we find a relative change of more than $300\%$ for $\fg$ at the PMS of $\Theta_G$,  it stays below $80\%$ for $\Theta_G$ at the PMS relative to $\fg$.  Hence, one could argue that the system at the PMS-point of $\fg$ is less regulator dependent overall than at the PMS-point of $\Theta_G$. \\
 	
\begin{figure}
	\includegraphics[width=\linewidth]{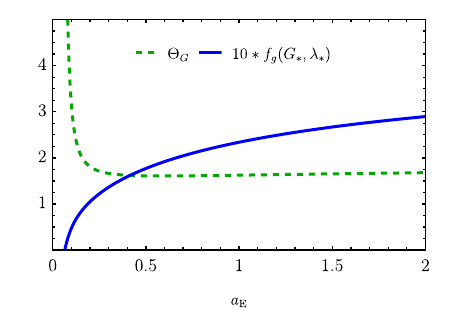}
	\caption{We show the gravitational critical exponent $\Theta_G$ and the gauge critical exponent $\fg(\GN_*,\CC_*)$ for the exponential shape function with minimal matter content at the Reuter fixed point as a function of the regulator parameter $\RegParExp$ at $\GFbeta=1$.}
	\label{fig:ExpCritExpSB1}
\end{figure}

\paragraph{Exponential shape function}\textcolor{white}{.}\\[3pt]
As for the Litim-type shape function, we find the Reuter fixed point for all values of $\RegParExp$. In \autoref{fig:ExpCritExpSB1} we show the two critical exponents $\Theta_G$ and $\fg$ for $\GFbeta=1$ as a function of the regulator parameter $\RegParExp$. In contrast to the case of the Litim-type shape function, $\fg$ does not feature a PMS-point, but increases towards $\RegParExp\to\infty$. However, $\Theta_G$ still features a local minimum, where the regulator dependence can be minimized. Hence, in contrast to the Litim-type shape function, for the exponential shape function, the choice for a PMS is unambiguous.  

\begin{figure}
	\includegraphics[width=\linewidth]{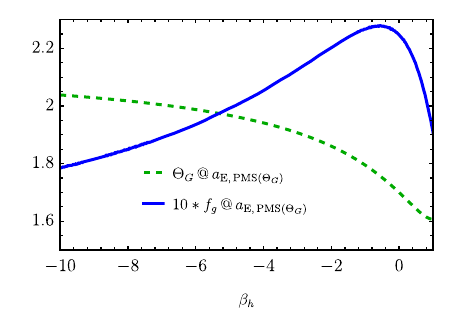}
	\caption{We show the critical exponents for minimal matter content at the PMS-point for $\Theta_G$ with exponential shape function as a function of $\GFbeta$.}
	\label{fig:ExpPMSThetGbetdepCE}
\end{figure}

We show both critical exponents evaluated at the PMS-value for $\RegParExp$ in \autoref{fig:ExpPMSThetGbetdepCE} as a function of the gauge-fixing parameter $\GFbeta$. As for the Litim-type shape function, we find a local maximum of $\fg$, where both  residual regulator and gauge-dependence of $\fg$ can be minimized. The gauge-parameter that maximizes $\fg$ is $\GFbeta\approx-0.56.$\\

In a nutshell, despite some differences between both shape-functions, we find a PMS-point for both for at least one critical exponent. Further, we find that the gauge-dependence of $\fg$ evaluated at this PMS-point for the regulator parameter $a_i$ can also be minimized. In particular, both the PMS-point for the regulator parameter $a_i$ and the PMS-point for the gauge-parameter $\GFbeta$ are local maxima of $\fg$. Notably, the gauge-parameter that maximizes $\fg$ is comparable between both shape functions, and to the case where $\GN$ and $\CC$ are treated as input parameters, see appendix \ref{app:GravCouplAsPar}. In light of a possible gravitational UV-completion of the Abelian gauge sector, this is important, since $\fg$ sets an upper bound on the low-energy value of the Abelian hypercharge that can be reached from a UV-completion, see the discussion in \autoref{sec:HyperchargeASQG}. 

The result that $\fg$ increases by minimizing residual gauge and regulator parameters provides tentative evidence that ASQG could offer a viable UV completion of the Abelian gauge sector. In the following, we will study how this feature changes under the impact of additional matter fields, to ultimately study the physically relevant case of SM-matter content.

\subsection{Adding matter degrees of freedom}  \label{sec:SMmatt}

For both shape functions, the Reuter fixed point persists in the presence of SM matter ($\Nf=22.5,$ $\Ns=4$ and $\Nv=12$). However, in both cases it features a large graviton-anomalous dimension (at least for $\alpha_i=1$), 
indicating that the current truncation might not be reliable. In particular, studies lifting the background-field approximation find a gravitational fixed point with SM-matter content and small anomalous dimensions, see \cite{Pastor-Gutierrez:2022nki}, and \cite{Eichhorn:2022gku} for a discussion of differences. To obtain more conclusive results, future studies should extend the truncation or go beyond the background-field approximation.

While these are relevant points, here we focus on the qualitative effect which matter has on the PMS-point and the resulting critical exponent. For completeness, we report the SM results in appendix \ref{app:SMmatt}, but caution, regarding their robustness. As before, we examine both shape functions individually, followed by a comparison of shared features.\\

\paragraph{Litim-type shape function}\textcolor{white}{.}\\[3pt]
We begin by examining how the inclusion of bosons and fermions alters the picture compared to minimal matter content, \ie, $\Ns=\Nf=\Nv=1$. We focus on $\GFbeta=1$.

When adding scalars and gauge fields to the system, we observe several changes: i) the maximum of $\fg$ in \autoref{fig:LitimCritExpSB0} is shifted towards larger $\RegParLit$, and the curve flattens such that $a_{\mathrm{Litim, PMS}}(\fg)$ increases, while the value $\fg(a_{\mathrm{Litim, PMS}})$ decreases; ii) the minimum of $\Theta_G$ in \autoref{fig:LitimCritExpSB0} moves towards larger $\RegParLit$ and the whole function is lifted such that $a_{\mathrm{Litim, PMS}}(\Theta_G)$  and $\Theta_G(a_{\mathrm{Litim, PMS}})$ increase together. In short, both PMS-points shift to larger $\RegParLit$, but the local extrema become more shallow; iii) the slope of $\Theta_G$ for large $\RegParLit$ is reduced; iv) the value of $\RegParLit$, where the gravitational critical exponents cease to form a complex pair, increases. 

Conversely, when adding fermions,\footnote{Specifically we add fermions at $\Ns=1$, and $\Nv=4$, where the Reuter fixed point exists for several fermions.} we observe the exact opposite effects, namely that i) local extrema of both $\Theta_G$ and $\fg$ in \autoref{fig:LitimCritExpSB0} are shifted to smaller $\RegParLit$ such that $a_{\mathrm{Litim, PMS}}$ is reduced for both critical exponents; ii) both curves in \autoref{fig:LitimCritExpSB0} are getting more pronounces such that local extrema are steeper. Already for $\Nf=2$, both PMS-points disappear, \ie, they are shifted beyond $\RegParLit\to0$. Furthermore, the gravitational critical exponents are no longer complex pairs for $\Nf\geq3$ for any value of $\RegParLit$. \\[12pt]

\paragraph{Exponential shape function}\textcolor{white}{.}\\[3pt]
We begin again with minimal matter content, $\Ns = \Nf = \Nv = 1$, where the gravitational critical exponents form a complex pair. The real part exhibits a PMS-point, see \autoref{fig:ExpCritExpSB1}. As before, we focus on $\GFbeta=1$

When increasing the number of scalars beyond the minimal matter content, i) the steep decrease of the critical exponent of the gauge coupling $\fg$ at small $\RegParExp$ in \autoref{fig:ExpCritExpSB1} is moved towards larger $\RegParExp$ while the overall value of $\fg$ decreases, \ie the curve $\fg(\RegParExp)$ shifts towards large $\RegParExp$ and downwards; ii) the pole of $\Theta_G$ at small $\RegParLit$ and the following minimum in \autoref{fig:ExpCritExpSB1} move to larger $\RegParExp$, and the overall curve becomes flatter. Finally, at $\Ns=2,$ the local extremum disappears.

Gauge fields have a qualitatively similar effect to scalars, but the shift of both $\Theta_G$ and $\fg$ is larger than for scalars.

In contrast, when increasing the number of fermions, i) the slope of $\fg$ at small $\RegParExp$ decreases, and $\fg$ itself slightly increases, while for large $\RegParExp$, $\fg$ remains unchanged; ii) the steep section of $\Theta_G$ moves to smaller $\RegParExp$, and the curve becomes more pronounced, deepening the local  minimum. For large $\RegParExp$, $\Theta_G$ remains unchanged. Between $\Nf=1.1$ and $\Nf=1.2$ the complex pair splits up; iii) The critical exponent mostly aligned with $\CC$ completely flattens and moves towards $\Theta_{\CC}=4$; iv) the critical exponent mostly aligned with $\GN$ moves to lower values. Already at $\Nf=1.2$, there is no local extremum in critical exponents present.

In short, for both shape functions with SM matter content, we do not find any PMS-point, see appendix \ref{app:SMmatt} for details. Nevertheless, we were able to investigate the qualitative effect of different matter-types on the existence and properties of PMS-points for both shape functions. To study the full SM-matter content in a robust way likely requires more converged gravitational beta-functions, potentially based on fluctuation computations, see \cite{Pawlowski:2020qer}.

\section{Conclusion\label{sec:Discussion}}
 
The triviality problem  challenges the consistency of renormalizable quantum field theories like the SM whenever they are neither asymptotically free nor asymptotically safe. Here, we investigate whether  ASQG, when coupled to matter, may also induce a UV completion in the matter sector. 

There is growing qualitative evidence from FRG computations that ASQG at the same time provides a UV fixed point for matter. We have investigated systematic uncertainties of the approximation scheme underlying this evidence. In the FRG, systematic uncertainties are encoded in unphysical dependences of universal quantities like critical exponents on the artificial regulator and the gravitational gauge-fixing. We have studied the regulator and the gauge dependence, by introducing a scaling parameter in the regulator and keeping one gravitational gauge-parameter free. Furthermore, we have fixed the shape function of the regulator to be either of Litim-type or exponential.

The dependence of universal quantities on unphysical parameters is minimized at extrema relative to these parameters, \ie points of minimal sensitivity. We have searched for such PMS-points for the Einstein-Hilbert term coupled to an Abelian gauge field.

From the perspective of the approximated RG-flow, there are two different ways of studying the influence of gravity on matter: Either the gravitational couplings are evaluated at their non-Gaussian fixed point in the UV. Alternatively, one can assume such a fixed point for gravity to exist, and leave the fixed-point values of the gravitational couplings as free parameters, and hence parameterize a possible UV-completion in the matter sector. We have focused on the first method here, and refer the reader to appendix \autoref{app:GravCouplAsPar} for a discussion of the second method. 

We find the following key results:
\begin{itemize}
    \item For minimal matter content ($\Nf=\Ns=\Nv=1$), and at the gravitational fixed point, for both the Litim-type and the exponential shape function, we find a point of minimal sensitivity of $\fg$ in both $\beta_h$ and $\RegParLit(\RegParExp).$ This point is a global maximum and in quantitative agreement between the two shape functions.
    \item Generically, adding bosons to the system tends to reduce the overall regulator dependence and shifts features such as PMS-points to larger values of $\RegParLit$ and $\RegParExp$. Conversely, fermions enhance the regulator dependence and shift such features to smaller values of the regulator parameters. The large number of fermions in the SM pushes the PMS-points to values of $\RegParLit,\RegParExp<0,$ removing them from the physical regime.
\end{itemize} 
In short, our results provide further evidence in favor of ASQG resolving the triviality problem. Yet, our analysis also indicates that systematic uncertainties are still large especially when adding many matter fields. Besides additional systematic uncertainties, the precision of the truncation we have studied is not sufficient to conclude that ASQG definitely resolves the triviality problem. Larger truncations and the connection to lattice studies will hopefully allow to resolve these issues. This is the subject of future work.

\section*{Acknowledgements}

We would like to thank Gustavo de Brito, Astrid Eichhorn, Jan Pawlowski, and Benjamin Knorr for interesting discussions, and Astrid Eichhorn and Benjamin Knorr for helpful comments on the manuscript.  M.\ S.\ acknowledges support by Perimeter Institute for Theoretical Physics during early stages of this work. Research at Perimeter Institute is supported in part by the Government of Canada through the Department of Innovation, Science and Economic Development and by the Province of Ontario through the Ministry of Colleges and Universities. The research of M.~S.~was in parts supported by a Radboud Excellence fellowship from Radboud University in Nijmegen, Netherlands. M.\ S.\ and F.\ W.\ would like to acknowledge the contribution of the COST Action CA23130 (“Bridging high and low energies in search of quantum gravity (BridgeQG)”). 

\appendix

\section{Gravitational couplings as regulator-independent input parameters\label{app:GravCouplAsPar}}

In this appendix, we treat the gravitational couplings as free parameters, independent of regulator and gauge. This contrasts with \autoref{sec:RegDepWithCC}, where we evaluate them at their respective fixed-point values. On the one hand, those fixed-point values are not physical and hence regulator and gauge-dependent themselves, which affects the overall regulator dependence of the critical exponents. On the other hand, the precise  fixed-point values of the gravitational couplings are truncation dependent and will change in larger truncations. Hence, the results presented in this appendix are complementary to our findings presented in \autoref{sec:RegDepWithCC}, and can be understood to parameterize systematic uncertainties related to extensions of the truncation in the gravitational sector. Both parameterized studies as well as the evaluation on gravitational fixed points have their respective merits.

As a first observation we note that $\fg\propto\GN$ holds in our approximation, such that we re-write
 \begin{equation}
 	\fg=\GN\fgtil(\CC)\,,\label{eq:Def_ftil}
 \end{equation} 
where $\fgtil$ only depends on the cosmological constant, and possibly on unphysical parameters.
 
When $\GN$ is treated as an input parameter, all regulator- and gauge-dependence of $\fg$ is contained in $\fgtil$. 
From \eqref{eq:etaAnonint}, we can see that for $\GFbeta=0$, only the transverse-traceless mode contributes to $\eta_A$. Furthermore, for  $\GFbeta=1$  both gravitational modes feature the same pole-structure, see  \eqref{eq:propTT} and \eqref{eq:propz0}. Since the scalar and transverse-traceless mode have the same relative prefactor in all three contributions in \eqref{eq:etaAnonint}, this implies that $\eta_A(\GFbeta=0)=\tfrac{10}{9}\eta_A(\GFbeta=1)$ for any regulator, if $\GN$ and $\CC$ are kept fixed. Therefore, when investigating the regulator dependence for different gauge choices, we do not discuss $\GFbeta=1$ and $\GFbeta=0$ separately.

We begin with the case $\CC = 0$, as it allows for an analytic integration over the loop momentum in \eqref{eq:etaAnonint} for both the Litim-type and exponential shape functions \eqref{eq: Litim} and \eqref{eq: Exp}, with a general gauge parameter $\GFbeta$. 
We then turn to the case $\CC\neq0.$ In this case, we treat the Limit-type and exponential shape functions separately. For each shape function, we apply the PMS to their respective parameters (\ie $\RegParLit,$ $\RegParExp$).

 \subsection{The case $\CC=0$}
 \label{app:ParamCC0}
 
\begin{figure}
 	\includegraphics[width=\linewidth]{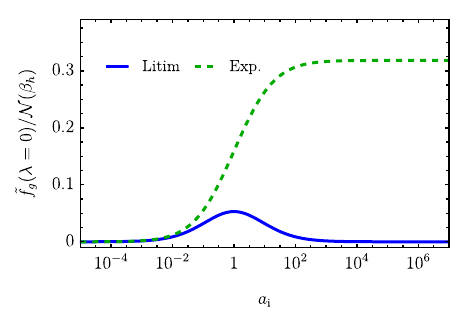}\\
 	\includegraphics[width=\linewidth]{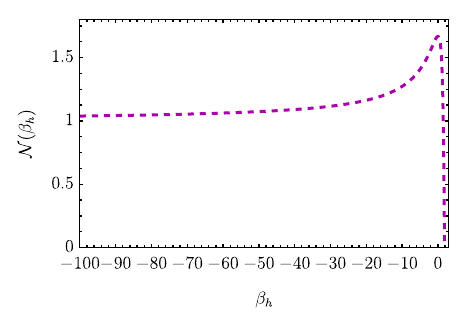}
 	\caption{We define $\mathcal{N}(\beta_h)=(15+(\GFbeta-10)\GFbeta)/(\GFbeta-3)^2$. Upper panel: We show the $a_i$-dependent part of $\fgtil$ for both Litim-type and exponential shape functions at $\CC=0$. We see that for the Litim-type shape function, $\fgtil$ always admits a local maximum at $\RegParLit=1$, while for the exponential shape function $\fgtil$ is maximized for $\RegParExp\to\infty$. Lower panel: We show the $\GFbeta$-dependent part of $\fgtil$, which is maximized for $\GFbeta=0$.}
 	\label{fig:LitimExpL0Expl}
 \end{figure}

For $\CC=0$, \eqref{eq:etaAnonint} implies that $\fgtil$ simplifies to:
\begin{equation}\label{eq:fgtilGenLamb0}
	\fgtil=2\frac{15+(\GFbeta-10)\GFbeta}{(\GFbeta-3)^2\pi}\int_0^\infty\mathrm{d}zz\frac{[z-r_k(z)][z r_k'(z)-r_k(z)]}{(z+r_k(z))^3}.
\end{equation}
Thus, the gauge dependence and the regulator dependence factorize. Independently of the regulator, the gauge dependence is minimized for $\beta_h=0$, where the gauge-dependent prefactor is maximized. This PMS-point is a global maximum of $\fgtil,$ which decreases by $40\%$ towards $\GFbeta\to-\infty$, and towards zero at $\GFbeta\approx 1.84$ (\cf \autoref{fig:LitimExpL0Expl}).

For the Litim-type and exponential shape functions (see \eqref{eq: Litim} and \eqref{eq: Exp}), we obtain
 \begin{alignat}{1}
 		\label{eq:fgtilLitl0}
 	\fgtilLit&=\frac{(15+(\GFbeta-10)\GFbeta)}{(\GFbeta-3)^2\pi}\frac{\RegParLit(2-2\RegParLit+(1+\RegParLit)\mathrm{Log}[\RegParLit])}{(\RegParLit-1)^3}\,,\\
 		\label{eq:fgtilExpl0}
 	\fgtilExp&=\frac{(15+(\GFbeta-10)\GFbeta)}{(\GFbeta-3)^2\pi}\frac{\RegParExp(\RegParExp-1-\mathrm{Log}[\RegParExp])}{(\RegParExp-1)^2}\,,
 \end{alignat}
respectively. These expressions have previously been studied for $\RegParLit\to0$ (and $\RegParExp\to0$) in \cite{deBrito:2022vbr}.

 From Eqs. \eqref{eq:fgtilLitl0} and \eqref{eq:fgtilExpl0}, we can already read off some key similarities and differences between the two shape functions:
 For the Litim-type function, $\fgtil$ has the limits
\begin{align}
    \fgtilLit&\to0\,,\quad&&\mathrm{for}\quad\RegParLit\to0\,,\\
 	\fgtilLit&\to0\,,\quad&&\mathrm{for}\quad\RegParLit\to\infty\,.
\end{align}
In contrast, for the exponential function $\fgtil$ has the limits
 \begin{align}
 	\fgtilExp&\to0\,,\quad&&\mathrm{for}\quad\RegParExp\to0\,,\\
 	\fgtilExp&\to\frac{15+(\GFbeta-10)\GFbeta}{(\GFbeta-3)^2\,\pi}\,,\quad&&\mathrm{for}\quad\RegParExp\to\infty\,.
 \end{align}

Combining this with $\fgtil\neq0$ for $\RegParLit>0$ automatically implies that there is at least one PMS-point at finite $\RegParLit$ for the Litim-type shape function, while this is not guaranteed for the exponential shape function. In fact, we find that
 \begin{alignat}{2}
 	\frac{\partial \fgtilLit}{\partial \RegParLit}\to0\,,\quad&&\mathrm{for}\quad\RegParLit\to1\,,\\
 	\frac{\partial \fgtilExp}{\partial \RegParExp}\to0\,,\quad&&\mathrm{for}\quad\RegParExp\to\infty\,,
 \end{alignat}
 which are the only extrema, as can be checked explicitly, see \autoref{fig:LitimExpL0Expl}. 
 As we will see below, the PMS-point for $\RegParLit$ varies with $\CC$ and $\GFbeta$, while the location of the maximum for $\RegParExp$ remains unchanged. While the limit $\RegParExp\to\infty$ minimizes the regulator dependence of  $\fgtilExp$, we do not consider this limit as a proper PMS-point, since the regulator parameters are constrained to $a_{i} \in(0,\infty)$, to provide a well-defined FRG-regulator.

 By inspecting the $\GFbeta$-independent part of \eqref{eq:fgtilLitl0} and \eqref{eq:fgtilExpl0}, we see that they do not change sign as a function of $\RegParLit$ ($\RegParExp$). Hence, at fixed $\GFbeta$, the sign of $\fgtilLit$ ($\fgtilExp$) is independent of the choices of shape function studied here. Thus, in the truncation in which $\CC=0$ and for the shape functions we consider, the qualitative properties of the gravitationally induced UV-completion are solely dependent on the gauge. In that vein, $\fgtil$ changes sign when  $(15~+~(\GFbeta~-~10)~\GFbeta)~=~0$. Hence, for $\GFbeta<5-\sqrt{10}\approx1.84$, $\fgtil$ is positive, allowing for a UV completion, for any $\RegParLit$ ($\RegParExp$), while it is negative otherwise. Note that $\GFbeta<3$ is additionally required, since the gauge-fixing condition is singular at $\GFbeta=3$. Hence, it is unclear if the sign-change at $\GFbeta\approx1.84$ is an artefact of the singularity, or stable. 
 
In short, while the value $\GFbeta=0$ is a PMS-point independent of the shape function, whether the regulator dependence can be minimized is shape-function dependent. Yet, our results indicate that, at least for $\CC=0$, the availability of a gravitationally induced UV-completion of the gauge coupling is (somewhat) gauge dependent, but does not vary between the shape functions we study in this paper. As we will see in the following, this picture is modified for non-vanishing $\CC$, due to more complicated gauge and regulator dependences.

\subsection{The case $\CC\neq 0$\label{app:ParamCCNeq0}}
 
We now generalize the results of the previous subsection to finite values of $\CC.$ While the corresponding generalization of \eqref{eq:fgtilGenLamb0} can be given in closed form, we do not display it here. By virtue of this closed-form expression, it can be shown analytically that $\fgtil$ still features a PMS point for the gauge dependence, independent of the choice of regulator. As for the case $\lambda=0,$ the gauge dependence is minimized at $\beta_h=0.$ Note that even when evaluated at the (regulator-dependent) gravitational fixed points, $\fg$ features a PMS-point at $\GFbeta\simeq0$ (\cf \autoref{fig:LitPMSfgbetdepCE}). Thus, this feature appears robust.

 Since the regulator dependence is more subtle and shape-function dependent, we treat the Litim-type shape function and exponential shape function separately in the following.\\
 
\paragraph{Litim-type shape function}\textcolor{white}{.}\\[3pt]
 For the Litim-type shape function \eqref{eq: Litim}, the integration over the loop momentum $q^2$ in \eqref{eq:etaAnonint} can be performed analytically for general regulator parameter $\RegParLit\in(0,\infty)$. 
 
\begin{figure}
 	\includegraphics[width=\linewidth]{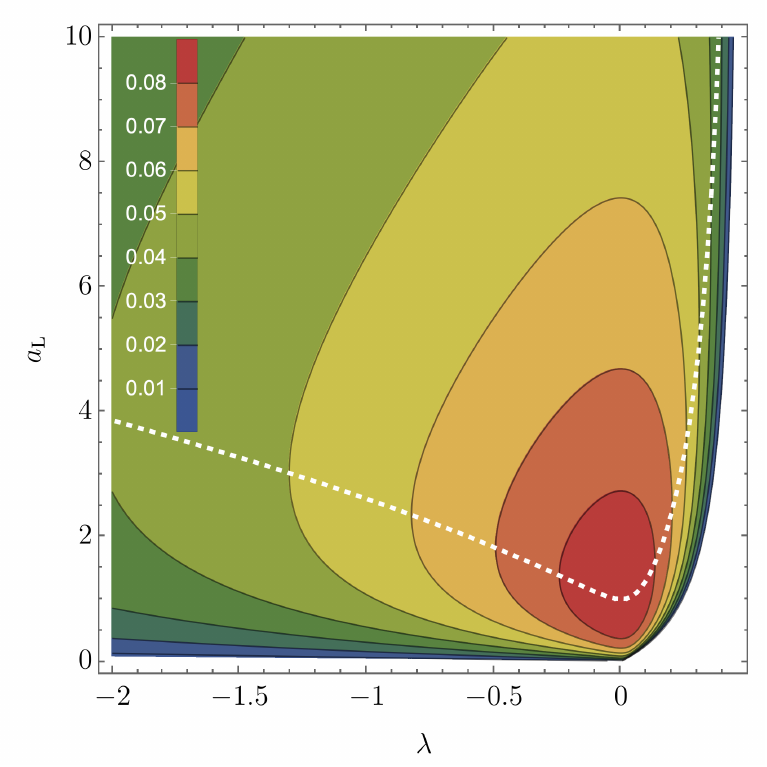}
 	\caption{We show the critical exponent $\fgtil$ of the Gaussian fixed point $\hypCFP=0$ for the Litim-type shape function as a function of the cosmological constant $\CC$ and the regulator parameter $\RegParLit$ for $\GFbeta=0$. The PMS of $\RegParLit$ as a function of $\CC$ is displayed as a white, dashed line. In all regions that are shown as white, the gauge coupling is irrelevant at $\hypCFP=0$.}
 	\label{fig:fgtilLitimb0Cont}
 \end{figure}

 In \autoref{fig:fgtilLitimb0Cont} we show the rescaled critical exponent $\fgtil$ of  the Gaussian fixed point $\hypCFP=0$ as a function of the cosmological constant $\CC$ and the regulator parameter $\RegParLit$ for $\GFbeta=0$. We see that $\fgtil$ is maximal in a region around $\CC\approx0$ and $\RegParLit\approx1$. We furthermore observe that $\fgtil$ decreases for large $\RegParLit$, when keeping $\CC$ fixed. Besides, we find a PMS-point at arbitrary but fixed $\CC$, which moves towards larger $\RegParLit$ when decreasing $\CC$. 
 In the regions marked white in \autoref{fig:fgtilLitimb0Cont}, the gauge coupling remains irrelevant at the Gaussian fixed point in our approximation\footnote{Note that \cite{Christiansen:2017cxa} have shown for $\RegParLit=1$ that this region can be removed when the momentum-dependence of $\fg$ is taken into account.}. Similarly to the previously studied cases at $\RegParLit=1$, we see that this happens for positive $\CC$. However, for $\RegParLit>1$, this region moves to larger $\CC$, \ie, closer to the pole at $\CC=1/2$.
 
\begin{figure}
	\includegraphics[width=\linewidth]{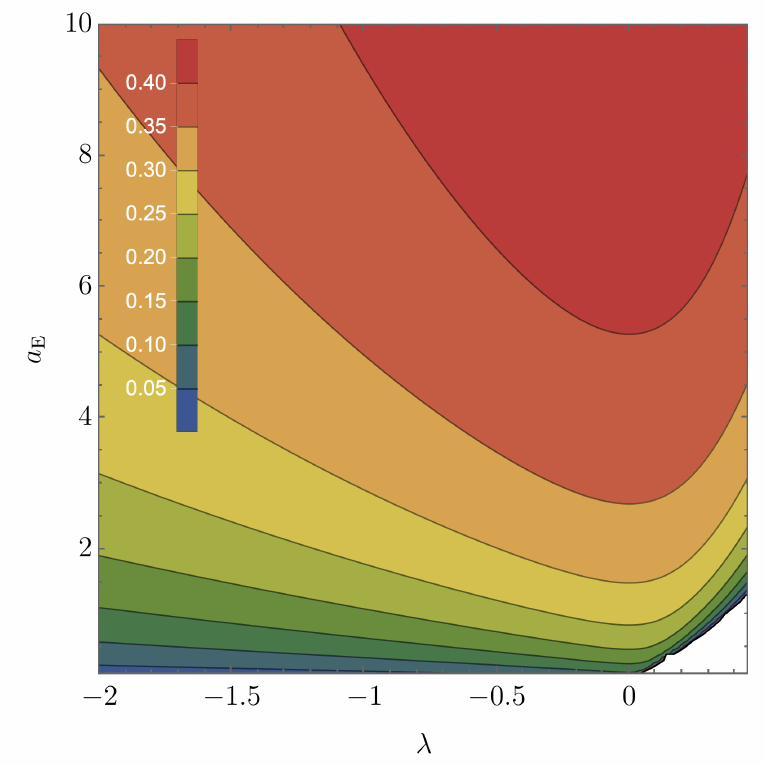}
	\caption{We show the critical exponent $\fgtil$ of the Gaussian fixed point $\hypCFP=0$ for the exponential shape function as a function of the cosmological constant $\CC$ and the regulator parameter $\RegParExp$ for $\GFbeta=0$. In all regions that are shown as white, the RG flow crosses a well known pole beyond which it cannot be continued.
	\label{fig:fgtilExpb0Cont}}
\end{figure}

 Besides the value of $\fgtil$ in the $\CC-\RegParLit$ plane, in \autoref{fig:fgtilLitimb0Cont} we also show the PMS value of $\RegParLit$,  and encode the value of the critical exponent $\fgtil$ at this point  in the colour. First,  we observe that for each value of the cosmological constant, there is a PMS-point. Especially towards negative values of the cosmological constant, the PMS-point $\RegParLitPMS$ differs significantly from $\RegParLit=1$, at which previous computations were performed. Furthermore,  we observe that $\fgtil$ remains positive all the way to $\CC=1/2$. This is achieved by an increasing $\RegParLitPMS$ for $\CC>0$. Therefore, we find that, if evaluated at a regulator that minimizes the residual regulator dependence, a UV-completion of the Abelian gauge sector can be achieved for any value of the cosmological constant. For fixed $\CC$, $\fgtil$ determines the lower bound on the Newton coupling $\GNcrit$, for which the UV completion is observationally viable.
 
The gauge dependence of the critical exponent $\fgtil$ at the PMS-point for $\RegParLit$ is mild, with a deviation at per-cent level between the choices $\GFbeta=0$ and $\GFbeta=-1$, and a $\sim10\%$ difference between the choices $\GFbeta = 0$ and $\GFbeta = 1$ -- corresponding to $\fg(\GFbeta = 0) = \tfrac{10}{9} \fg(\GFbeta = 1)$ as motivated in the introduction of this appendix.\\

\paragraph{Exponential shape function}\textcolor{white}{.}\\[3pt]
In contrast to the Litim-type shape function, the exponential shape function \eqref{eq: Exp} does not allow to integrate over the loop momentum $q^2$ in \eqref{eq:etaAnonint} analytically. Instead, we revert to numerical techniques.

Focussing on the gauge $\GFbeta=0,$ we display the rescaled critical exponent $\fgtil$ of the Gaussian fixed point $\hypCFP=0$ as a function of the cosmological constant $\CC$ and the regulator parameter $\RegParExp$ in Figure \ref{fig:fgtilExpb0Cont}. In contrast to the Litim-type shape function, $\fgtil$ does not exhibit extrema for finite $\RegParExp$ irrespective of the value of $\lambda$. Thus, there is no PMS-point. 

Similarly to the Litim-type shape function, the white region, indicating the UV-incomplete sector tends to larger values of $\lambda$ for increasing $\RegParExp.$ Indeed, the pole itself is shifted beyond $\lambda=1/2$ for $\RegParExp>1.$ Since we do not find any PMS point for $\RegParExp$, we do not further investigate gauge-dependence of $\fgtil$ in this setting.

\section{PMS at $G_*$ for $\CC=0$} \label{sec:lam0FPPMS}
In this appendix, we specify to ASQG as the UV-completion of the theory, where the fixed-point value of $\GN$ is obtained from an FRG equation, and hence regulator-dependent. However, in contrast to \autoref{sec:minmatt}, we restrict our analysis to $\CC=0$. This marks a step between our study in appendix \ref{app:GravCouplAsPar}, where no gravitational coupling was evaluated at the fixed point, and \autoref{sec:minmatt}, where both gravitational couplings are evaluated at their fixed points. With this intermediate step we can retain some analytical features discussed in appendix \ref{app:GravCouplAsPar}, while moving towards the physically more interesting case discussed in the main part of the paper. 

As before, we use the beta-function $\beta_G$ from \cite{deBrito:2022vbr}. The system at hand now features two couplings, $\GN$ and $\hypC$, and hence there are two critical exponents $\Theta_{G}$ and $\fg$ which in this truncation depend on the regulator parameters. However, due to the specific setup used in \cite{deBrito:2022vbr} to compute $\beta_G$, in particular, the use of the background-field approximation, $\Theta_G=2$ is fixed. Therefore, $\fg$ is the only critical exponent with non-trivial regulator and gauge dependence, to which we will apply the PMS. 

\begin{figure}
 	\includegraphics[width=\linewidth]{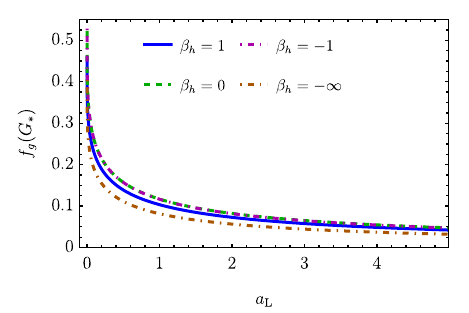}
 	\includegraphics[width=\linewidth]{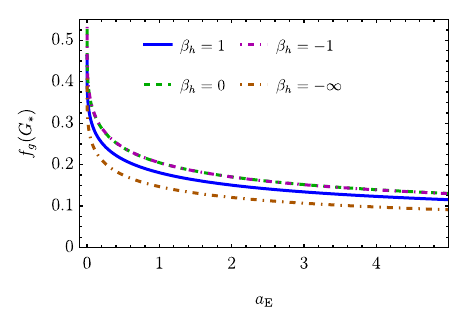}
 	\caption{We show the critical exponent $\fg$ of the Abelian gauge coupling evaluated at the asymptotically safe fixed point $\GN_*$ for the Litim-type (upper panel) and exponential shape functions (lower panel) and for a choice of gauge-parameters $\GFbeta$.}
 	\label{fig:Theta_eNoLam}
 \end{figure}

In \autoref{fig:Theta_eNoLam} we show the critical exponent $\fg$ both for the Litim-type (upper panel) and exponential shape function (lower panel) for different choices of the gauge parameter $\GFbeta$ for $\Nf=\Ns=\Nv=1$, see also \cite{deBrito:2022vbr}. For given $\GFbeta$, $\fg$ agrees between both shape functions in the limit $a_i\to0$, see \cite{deBrito:2022vbr}, and $\fg\to0$ for $a_i\to\infty$ for any gauge. We see that the (shape-function independent) limit $a_{i}\to0$ is a maximum for $\fg$  (and a minimum for $\GFbeta\gtrapprox1.84$, as discussed above) and that there are no local extrema. In fact, one can show that $\fg(\RegParLit)$ is monotonically decreasing  for $\GFbeta<2$, indicating that there is no PMS-point for $\fg$, when the regulator-dependence of $\GN_*$ is also considered. Accordingly, in  the gauge-gravity system with $\CC=0$ fixed, only $\Theta_{G}$ is (trivially) regulator independent, while $\fg$ remains regulator dependent, for any choice of $a_i\in(0,\infty)$.

\section{Fixed-point study for Standard-Model matter content\label{app:SMmatt}} 

In this appendix, we extrapolate the analysis of \autoref{sec:SMmatt} to the full matter content of the SM ($\Nf=22.5,$ $\Ns=4$ and $\Nv=12$).  We examine both shape functions individually. However, as indicated in \autoref{sec:SMmatt}, we caution regarding the robustness of these results due to the large graviton anomalous dimension at the Reuter fixed point in our truncation and the background-field approximation.\\

\begin{figure}
	\includegraphics[width=\linewidth]{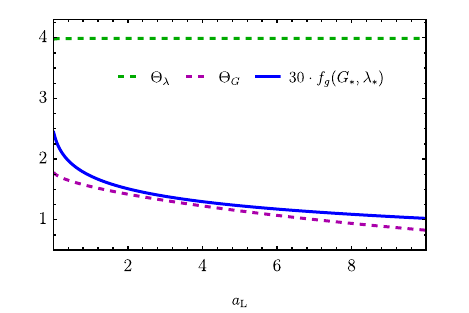}
	\caption{We show all critical exponents for the Litim-type shape function with SM matter content at $\GFbeta=1$.}
	\label{fig:LitimCritExpSB1SM}
\end{figure}

\paragraph{Litim-type shape function}\textcolor{white}{.}\\[3pt]
Due to the large number of fermions in the SM, the fermionic effect discussed in \autoref{sec:SMmatt} dominates over the bosonic one. In \autoref{fig:LitimCritExpSB1SM} we show all three critical exponents at the SM matter content, as a function of $\RegParLit$, and for $\GFbeta=1$. We see that neither $\fg$ nor $\Theta_{\GN}$ feature any local extrema, and behave qualitatively similar. However, the opposite effect of bosonic and fermionic matter on $\Theta_{\CC}$ at large $\RegParLit$ balances, to give rise to a new local maximum. For $\GFbeta=1$ it is located at $\RegParLit\approx4.5$. 

\begin{figure}
	\includegraphics[width=\linewidth]{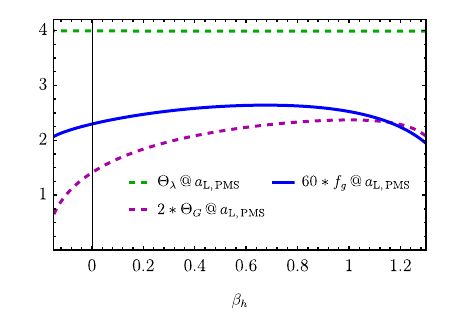}
	\caption{We show all critical exponents for the Litim-type shape function with SM matter content at the PMS-point for $\Theta_{\CC}$ as a function of $\GFbeta$.}
	\label{fig:LitimCritExpPMSBet}
\end{figure}

 Next, we track the behaviour of the PMS-point under changes in $\GFbeta$. This reveals that only for some values of $\GFbeta$ the PMS-point exists, while for others the fixed point diverges or moves into the complex plane before any local extremum is reached. In \autoref{fig:LitimCritExpPMSBet} we show all three critical exponents at the PMS-point of $\Theta_{\CC}$ in $\RegParLit$. As we can see, $\Theta_{\CC}$ varies only marginally with $\GFbeta$. Furthermore, both $\fg$ and $\Theta_{\GN}$ feature a local maximum as a function of $\GFbeta$, which is located at $\GFbeta\approx0.66$ and $\GFbeta\approx1$, respectively. Hence, also for the SM matter content, we can significantly reduce the regulator and gauge dependence of the system by choosing preferred parameters. However, in contrast to the minimal matter content, we do not have the choice to minimize the regulator dependence of $\fg$.\\

\begin{figure}
	\includegraphics[width=\linewidth]{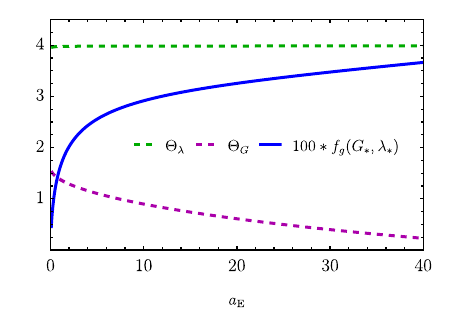}
	\caption{We show all critical exponents for the exponential shape function with SM matter content at $\GFbeta=1$.}
	\label{fig:ExpCritExpSB1SM}
\end{figure}

\paragraph{Exponential shape function}\textcolor{white}{.}\\[3pt]
As for the Litim-type shape function, the effect of fermions dominates due to their large number. In \autoref{fig:ExpCritExpSB1SM} we show the resulting critical exponents: $\Theta_{\CC}\approx4$ is almost constant for any $\RegParExp$. Conversely, $\Theta_{\GN}<2$ for any $\RegParExp$, and monotonically decreases with $\RegParExp$. The change of slope of $\fg$ caused by fermions also dominates, such that $\fg$ is monotonically increasing for the SM matter content. Ultimately, we do not find any local extremum of any of the critical exponents of the system.

Furthermore, $\Theta_{\GN}$ changes sign for $\RegParExp\sim 50$, beyond which and $\lambda$ is an irrelevant direction. If taken at face value, this would imply that physical properties depend on the choice of regulator. We do not interpret this as a genuine physical effect, but rather as a further sign that the truncation is unreliable at SM matter content — consistent with the large anomalous dimensions observed.

\bibliographystyle{apsrev4-1}
\bibliography{references.bib, manuals.bib}

\begin{thebibliography}{146}%
\makeatletter
\providecommand \@ifxundefined [1]{%
 \@ifx{#1\undefined}
}%
\providecommand \@ifnum [1]{%
 \ifnum #1\expandafter \@firstoftwo
 \else \expandafter \@secondoftwo
 \fi
}%
\providecommand \@ifx [1]{%
 \ifx #1\expandafter \@firstoftwo
 \else \expandafter \@secondoftwo
 \fi
}%
\providecommand \natexlab [1]{#1}%
\providecommand \enquote  [1]{``#1''}%
\providecommand \bibnamefont  [1]{#1}%
\providecommand \bibfnamefont [1]{#1}%
\providecommand \citenamefont [1]{#1}%
\providecommand \href@noop [0]{\@secondoftwo}%
\providecommand \href [0]{\begingroup \@sanitize@url \@href}%
\providecommand \@href[1]{\@@startlink{#1}\@@href}%
\providecommand \@@href[1]{\endgroup#1\@@endlink}%
\providecommand \@sanitize@url [0]{\catcode `\\12\catcode `\$12\catcode
  `\&12\catcode `\#12\catcode `\^12\catcode `\_12\catcode `\%12\relax}%
\providecommand \@@startlink[1]{}%
\providecommand \@@endlink[0]{}%
\providecommand \url  [0]{\begingroup\@sanitize@url \@url }%
\providecommand \@url [1]{\endgroup\@href {#1}{\urlprefix }}%
\providecommand \urlprefix  [0]{URL }%
\providecommand \Eprint [0]{\href }%
\providecommand \doibase [0]{http://dx.doi.org/}%
\providecommand \selectlanguage [0]{\@gobble}%
\providecommand \bibinfo  [0]{\@secondoftwo}%
\providecommand \bibfield  [0]{\@secondoftwo}%
\providecommand \translation [1]{[#1]}%
\providecommand \BibitemOpen [0]{}%
\providecommand \bibitemStop [0]{}%
\providecommand \bibitemNoStop [0]{.\EOS\space}%
\providecommand \EOS [0]{\spacefactor3000\relax}%
\providecommand \BibitemShut  [1]{\csname bibitem#1\endcsname}%
\let\auto@bib@innerbib\@empty
\bibitem [{\citenamefont {Landau}\ \emph {et~al.}(1954)\citenamefont {Landau},
  \citenamefont {Abrikosov},\ and\ \citenamefont
  {Khalatnikov}}]{Landau:1954nau}%
  \BibitemOpen
  \bibfield  {author} {\bibinfo {author} {\bibfnamefont {L.~D.}\ \bibnamefont
  {Landau}}, \bibinfo {author} {\bibfnamefont {A.~A.}\ \bibnamefont
  {Abrikosov}}, \ and\ \bibinfo {author} {\bibfnamefont {I.~M.}\ \bibnamefont
  {Khalatnikov}},\ }\href {\doibase 10.1016/b978-0-08-010586-4.50083-3}
  {\bibfield  {journal} {\bibinfo  {journal} {Dokl. Akad. Nauk SSSR}\ }\textbf
  {\bibinfo {volume} {95}} (\bibinfo {year} {1954}),\
  10.1016/b978-0-08-010586-4.50083-3}\BibitemShut {NoStop}%
\bibitem [{\citenamefont {Landau}\ and\ \citenamefont
  {Pomeranchuk}(1955)}]{Landau:1955ip}%
  \BibitemOpen
  \bibfield  {author} {\bibinfo {author} {\bibfnamefont {L.~D.}\ \bibnamefont
  {Landau}}\ and\ \bibinfo {author} {\bibfnamefont {I.~Y.}\ \bibnamefont
  {Pomeranchuk}},\ }\href {\doibase 10.1016/B978-0-08-010586-4.50091-2}
  {\bibfield  {journal} {\bibinfo  {journal} {Dokl. Akad. Nauk SSSR}\ }\textbf
  {\bibinfo {volume} {102}},\ \bibinfo {pages} {489} (\bibinfo {year}
  {1955})}\BibitemShut {NoStop}%
\bibitem [{\citenamefont {Landau}\ \emph {et~al.}(1956)\citenamefont {Landau},
  \citenamefont {Abrikosov},\ and\ \citenamefont {Halatnikov}}]{Landau:1956zr}%
  \BibitemOpen
  \bibfield  {author} {\bibinfo {author} {\bibfnamefont {L.~D.}\ \bibnamefont
  {Landau}}, \bibinfo {author} {\bibfnamefont {A.}~\bibnamefont {Abrikosov}}, \
  and\ \bibinfo {author} {\bibfnamefont {L.}~\bibnamefont {Halatnikov}},\
  }\href {\doibase 10.1007/BF02745513} {\bibfield  {journal} {\bibinfo
  {journal} {Nuovo Cim. Suppl.}\ }\textbf {\bibinfo {volume} {3}},\ \bibinfo
  {pages} {80} (\bibinfo {year} {1956})}\BibitemShut {NoStop}%
\bibitem [{\citenamefont {Yndurain}(1991)}]{Yndurain:1991ez}%
  \BibitemOpen
  \bibfield  {author} {\bibinfo {author} {\bibfnamefont {F.~J.}\ \bibnamefont
  {Yndurain}},\ }\href@noop {} {\  (\bibinfo {year} {1991})}\BibitemShut
  {NoStop}%
\bibitem [{\citenamefont {Pomeranchuk}\ \emph {et~al.}(1956)\citenamefont
  {Pomeranchuk}, \citenamefont {Sudakov},\ and\ \citenamefont
  {Ter-Martirosyan}}]{Pomeranchuk:1956zz}%
  \BibitemOpen
  \bibfield  {author} {\bibinfo {author} {\bibfnamefont {I.~Y.}\ \bibnamefont
  {Pomeranchuk}}, \bibinfo {author} {\bibfnamefont {V.~V.}\ \bibnamefont
  {Sudakov}}, \ and\ \bibinfo {author} {\bibfnamefont {K.~A.}\ \bibnamefont
  {Ter-Martirosyan}},\ }\href {\doibase 10.1103/PhysRev.103.784} {\bibfield
  {journal} {\bibinfo  {journal} {Phys. Rev.}\ }\textbf {\bibinfo {volume}
  {103}},\ \bibinfo {pages} {784} (\bibinfo {year} {1956})}\BibitemShut
  {NoStop}%
\bibitem [{\citenamefont {Wilson}\ and\ \citenamefont
  {Kogut}(1974)}]{Wilson:1973jj}%
  \BibitemOpen
  \bibfield  {author} {\bibinfo {author} {\bibfnamefont {K.~G.}\ \bibnamefont
  {Wilson}}\ and\ \bibinfo {author} {\bibfnamefont {J.~B.}\ \bibnamefont
  {Kogut}},\ }\href {\doibase 10.1016/0370-1573(74)90023-4} {\bibfield
  {journal} {\bibinfo  {journal} {Phys. Rept.}\ }\textbf {\bibinfo {volume}
  {12}},\ \bibinfo {pages} {75} (\bibinfo {year} {1974})}\BibitemShut {NoStop}%
\bibitem [{\citenamefont {Aizenman}(1982)}]{Aizenman:1982ze}%
  \BibitemOpen
  \bibfield  {author} {\bibinfo {author} {\bibfnamefont {M.}~\bibnamefont
  {Aizenman}},\ }\href {\doibase 10.1007/BF01205659} {\bibfield  {journal}
  {\bibinfo  {journal} {Commun. Math. Phys.}\ }\textbf {\bibinfo {volume}
  {86}},\ \bibinfo {pages} {1} (\bibinfo {year} {1982})}\BibitemShut {NoStop}%
\bibitem [{\citenamefont {Frohlich}(1982)}]{Frohlich:1982tw}%
  \BibitemOpen
  \bibfield  {author} {\bibinfo {author} {\bibfnamefont {J.}~\bibnamefont
  {Frohlich}},\ }\href {\doibase 10.1016/0550-3213(82)90088-8} {\bibfield
  {journal} {\bibinfo  {journal} {Nucl. Phys. B}\ }\textbf {\bibinfo {volume}
  {200}},\ \bibinfo {pages} {281} (\bibinfo {year} {1982})}\BibitemShut
  {NoStop}%
\bibitem [{\citenamefont {Luscher}\ and\ \citenamefont
  {Weisz}(1987)}]{Luscher:1987ay}%
  \BibitemOpen
  \bibfield  {author} {\bibinfo {author} {\bibfnamefont {M.}~\bibnamefont
  {Luscher}}\ and\ \bibinfo {author} {\bibfnamefont {P.}~\bibnamefont
  {Weisz}},\ }\href {\doibase 10.1016/0550-3213(87)90177-5} {\bibfield
  {journal} {\bibinfo  {journal} {Nucl. Phys. B}\ }\textbf {\bibinfo {volume}
  {290}},\ \bibinfo {pages} {25} (\bibinfo {year} {1987})}\BibitemShut
  {NoStop}%
\bibitem [{\citenamefont {Gockeler}\ \emph {et~al.}(1998)\citenamefont
  {Gockeler}, \citenamefont {Horsley}, \citenamefont {Linke}, \citenamefont
  {Rakow}, \citenamefont {Schierholz},\ and\ \citenamefont
  {Stuben}}]{Gockeler:1997dn}%
  \BibitemOpen
  \bibfield  {author} {\bibinfo {author} {\bibfnamefont {M.}~\bibnamefont
  {Gockeler}}, \bibinfo {author} {\bibfnamefont {R.}~\bibnamefont {Horsley}},
  \bibinfo {author} {\bibfnamefont {V.}~\bibnamefont {Linke}}, \bibinfo
  {author} {\bibfnamefont {P.~E.~L.}\ \bibnamefont {Rakow}}, \bibinfo {author}
  {\bibfnamefont {G.}~\bibnamefont {Schierholz}}, \ and\ \bibinfo {author}
  {\bibfnamefont {H.}~\bibnamefont {Stuben}},\ }\href {\doibase
  10.1103/PhysRevLett.80.4119} {\bibfield  {journal} {\bibinfo  {journal}
  {Phys. Rev. Lett.}\ }\textbf {\bibinfo {volume} {80}},\ \bibinfo {pages}
  {4119} (\bibinfo {year} {1998})},\ \Eprint
  {http://arxiv.org/abs/hep-th/9712244} {arXiv:hep-th/9712244} \BibitemShut
  {NoStop}%
\bibitem [{\citenamefont {Gies}\ and\ \citenamefont
  {Jaeckel}(2004)}]{Gies:2004hy}%
  \BibitemOpen
  \bibfield  {author} {\bibinfo {author} {\bibfnamefont {H.}~\bibnamefont
  {Gies}}\ and\ \bibinfo {author} {\bibfnamefont {J.}~\bibnamefont {Jaeckel}},\
  }\href {\doibase 10.1103/PhysRevLett.93.110405} {\bibfield  {journal}
  {\bibinfo  {journal} {Phys. Rev. Lett.}\ }\textbf {\bibinfo {volume} {93}},\
  \bibinfo {pages} {110405} (\bibinfo {year} {2004})},\ \Eprint
  {http://arxiv.org/abs/hep-ph/0405183} {arXiv:hep-ph/0405183} \BibitemShut
  {NoStop}%
\bibitem [{\citenamefont {Polchinski}(2007)}]{Polchinski:1998rq}%
  \BibitemOpen
  \bibfield  {author} {\bibinfo {author} {\bibfnamefont {J.}~\bibnamefont
  {Polchinski}},\ }\href {\doibase 10.1017/CBO9780511816079} {\emph {\bibinfo
  {title} {{String theory. Vol. 1: An introduction to the bosonic string}}}},\
  Cambridge Monographs on Mathematical Physics\ (\bibinfo  {publisher}
  {Cambridge University Press},\ \bibinfo {year} {2007})\BibitemShut {NoStop}%
\bibitem [{\citenamefont {Frasca}\ \emph {et~al.}(2021)\citenamefont {Frasca},
  \citenamefont {Ghoshal},\ and\ \citenamefont {Okada}}]{Frasca:2021iip}%
  \BibitemOpen
  \bibfield  {author} {\bibinfo {author} {\bibfnamefont {M.}~\bibnamefont
  {Frasca}}, \bibinfo {author} {\bibfnamefont {A.}~\bibnamefont {Ghoshal}}, \
  and\ \bibinfo {author} {\bibfnamefont {N.}~\bibnamefont {Okada}},\ }\href
  {\doibase 10.1103/PhysRevD.104.096010} {\bibfield  {journal} {\bibinfo
  {journal} {Phys. Rev. D}\ }\textbf {\bibinfo {volume} {104}},\ \bibinfo
  {pages} {096010} (\bibinfo {year} {2021})},\ \Eprint
  {http://arxiv.org/abs/2106.07629} {arXiv:2106.07629 [hep-th]} \BibitemShut
  {NoStop}%
\bibitem [{\citenamefont {Abel}\ \emph {et~al.}(2024)\citenamefont {Abel},
  \citenamefont {Dienes},\ and\ \citenamefont {Nutricati}}]{Abel:2024twz}%
  \BibitemOpen
  \bibfield  {author} {\bibinfo {author} {\bibfnamefont {S.}~\bibnamefont
  {Abel}}, \bibinfo {author} {\bibfnamefont {K.~R.}\ \bibnamefont {Dienes}}, \
  and\ \bibinfo {author} {\bibfnamefont {L.~A.}\ \bibnamefont {Nutricati}},\
  }\href {\doibase 10.1103/PhysRevD.110.126021} {\bibfield  {journal} {\bibinfo
   {journal} {Phys. Rev. D}\ }\textbf {\bibinfo {volume} {110}},\ \bibinfo
  {pages} {126021} (\bibinfo {year} {2024})},\ \Eprint
  {http://arxiv.org/abs/2407.11160} {arXiv:2407.11160 [hep-th]} \BibitemShut
  {NoStop}%
\bibitem [{\citenamefont {Bombelli}\ \emph {et~al.}(1987)\citenamefont
  {Bombelli}, \citenamefont {Lee}, \citenamefont {Meyer},\ and\ \citenamefont
  {Sorkin}}]{Bombelli:1987aa}%
  \BibitemOpen
  \bibfield  {author} {\bibinfo {author} {\bibfnamefont {L.}~\bibnamefont
  {Bombelli}}, \bibinfo {author} {\bibfnamefont {J.}~\bibnamefont {Lee}},
  \bibinfo {author} {\bibfnamefont {D.}~\bibnamefont {Meyer}}, \ and\ \bibinfo
  {author} {\bibfnamefont {R.}~\bibnamefont {Sorkin}},\ }\href {\doibase
  10.1103/PhysRevLett.59.521} {\bibfield  {journal} {\bibinfo  {journal} {Phys.
  Rev. Lett.}\ }\textbf {\bibinfo {volume} {59}},\ \bibinfo {pages} {521}
  (\bibinfo {year} {1987})}\BibitemShut {NoStop}%
\bibitem [{\citenamefont {de~Brito}\ \emph {et~al.}(2023)\citenamefont
  {de~Brito}, \citenamefont {Eichhorn},\ and\ \citenamefont
  {Fausten}}]{deBrito:2023nie}%
  \BibitemOpen
  \bibfield  {author} {\bibinfo {author} {\bibfnamefont {G.~P.}\ \bibnamefont
  {de~Brito}}, \bibinfo {author} {\bibfnamefont {A.}~\bibnamefont {Eichhorn}},
  \ and\ \bibinfo {author} {\bibfnamefont {L.}~\bibnamefont {Fausten}},\ }\href
  {\doibase 10.1007/s10714-023-03177-6} {\bibfield  {journal} {\bibinfo
  {journal} {Gen. Rel. Grav.}\ }\textbf {\bibinfo {volume} {55}},\ \bibinfo
  {pages} {128} (\bibinfo {year} {2023})},\ \Eprint
  {http://arxiv.org/abs/2305.07595} {arXiv:2305.07595 [gr-qc]} \BibitemShut
  {NoStop}%
\bibitem [{\citenamefont {Perez}(2013)}]{Perez:2012wv}%
  \BibitemOpen
  \bibfield  {author} {\bibinfo {author} {\bibfnamefont {A.}~\bibnamefont
  {Perez}},\ }\href {\doibase 10.12942/lrr-2013-3} {\bibfield  {journal}
  {\bibinfo  {journal} {Living Rev. Rel.}\ }\textbf {\bibinfo {volume} {16}},\
  \bibinfo {pages} {3} (\bibinfo {year} {2013})},\ \Eprint
  {http://arxiv.org/abs/1205.2019} {arXiv:1205.2019 [gr-qc]} \BibitemShut
  {NoStop}%
\bibitem [{\citenamefont {Daum}\ \emph {et~al.}(2010)\citenamefont {Daum},
  \citenamefont {Harst},\ and\ \citenamefont {Reuter}}]{Daum:2009dn}%
  \BibitemOpen
  \bibfield  {author} {\bibinfo {author} {\bibfnamefont {J.-E.}\ \bibnamefont
  {Daum}}, \bibinfo {author} {\bibfnamefont {U.}~\bibnamefont {Harst}}, \ and\
  \bibinfo {author} {\bibfnamefont {M.}~\bibnamefont {Reuter}},\ }\href
  {\doibase 10.1007/JHEP01(2010)084} {\bibfield  {journal} {\bibinfo  {journal}
  {JHEP}\ }\textbf {\bibinfo {volume} {01}},\ \bibinfo {pages} {084} (\bibinfo
  {year} {2010})},\ \Eprint {http://arxiv.org/abs/0910.4938} {arXiv:0910.4938
  [hep-th]} \BibitemShut {NoStop}%
\bibitem [{\citenamefont {Harst}\ and\ \citenamefont
  {Reuter}(2011)}]{Harst:2011zx}%
  \BibitemOpen
  \bibfield  {author} {\bibinfo {author} {\bibfnamefont {U.}~\bibnamefont
  {Harst}}\ and\ \bibinfo {author} {\bibfnamefont {M.}~\bibnamefont {Reuter}},\
  }\href {\doibase 10.1007/JHEP05(2011)119} {\bibfield  {journal} {\bibinfo
  {journal} {JHEP}\ }\textbf {\bibinfo {volume} {05}},\ \bibinfo {pages} {119}
  (\bibinfo {year} {2011})},\ \Eprint {http://arxiv.org/abs/1101.6007}
  {arXiv:1101.6007 [hep-th]} \BibitemShut {NoStop}%
\bibitem [{\citenamefont {Folkerts}\ \emph {et~al.}(2012)\citenamefont
  {Folkerts}, \citenamefont {Litim},\ and\ \citenamefont
  {Pawlowski}}]{Folkerts:2011jz}%
  \BibitemOpen
  \bibfield  {author} {\bibinfo {author} {\bibfnamefont {S.}~\bibnamefont
  {Folkerts}}, \bibinfo {author} {\bibfnamefont {D.~F.}\ \bibnamefont {Litim}},
  \ and\ \bibinfo {author} {\bibfnamefont {J.~M.}\ \bibnamefont {Pawlowski}},\
  }\href {\doibase 10.1016/j.physletb.2012.02.002} {\bibfield  {journal}
  {\bibinfo  {journal} {Phys. Lett. B}\ }\textbf {\bibinfo {volume} {709}},\
  \bibinfo {pages} {234} (\bibinfo {year} {2012})},\ \Eprint
  {http://arxiv.org/abs/1101.5552} {arXiv:1101.5552 [hep-th]} \BibitemShut
  {NoStop}%
\bibitem [{\citenamefont {Christiansen}\ and\ \citenamefont
  {Eichhorn}(2017)}]{Christiansen:2017gtg}%
  \BibitemOpen
  \bibfield  {author} {\bibinfo {author} {\bibfnamefont {N.}~\bibnamefont
  {Christiansen}}\ and\ \bibinfo {author} {\bibfnamefont {A.}~\bibnamefont
  {Eichhorn}},\ }\href {\doibase 10.1016/j.physletb.2017.04.047} {\bibfield
  {journal} {\bibinfo  {journal} {Phys. Lett. B}\ }\textbf {\bibinfo {volume}
  {770}},\ \bibinfo {pages} {154} (\bibinfo {year} {2017})},\ \Eprint
  {http://arxiv.org/abs/1702.07724} {arXiv:1702.07724 [hep-th]} \BibitemShut
  {NoStop}%
\bibitem [{\citenamefont {Eichhorn}\ and\ \citenamefont
  {Versteegen}(2018)}]{Eichhorn:2017lry}%
  \BibitemOpen
  \bibfield  {author} {\bibinfo {author} {\bibfnamefont {A.}~\bibnamefont
  {Eichhorn}}\ and\ \bibinfo {author} {\bibfnamefont {F.}~\bibnamefont
  {Versteegen}},\ }\href {\doibase 10.1007/JHEP01(2018)030} {\bibfield
  {journal} {\bibinfo  {journal} {JHEP}\ }\textbf {\bibinfo {volume} {01}},\
  \bibinfo {pages} {030} (\bibinfo {year} {2018})},\ \Eprint
  {http://arxiv.org/abs/1709.07252} {arXiv:1709.07252 [hep-th]} \BibitemShut
  {NoStop}%
\bibitem [{\citenamefont {Christiansen}\ \emph
  {et~al.}(2018{\natexlab{a}})\citenamefont {Christiansen}, \citenamefont
  {Litim}, \citenamefont {Pawlowski},\ and\ \citenamefont
  {Reichert}}]{Christiansen:2017cxa}%
  \BibitemOpen
  \bibfield  {author} {\bibinfo {author} {\bibfnamefont {N.}~\bibnamefont
  {Christiansen}}, \bibinfo {author} {\bibfnamefont {D.~F.}\ \bibnamefont
  {Litim}}, \bibinfo {author} {\bibfnamefont {J.~M.}\ \bibnamefont
  {Pawlowski}}, \ and\ \bibinfo {author} {\bibfnamefont {M.}~\bibnamefont
  {Reichert}},\ }\href {\doibase 10.1103/PhysRevD.97.106012} {\bibfield
  {journal} {\bibinfo  {journal} {Phys. Rev. D}\ }\textbf {\bibinfo {volume}
  {97}},\ \bibinfo {pages} {106012} (\bibinfo {year} {2018}{\natexlab{a}})},\
  \Eprint {http://arxiv.org/abs/1710.04669} {arXiv:1710.04669 [hep-th]}
  \BibitemShut {NoStop}%
\bibitem [{\citenamefont {Eichhorn}\ and\ \citenamefont
  {Schiffer}(2019)}]{Eichhorn:2019yzm}%
  \BibitemOpen
  \bibfield  {author} {\bibinfo {author} {\bibfnamefont {A.}~\bibnamefont
  {Eichhorn}}\ and\ \bibinfo {author} {\bibfnamefont {M.}~\bibnamefont
  {Schiffer}},\ }\href {\doibase 10.1016/j.physletb.2019.05.005} {\bibfield
  {journal} {\bibinfo  {journal} {Phys. Lett. B}\ }\textbf {\bibinfo {volume}
  {793}},\ \bibinfo {pages} {383} (\bibinfo {year} {2019})},\ \Eprint
  {http://arxiv.org/abs/1902.06479} {arXiv:1902.06479 [hep-th]} \BibitemShut
  {NoStop}%
\bibitem [{\citenamefont {Reuter}(1998)}]{Reuter:1996cp}%
  \BibitemOpen
  \bibfield  {author} {\bibinfo {author} {\bibfnamefont {M.}~\bibnamefont
  {Reuter}},\ }\href {\doibase 10.1103/PhysRevD.57.971} {\bibfield  {journal}
  {\bibinfo  {journal} {Phys. Rev. D}\ }\textbf {\bibinfo {volume} {57}},\
  \bibinfo {pages} {971} (\bibinfo {year} {1998})},\ \Eprint
  {http://arxiv.org/abs/hep-th/9605030} {arXiv:hep-th/9605030} \BibitemShut
  {NoStop}%
\bibitem [{\citenamefont {Souma}(1999)}]{Souma:1999at}%
  \BibitemOpen
  \bibfield  {author} {\bibinfo {author} {\bibfnamefont {W.}~\bibnamefont
  {Souma}},\ }\href {\doibase 10.1143/PTP.102.181} {\bibfield  {journal}
  {\bibinfo  {journal} {Prog. Theor. Phys.}\ }\textbf {\bibinfo {volume}
  {102}},\ \bibinfo {pages} {181} (\bibinfo {year} {1999})},\ \Eprint
  {http://arxiv.org/abs/hep-th/9907027} {arXiv:hep-th/9907027} \BibitemShut
  {NoStop}%
\bibitem [{\citenamefont {Lauscher}\ and\ \citenamefont
  {Reuter}(2002{\natexlab{a}})}]{Lauscher:2001ya}%
  \BibitemOpen
  \bibfield  {author} {\bibinfo {author} {\bibfnamefont {O.}~\bibnamefont
  {Lauscher}}\ and\ \bibinfo {author} {\bibfnamefont {M.}~\bibnamefont
  {Reuter}},\ }\href {\doibase 10.1103/PhysRevD.65.025013} {\bibfield
  {journal} {\bibinfo  {journal} {Phys. Rev. D}\ }\textbf {\bibinfo {volume}
  {65}},\ \bibinfo {pages} {025013} (\bibinfo {year} {2002}{\natexlab{a}})},\
  \Eprint {http://arxiv.org/abs/hep-th/0108040} {arXiv:hep-th/0108040}
  \BibitemShut {NoStop}%
\bibitem [{\citenamefont {Reuter}\ and\ \citenamefont
  {Saueressig}(2002)}]{Reuter:2001ag}%
  \BibitemOpen
  \bibfield  {author} {\bibinfo {author} {\bibfnamefont {M.}~\bibnamefont
  {Reuter}}\ and\ \bibinfo {author} {\bibfnamefont {F.}~\bibnamefont
  {Saueressig}},\ }\href {\doibase 10.1103/PhysRevD.65.065016} {\bibfield
  {journal} {\bibinfo  {journal} {Phys. Rev. D}\ }\textbf {\bibinfo {volume}
  {65}},\ \bibinfo {pages} {065016} (\bibinfo {year} {2002})},\ \Eprint
  {http://arxiv.org/abs/hep-th/0110054} {arXiv:hep-th/0110054} \BibitemShut
  {NoStop}%
\bibitem [{\citenamefont {Lauscher}\ and\ \citenamefont
  {Reuter}(2002{\natexlab{b}})}]{Lauscher:2002sq}%
  \BibitemOpen
  \bibfield  {author} {\bibinfo {author} {\bibfnamefont {O.}~\bibnamefont
  {Lauscher}}\ and\ \bibinfo {author} {\bibfnamefont {M.}~\bibnamefont
  {Reuter}},\ }\href {\doibase 10.1103/PhysRevD.66.025026} {\bibfield
  {journal} {\bibinfo  {journal} {Phys. Rev. D}\ }\textbf {\bibinfo {volume}
  {66}},\ \bibinfo {pages} {025026} (\bibinfo {year} {2002}{\natexlab{b}})},\
  \Eprint {http://arxiv.org/abs/hep-th/0205062} {arXiv:hep-th/0205062}
  \BibitemShut {NoStop}%
\bibitem [{\citenamefont {Litim}(2004)}]{Litim:2003vp}%
  \BibitemOpen
  \bibfield  {author} {\bibinfo {author} {\bibfnamefont {D.~F.}\ \bibnamefont
  {Litim}},\ }\href {\doibase 10.1103/PhysRevLett.92.201301} {\bibfield
  {journal} {\bibinfo  {journal} {Phys. Rev. Lett.}\ }\textbf {\bibinfo
  {volume} {92}},\ \bibinfo {pages} {201301} (\bibinfo {year} {2004})},\
  \Eprint {http://arxiv.org/abs/hep-th/0312114} {arXiv:hep-th/0312114}
  \BibitemShut {NoStop}%
\bibitem [{\citenamefont {Codello}\ and\ \citenamefont
  {Percacci}(2006)}]{Codello:2006in}%
  \BibitemOpen
  \bibfield  {author} {\bibinfo {author} {\bibfnamefont {A.}~\bibnamefont
  {Codello}}\ and\ \bibinfo {author} {\bibfnamefont {R.}~\bibnamefont
  {Percacci}},\ }\href {\doibase 10.1103/PhysRevLett.97.221301} {\bibfield
  {journal} {\bibinfo  {journal} {Phys. Rev. Lett.}\ }\textbf {\bibinfo
  {volume} {97}},\ \bibinfo {pages} {221301} (\bibinfo {year} {2006})},\
  \Eprint {http://arxiv.org/abs/hep-th/0607128} {arXiv:hep-th/0607128}
  \BibitemShut {NoStop}%
\bibitem [{\citenamefont {Machado}\ and\ \citenamefont
  {Saueressig}(2008)}]{Machado:2007ea}%
  \BibitemOpen
  \bibfield  {author} {\bibinfo {author} {\bibfnamefont {P.~F.}\ \bibnamefont
  {Machado}}\ and\ \bibinfo {author} {\bibfnamefont {F.}~\bibnamefont
  {Saueressig}},\ }\href {\doibase 10.1103/PhysRevD.77.124045} {\bibfield
  {journal} {\bibinfo  {journal} {Phys. Rev. D}\ }\textbf {\bibinfo {volume}
  {77}},\ \bibinfo {pages} {124045} (\bibinfo {year} {2008})},\ \Eprint
  {http://arxiv.org/abs/0712.0445} {arXiv:0712.0445 [hep-th]} \BibitemShut
  {NoStop}%
\bibitem [{\citenamefont {Codello}\ \emph {et~al.}(2009)\citenamefont
  {Codello}, \citenamefont {Percacci},\ and\ \citenamefont
  {Rahmede}}]{Codello:2008vh}%
  \BibitemOpen
  \bibfield  {author} {\bibinfo {author} {\bibfnamefont {A.}~\bibnamefont
  {Codello}}, \bibinfo {author} {\bibfnamefont {R.}~\bibnamefont {Percacci}}, \
  and\ \bibinfo {author} {\bibfnamefont {C.}~\bibnamefont {Rahmede}},\ }\href
  {\doibase 10.1016/j.aop.2008.08.008} {\bibfield  {journal} {\bibinfo
  {journal} {Annals Phys.}\ }\textbf {\bibinfo {volume} {324}},\ \bibinfo
  {pages} {414} (\bibinfo {year} {2009})},\ \Eprint
  {http://arxiv.org/abs/0805.2909} {arXiv:0805.2909 [hep-th]} \BibitemShut
  {NoStop}%
\bibitem [{\citenamefont {Benedetti}\ \emph {et~al.}(2009)\citenamefont
  {Benedetti}, \citenamefont {Machado},\ and\ \citenamefont
  {Saueressig}}]{Benedetti:2009rx}%
  \BibitemOpen
  \bibfield  {author} {\bibinfo {author} {\bibfnamefont {D.}~\bibnamefont
  {Benedetti}}, \bibinfo {author} {\bibfnamefont {P.~F.}\ \bibnamefont
  {Machado}}, \ and\ \bibinfo {author} {\bibfnamefont {F.}~\bibnamefont
  {Saueressig}},\ }\href {\doibase 10.1142/S0217732309031521} {\bibfield
  {journal} {\bibinfo  {journal} {Mod. Phys. Lett. A}\ }\textbf {\bibinfo
  {volume} {24}},\ \bibinfo {pages} {2233} (\bibinfo {year} {2009})},\ \Eprint
  {http://arxiv.org/abs/0901.2984} {arXiv:0901.2984 [hep-th]} \BibitemShut
  {NoStop}%
\bibitem [{\citenamefont {Eichhorn}\ \emph {et~al.}(2009)\citenamefont
  {Eichhorn}, \citenamefont {Gies},\ and\ \citenamefont
  {Scherer}}]{Eichhorn:2009ah}%
  \BibitemOpen
  \bibfield  {author} {\bibinfo {author} {\bibfnamefont {A.}~\bibnamefont
  {Eichhorn}}, \bibinfo {author} {\bibfnamefont {H.}~\bibnamefont {Gies}}, \
  and\ \bibinfo {author} {\bibfnamefont {M.~M.}\ \bibnamefont {Scherer}},\
  }\href {\doibase 10.1103/PhysRevD.80.104003} {\bibfield  {journal} {\bibinfo
  {journal} {Phys. Rev. D}\ }\textbf {\bibinfo {volume} {80}},\ \bibinfo
  {pages} {104003} (\bibinfo {year} {2009})},\ \Eprint
  {http://arxiv.org/abs/0907.1828} {arXiv:0907.1828 [hep-th]} \BibitemShut
  {NoStop}%
\bibitem [{\citenamefont {Manrique}\ \emph {et~al.}(2011)\citenamefont
  {Manrique}, \citenamefont {Reuter},\ and\ \citenamefont
  {Saueressig}}]{Manrique:2010am}%
  \BibitemOpen
  \bibfield  {author} {\bibinfo {author} {\bibfnamefont {E.}~\bibnamefont
  {Manrique}}, \bibinfo {author} {\bibfnamefont {M.}~\bibnamefont {Reuter}}, \
  and\ \bibinfo {author} {\bibfnamefont {F.}~\bibnamefont {Saueressig}},\
  }\href {\doibase 10.1016/j.aop.2010.11.006} {\bibfield  {journal} {\bibinfo
  {journal} {Annals Phys.}\ }\textbf {\bibinfo {volume} {326}},\ \bibinfo
  {pages} {463} (\bibinfo {year} {2011})},\ \Eprint
  {http://arxiv.org/abs/1006.0099} {arXiv:1006.0099 [hep-th]} \BibitemShut
  {NoStop}%
\bibitem [{\citenamefont {Eichhorn}\ and\ \citenamefont
  {Gies}(2010)}]{Eichhorn:2010tb}%
  \BibitemOpen
  \bibfield  {author} {\bibinfo {author} {\bibfnamefont {A.}~\bibnamefont
  {Eichhorn}}\ and\ \bibinfo {author} {\bibfnamefont {H.}~\bibnamefont
  {Gies}},\ }\href {\doibase 10.1103/PhysRevD.81.104010} {\bibfield  {journal}
  {\bibinfo  {journal} {Phys. Rev. D}\ }\textbf {\bibinfo {volume} {81}},\
  \bibinfo {pages} {104010} (\bibinfo {year} {2010})},\ \Eprint
  {http://arxiv.org/abs/1001.5033} {arXiv:1001.5033 [hep-th]} \BibitemShut
  {NoStop}%
\bibitem [{\citenamefont {Groh}\ and\ \citenamefont
  {Saueressig}(2010)}]{Groh:2010ta}%
  \BibitemOpen
  \bibfield  {author} {\bibinfo {author} {\bibfnamefont {K.}~\bibnamefont
  {Groh}}\ and\ \bibinfo {author} {\bibfnamefont {F.}~\bibnamefont
  {Saueressig}},\ }\href {\doibase 10.1088/1751-8113/43/36/365403} {\bibfield
  {journal} {\bibinfo  {journal} {J. Phys. A}\ }\textbf {\bibinfo {volume}
  {43}},\ \bibinfo {pages} {365403} (\bibinfo {year} {2010})},\ \Eprint
  {http://arxiv.org/abs/1001.5032} {arXiv:1001.5032 [hep-th]} \BibitemShut
  {NoStop}%
\bibitem [{\citenamefont {Dietz}\ and\ \citenamefont
  {Morris}(2013)}]{Dietz:2012ic}%
  \BibitemOpen
  \bibfield  {author} {\bibinfo {author} {\bibfnamefont {J.~A.}\ \bibnamefont
  {Dietz}}\ and\ \bibinfo {author} {\bibfnamefont {T.~R.}\ \bibnamefont
  {Morris}},\ }\href {\doibase 10.1007/JHEP01(2013)108} {\bibfield  {journal}
  {\bibinfo  {journal} {JHEP}\ }\textbf {\bibinfo {volume} {01}},\ \bibinfo
  {pages} {108} (\bibinfo {year} {2013})},\ \Eprint
  {http://arxiv.org/abs/1211.0955} {arXiv:1211.0955 [hep-th]} \BibitemShut
  {NoStop}%
\bibitem [{\citenamefont {Christiansen}\ \emph {et~al.}(2014)\citenamefont
  {Christiansen}, \citenamefont {Litim}, \citenamefont {Pawlowski},\ and\
  \citenamefont {Rodigast}}]{Christiansen:2012rx}%
  \BibitemOpen
  \bibfield  {author} {\bibinfo {author} {\bibfnamefont {N.}~\bibnamefont
  {Christiansen}}, \bibinfo {author} {\bibfnamefont {D.~F.}\ \bibnamefont
  {Litim}}, \bibinfo {author} {\bibfnamefont {J.~M.}\ \bibnamefont
  {Pawlowski}}, \ and\ \bibinfo {author} {\bibfnamefont {A.}~\bibnamefont
  {Rodigast}},\ }\href {\doibase 10.1016/j.physletb.2013.11.025} {\bibfield
  {journal} {\bibinfo  {journal} {Phys. Lett. B}\ }\textbf {\bibinfo {volume}
  {728}},\ \bibinfo {pages} {114} (\bibinfo {year} {2014})},\ \Eprint
  {http://arxiv.org/abs/1209.4038} {arXiv:1209.4038 [hep-th]} \BibitemShut
  {NoStop}%
\bibitem [{\citenamefont {Rechenberger}\ and\ \citenamefont
  {Saueressig}(2012)}]{Rechenberger:2012pm}%
  \BibitemOpen
  \bibfield  {author} {\bibinfo {author} {\bibfnamefont {S.}~\bibnamefont
  {Rechenberger}}\ and\ \bibinfo {author} {\bibfnamefont {F.}~\bibnamefont
  {Saueressig}},\ }\href {\doibase 10.1103/PhysRevD.86.024018} {\bibfield
  {journal} {\bibinfo  {journal} {Phys. Rev. D}\ }\textbf {\bibinfo {volume}
  {86}},\ \bibinfo {pages} {024018} (\bibinfo {year} {2012})},\ \Eprint
  {http://arxiv.org/abs/1206.0657} {arXiv:1206.0657 [hep-th]} \BibitemShut
  {NoStop}%
\bibitem [{\citenamefont {Falls}\ \emph {et~al.}(2013)\citenamefont {Falls},
  \citenamefont {Litim}, \citenamefont {Nikolakopoulos},\ and\ \citenamefont
  {Rahmede}}]{Falls:2013bv}%
  \BibitemOpen
  \bibfield  {author} {\bibinfo {author} {\bibfnamefont {K.}~\bibnamefont
  {Falls}}, \bibinfo {author} {\bibfnamefont {D.~F.}\ \bibnamefont {Litim}},
  \bibinfo {author} {\bibfnamefont {K.}~\bibnamefont {Nikolakopoulos}}, \ and\
  \bibinfo {author} {\bibfnamefont {C.}~\bibnamefont {Rahmede}},\ }\href@noop
  {} {\  (\bibinfo {year} {2013})},\ \Eprint {http://arxiv.org/abs/1301.4191}
  {arXiv:1301.4191 [hep-th]} \BibitemShut {NoStop}%
\bibitem [{\citenamefont {Ohta}\ and\ \citenamefont
  {Percacci}(2014)}]{Ohta:2013uca}%
  \BibitemOpen
  \bibfield  {author} {\bibinfo {author} {\bibfnamefont {N.}~\bibnamefont
  {Ohta}}\ and\ \bibinfo {author} {\bibfnamefont {R.}~\bibnamefont
  {Percacci}},\ }\href {\doibase 10.1088/0264-9381/31/1/015024} {\bibfield
  {journal} {\bibinfo  {journal} {Class. Quant. Grav.}\ }\textbf {\bibinfo
  {volume} {31}},\ \bibinfo {pages} {015024} (\bibinfo {year} {2014})},\
  \Eprint {http://arxiv.org/abs/1308.3398} {arXiv:1308.3398 [hep-th]}
  \BibitemShut {NoStop}%
\bibitem [{\citenamefont {Eichhorn}(2013)}]{Eichhorn:2013xr}%
  \BibitemOpen
  \bibfield  {author} {\bibinfo {author} {\bibfnamefont {A.}~\bibnamefont
  {Eichhorn}},\ }\href {\doibase 10.1088/0264-9381/30/11/115016} {\bibfield
  {journal} {\bibinfo  {journal} {Class. Quant. Grav.}\ }\textbf {\bibinfo
  {volume} {30}},\ \bibinfo {pages} {115016} (\bibinfo {year} {2013})},\
  \Eprint {http://arxiv.org/abs/1301.0879} {arXiv:1301.0879 [gr-qc]}
  \BibitemShut {NoStop}%
\bibitem [{\citenamefont {Falls}\ \emph {et~al.}(2016)\citenamefont {Falls},
  \citenamefont {Litim}, \citenamefont {Nikolakopoulos},\ and\ \citenamefont
  {Rahmede}}]{Falls:2014tra}%
  \BibitemOpen
  \bibfield  {author} {\bibinfo {author} {\bibfnamefont {K.}~\bibnamefont
  {Falls}}, \bibinfo {author} {\bibfnamefont {D.~F.}\ \bibnamefont {Litim}},
  \bibinfo {author} {\bibfnamefont {K.}~\bibnamefont {Nikolakopoulos}}, \ and\
  \bibinfo {author} {\bibfnamefont {C.}~\bibnamefont {Rahmede}},\ }\href
  {\doibase 10.1103/PhysRevD.93.104022} {\bibfield  {journal} {\bibinfo
  {journal} {Phys. Rev. D}\ }\textbf {\bibinfo {volume} {93}},\ \bibinfo
  {pages} {104022} (\bibinfo {year} {2016})},\ \Eprint
  {http://arxiv.org/abs/1410.4815} {arXiv:1410.4815 [hep-th]} \BibitemShut
  {NoStop}%
\bibitem [{\citenamefont {Codello}\ \emph {et~al.}(2014)\citenamefont
  {Codello}, \citenamefont {D'Odorico},\ and\ \citenamefont
  {Pagani}}]{Codello:2013fpa}%
  \BibitemOpen
  \bibfield  {author} {\bibinfo {author} {\bibfnamefont {A.}~\bibnamefont
  {Codello}}, \bibinfo {author} {\bibfnamefont {G.}~\bibnamefont {D'Odorico}},
  \ and\ \bibinfo {author} {\bibfnamefont {C.}~\bibnamefont {Pagani}},\ }\href
  {\doibase 10.1103/PhysRevD.89.081701} {\bibfield  {journal} {\bibinfo
  {journal} {Phys. Rev. D}\ }\textbf {\bibinfo {volume} {89}},\ \bibinfo
  {pages} {081701} (\bibinfo {year} {2014})},\ \Eprint
  {http://arxiv.org/abs/1304.4777} {arXiv:1304.4777 [gr-qc]} \BibitemShut
  {NoStop}%
\bibitem [{\citenamefont {Christiansen}\ \emph {et~al.}(2016)\citenamefont
  {Christiansen}, \citenamefont {Knorr}, \citenamefont {Pawlowski},\ and\
  \citenamefont {Rodigast}}]{Christiansen:2014raa}%
  \BibitemOpen
  \bibfield  {author} {\bibinfo {author} {\bibfnamefont {N.}~\bibnamefont
  {Christiansen}}, \bibinfo {author} {\bibfnamefont {B.}~\bibnamefont {Knorr}},
  \bibinfo {author} {\bibfnamefont {J.~M.}\ \bibnamefont {Pawlowski}}, \ and\
  \bibinfo {author} {\bibfnamefont {A.}~\bibnamefont {Rodigast}},\ }\href
  {\doibase 10.1103/PhysRevD.93.044036} {\bibfield  {journal} {\bibinfo
  {journal} {Phys. Rev. D}\ }\textbf {\bibinfo {volume} {93}},\ \bibinfo
  {pages} {044036} (\bibinfo {year} {2016})},\ \Eprint
  {http://arxiv.org/abs/1403.1232} {arXiv:1403.1232 [hep-th]} \BibitemShut
  {NoStop}%
\bibitem [{\citenamefont {Demmel}\ \emph {et~al.}(2015)\citenamefont {Demmel},
  \citenamefont {Saueressig},\ and\ \citenamefont {Zanusso}}]{Demmel:2015oqa}%
  \BibitemOpen
  \bibfield  {author} {\bibinfo {author} {\bibfnamefont {M.}~\bibnamefont
  {Demmel}}, \bibinfo {author} {\bibfnamefont {F.}~\bibnamefont {Saueressig}},
  \ and\ \bibinfo {author} {\bibfnamefont {O.}~\bibnamefont {Zanusso}},\ }\href
  {\doibase 10.1007/JHEP08(2015)113} {\bibfield  {journal} {\bibinfo  {journal}
  {JHEP}\ }\textbf {\bibinfo {volume} {08}},\ \bibinfo {pages} {113} (\bibinfo
  {year} {2015})},\ \Eprint {http://arxiv.org/abs/1504.07656} {arXiv:1504.07656
  [hep-th]} \BibitemShut {NoStop}%
\bibitem [{\citenamefont {Gies}\ \emph {et~al.}(2015)\citenamefont {Gies},
  \citenamefont {Knorr},\ and\ \citenamefont {Lippoldt}}]{Gies:2015tca}%
  \BibitemOpen
  \bibfield  {author} {\bibinfo {author} {\bibfnamefont {H.}~\bibnamefont
  {Gies}}, \bibinfo {author} {\bibfnamefont {B.}~\bibnamefont {Knorr}}, \ and\
  \bibinfo {author} {\bibfnamefont {S.}~\bibnamefont {Lippoldt}},\ }\href
  {\doibase 10.1103/PhysRevD.92.084020} {\bibfield  {journal} {\bibinfo
  {journal} {Phys. Rev. D}\ }\textbf {\bibinfo {volume} {92}},\ \bibinfo
  {pages} {084020} (\bibinfo {year} {2015})},\ \Eprint
  {http://arxiv.org/abs/1507.08859} {arXiv:1507.08859 [hep-th]} \BibitemShut
  {NoStop}%
\bibitem [{\citenamefont {Christiansen}\ \emph {et~al.}(2015)\citenamefont
  {Christiansen}, \citenamefont {Knorr}, \citenamefont {Meibohm}, \citenamefont
  {Pawlowski},\ and\ \citenamefont {Reichert}}]{Christiansen:2015rva}%
  \BibitemOpen
  \bibfield  {author} {\bibinfo {author} {\bibfnamefont {N.}~\bibnamefont
  {Christiansen}}, \bibinfo {author} {\bibfnamefont {B.}~\bibnamefont {Knorr}},
  \bibinfo {author} {\bibfnamefont {J.}~\bibnamefont {Meibohm}}, \bibinfo
  {author} {\bibfnamefont {J.~M.}\ \bibnamefont {Pawlowski}}, \ and\ \bibinfo
  {author} {\bibfnamefont {M.}~\bibnamefont {Reichert}},\ }\href {\doibase
  10.1103/PhysRevD.92.121501} {\bibfield  {journal} {\bibinfo  {journal} {Phys.
  Rev. D}\ }\textbf {\bibinfo {volume} {92}},\ \bibinfo {pages} {121501}
  (\bibinfo {year} {2015})},\ \Eprint {http://arxiv.org/abs/1506.07016}
  {arXiv:1506.07016 [hep-th]} \BibitemShut {NoStop}%
\bibitem [{\citenamefont {Ohta}\ \emph {et~al.}(2016)\citenamefont {Ohta},
  \citenamefont {Percacci},\ and\ \citenamefont {Vacca}}]{Ohta:2015fcu}%
  \BibitemOpen
  \bibfield  {author} {\bibinfo {author} {\bibfnamefont {N.}~\bibnamefont
  {Ohta}}, \bibinfo {author} {\bibfnamefont {R.}~\bibnamefont {Percacci}}, \
  and\ \bibinfo {author} {\bibfnamefont {G.~P.}\ \bibnamefont {Vacca}},\ }\href
  {\doibase 10.1140/epjc/s10052-016-3895-1} {\bibfield  {journal} {\bibinfo
  {journal} {Eur. Phys. J. C}\ }\textbf {\bibinfo {volume} {76}},\ \bibinfo
  {pages} {46} (\bibinfo {year} {2016})},\ \Eprint
  {http://arxiv.org/abs/1511.09393} {arXiv:1511.09393 [hep-th]} \BibitemShut
  {NoStop}%
\bibitem [{\citenamefont {Ohta}\ \emph {et~al.}(2015)\citenamefont {Ohta},
  \citenamefont {Percacci},\ and\ \citenamefont {Vacca}}]{Ohta:2015efa}%
  \BibitemOpen
  \bibfield  {author} {\bibinfo {author} {\bibfnamefont {N.}~\bibnamefont
  {Ohta}}, \bibinfo {author} {\bibfnamefont {R.}~\bibnamefont {Percacci}}, \
  and\ \bibinfo {author} {\bibfnamefont {G.~P.}\ \bibnamefont {Vacca}},\ }\href
  {\doibase 10.1103/PhysRevD.92.061501} {\bibfield  {journal} {\bibinfo
  {journal} {Phys. Rev. D}\ }\textbf {\bibinfo {volume} {92}},\ \bibinfo
  {pages} {061501} (\bibinfo {year} {2015})},\ \Eprint
  {http://arxiv.org/abs/1507.00968} {arXiv:1507.00968 [hep-th]} \BibitemShut
  {NoStop}%
\bibitem [{\citenamefont {Falls}(2015)}]{Falls:2015qga}%
  \BibitemOpen
  \bibfield  {author} {\bibinfo {author} {\bibfnamefont {K.}~\bibnamefont
  {Falls}},\ }\href {\doibase 10.1103/PhysRevD.92.124057} {\bibfield  {journal}
  {\bibinfo  {journal} {Phys. Rev. D}\ }\textbf {\bibinfo {volume} {92}},\
  \bibinfo {pages} {124057} (\bibinfo {year} {2015})},\ \Eprint
  {http://arxiv.org/abs/1501.05331} {arXiv:1501.05331 [hep-th]} \BibitemShut
  {NoStop}%
\bibitem [{\citenamefont {Eichhorn}(2015)}]{Eichhorn:2015bna}%
  \BibitemOpen
  \bibfield  {author} {\bibinfo {author} {\bibfnamefont {A.}~\bibnamefont
  {Eichhorn}},\ }\href {\doibase 10.1007/JHEP04(2015)096} {\bibfield  {journal}
  {\bibinfo  {journal} {JHEP}\ }\textbf {\bibinfo {volume} {04}},\ \bibinfo
  {pages} {096} (\bibinfo {year} {2015})},\ \Eprint
  {http://arxiv.org/abs/1501.05848} {arXiv:1501.05848 [gr-qc]} \BibitemShut
  {NoStop}%
\bibitem [{\citenamefont {Gies}\ \emph {et~al.}(2016)\citenamefont {Gies},
  \citenamefont {Knorr}, \citenamefont {Lippoldt},\ and\ \citenamefont
  {Saueressig}}]{Gies:2016con}%
  \BibitemOpen
  \bibfield  {author} {\bibinfo {author} {\bibfnamefont {H.}~\bibnamefont
  {Gies}}, \bibinfo {author} {\bibfnamefont {B.}~\bibnamefont {Knorr}},
  \bibinfo {author} {\bibfnamefont {S.}~\bibnamefont {Lippoldt}}, \ and\
  \bibinfo {author} {\bibfnamefont {F.}~\bibnamefont {Saueressig}},\ }\href
  {\doibase 10.1103/PhysRevLett.116.211302} {\bibfield  {journal} {\bibinfo
  {journal} {Phys. Rev. Lett.}\ }\textbf {\bibinfo {volume} {116}},\ \bibinfo
  {pages} {211302} (\bibinfo {year} {2016})},\ \Eprint
  {http://arxiv.org/abs/1601.01800} {arXiv:1601.01800 [hep-th]} \BibitemShut
  {NoStop}%
\bibitem [{\citenamefont {Denz}\ \emph {et~al.}(2018)\citenamefont {Denz},
  \citenamefont {Pawlowski},\ and\ \citenamefont {Reichert}}]{Denz:2016qks}%
  \BibitemOpen
  \bibfield  {author} {\bibinfo {author} {\bibfnamefont {T.}~\bibnamefont
  {Denz}}, \bibinfo {author} {\bibfnamefont {J.~M.}\ \bibnamefont {Pawlowski}},
  \ and\ \bibinfo {author} {\bibfnamefont {M.}~\bibnamefont {Reichert}},\
  }\href {\doibase 10.1140/epjc/s10052-018-5806-0} {\bibfield  {journal}
  {\bibinfo  {journal} {Eur. Phys. J. C}\ }\textbf {\bibinfo {volume} {78}},\
  \bibinfo {pages} {336} (\bibinfo {year} {2018})},\ \Eprint
  {http://arxiv.org/abs/1612.07315} {arXiv:1612.07315 [hep-th]} \BibitemShut
  {NoStop}%
\bibitem [{\citenamefont {Biemans}\ \emph
  {et~al.}(2017{\natexlab{a}})\citenamefont {Biemans}, \citenamefont
  {Platania},\ and\ \citenamefont {Saueressig}}]{Biemans:2016rvp}%
  \BibitemOpen
  \bibfield  {author} {\bibinfo {author} {\bibfnamefont {J.}~\bibnamefont
  {Biemans}}, \bibinfo {author} {\bibfnamefont {A.}~\bibnamefont {Platania}}, \
  and\ \bibinfo {author} {\bibfnamefont {F.}~\bibnamefont {Saueressig}},\
  }\href {\doibase 10.1103/PhysRevD.95.086013} {\bibfield  {journal} {\bibinfo
  {journal} {Phys. Rev. D}\ }\textbf {\bibinfo {volume} {95}},\ \bibinfo
  {pages} {086013} (\bibinfo {year} {2017}{\natexlab{a}})},\ \Eprint
  {http://arxiv.org/abs/1609.04813} {arXiv:1609.04813 [hep-th]} \BibitemShut
  {NoStop}%
\bibitem [{\citenamefont {Falls}\ and\ \citenamefont
  {Ohta}(2016)}]{Falls:2016msz}%
  \BibitemOpen
  \bibfield  {author} {\bibinfo {author} {\bibfnamefont {K.}~\bibnamefont
  {Falls}}\ and\ \bibinfo {author} {\bibfnamefont {N.}~\bibnamefont {Ohta}},\
  }\href {\doibase 10.1103/PhysRevD.94.084005} {\bibfield  {journal} {\bibinfo
  {journal} {Phys. Rev. D}\ }\textbf {\bibinfo {volume} {94}},\ \bibinfo
  {pages} {084005} (\bibinfo {year} {2016})},\ \Eprint
  {http://arxiv.org/abs/1607.08460} {arXiv:1607.08460 [hep-th]} \BibitemShut
  {NoStop}%
\bibitem [{\citenamefont {Falls}\ \emph
  {et~al.}(2018{\natexlab{a}})\citenamefont {Falls}, \citenamefont {Litim},
  \citenamefont {Nikolakopoulos},\ and\ \citenamefont
  {Rahmede}}]{Falls:2016wsa}%
  \BibitemOpen
  \bibfield  {author} {\bibinfo {author} {\bibfnamefont {K.}~\bibnamefont
  {Falls}}, \bibinfo {author} {\bibfnamefont {D.~F.}\ \bibnamefont {Litim}},
  \bibinfo {author} {\bibfnamefont {K.}~\bibnamefont {Nikolakopoulos}}, \ and\
  \bibinfo {author} {\bibfnamefont {C.}~\bibnamefont {Rahmede}},\ }\href
  {\doibase 10.1088/1361-6382/aac440} {\bibfield  {journal} {\bibinfo
  {journal} {Class. Quant. Grav.}\ }\textbf {\bibinfo {volume} {35}},\ \bibinfo
  {pages} {135006} (\bibinfo {year} {2018}{\natexlab{a}})},\ \Eprint
  {http://arxiv.org/abs/1607.04962} {arXiv:1607.04962 [gr-qc]} \BibitemShut
  {NoStop}%
\bibitem [{\citenamefont {de~Alwis}(2018)}]{deAlwis:2017ysy}%
  \BibitemOpen
  \bibfield  {author} {\bibinfo {author} {\bibfnamefont {S.~P.}\ \bibnamefont
  {de~Alwis}},\ }\href {\doibase 10.1007/JHEP03(2018)118} {\bibfield  {journal}
  {\bibinfo  {journal} {JHEP}\ }\textbf {\bibinfo {volume} {03}},\ \bibinfo
  {pages} {118} (\bibinfo {year} {2018})},\ \Eprint
  {http://arxiv.org/abs/1707.09298} {arXiv:1707.09298 [hep-th]} \BibitemShut
  {NoStop}%
\bibitem [{\citenamefont {Christiansen}\ \emph
  {et~al.}(2018{\natexlab{b}})\citenamefont {Christiansen}, \citenamefont
  {Falls}, \citenamefont {Pawlowski},\ and\ \citenamefont
  {Reichert}}]{Christiansen:2017bsy}%
  \BibitemOpen
  \bibfield  {author} {\bibinfo {author} {\bibfnamefont {N.}~\bibnamefont
  {Christiansen}}, \bibinfo {author} {\bibfnamefont {K.}~\bibnamefont {Falls}},
  \bibinfo {author} {\bibfnamefont {J.~M.}\ \bibnamefont {Pawlowski}}, \ and\
  \bibinfo {author} {\bibfnamefont {M.}~\bibnamefont {Reichert}},\ }\href
  {\doibase 10.1103/PhysRevD.97.046007} {\bibfield  {journal} {\bibinfo
  {journal} {Phys. Rev. D}\ }\textbf {\bibinfo {volume} {97}},\ \bibinfo
  {pages} {046007} (\bibinfo {year} {2018}{\natexlab{b}})},\ \Eprint
  {http://arxiv.org/abs/1711.09259} {arXiv:1711.09259 [hep-th]} \BibitemShut
  {NoStop}%
\bibitem [{\citenamefont {Falls}\ \emph
  {et~al.}(2018{\natexlab{b}})\citenamefont {Falls}, \citenamefont {King},
  \citenamefont {Litim}, \citenamefont {Nikolakopoulos},\ and\ \citenamefont
  {Rahmede}}]{Falls:2017lst}%
  \BibitemOpen
  \bibfield  {author} {\bibinfo {author} {\bibfnamefont {K.}~\bibnamefont
  {Falls}}, \bibinfo {author} {\bibfnamefont {C.~R.}\ \bibnamefont {King}},
  \bibinfo {author} {\bibfnamefont {D.~F.}\ \bibnamefont {Litim}}, \bibinfo
  {author} {\bibfnamefont {K.}~\bibnamefont {Nikolakopoulos}}, \ and\ \bibinfo
  {author} {\bibfnamefont {C.}~\bibnamefont {Rahmede}},\ }\href {\doibase
  10.1103/PhysRevD.97.086006} {\bibfield  {journal} {\bibinfo  {journal} {Phys.
  Rev. D}\ }\textbf {\bibinfo {volume} {97}},\ \bibinfo {pages} {086006}
  (\bibinfo {year} {2018}{\natexlab{b}})},\ \Eprint
  {http://arxiv.org/abs/1801.00162} {arXiv:1801.00162 [hep-th]} \BibitemShut
  {NoStop}%
\bibitem [{\citenamefont {Houthoff}\ \emph {et~al.}(2017)\citenamefont
  {Houthoff}, \citenamefont {Kurov},\ and\ \citenamefont
  {Saueressig}}]{Houthoff:2017oam}%
  \BibitemOpen
  \bibfield  {author} {\bibinfo {author} {\bibfnamefont {W.~B.}\ \bibnamefont
  {Houthoff}}, \bibinfo {author} {\bibfnamefont {A.}~\bibnamefont {Kurov}}, \
  and\ \bibinfo {author} {\bibfnamefont {F.}~\bibnamefont {Saueressig}},\
  }\href {\doibase 10.1140/epjc/s10052-017-5046-8} {\bibfield  {journal}
  {\bibinfo  {journal} {Eur. Phys. J. C}\ }\textbf {\bibinfo {volume} {77}},\
  \bibinfo {pages} {491} (\bibinfo {year} {2017})},\ \Eprint
  {http://arxiv.org/abs/1705.01848} {arXiv:1705.01848 [hep-th]} \BibitemShut
  {NoStop}%
\bibitem [{\citenamefont {Falls}(2017)}]{Falls:2017cze}%
  \BibitemOpen
  \bibfield  {author} {\bibinfo {author} {\bibfnamefont {K.}~\bibnamefont
  {Falls}},\ }\href {\doibase 10.1103/PhysRevD.96.126016} {\bibfield  {journal}
  {\bibinfo  {journal} {Phys. Rev. D}\ }\textbf {\bibinfo {volume} {96}},\
  \bibinfo {pages} {126016} (\bibinfo {year} {2017})},\ \Eprint
  {http://arxiv.org/abs/1702.03577} {arXiv:1702.03577 [hep-th]} \BibitemShut
  {NoStop}%
\bibitem [{\citenamefont {Becker}\ \emph {et~al.}(2017)\citenamefont {Becker},
  \citenamefont {Ripken},\ and\ \citenamefont {Saueressig}}]{Becker:2017tcx}%
  \BibitemOpen
  \bibfield  {author} {\bibinfo {author} {\bibfnamefont {D.}~\bibnamefont
  {Becker}}, \bibinfo {author} {\bibfnamefont {C.}~\bibnamefont {Ripken}}, \
  and\ \bibinfo {author} {\bibfnamefont {F.}~\bibnamefont {Saueressig}},\
  }\href {\doibase 10.1007/JHEP12(2017)121} {\bibfield  {journal} {\bibinfo
  {journal} {JHEP}\ }\textbf {\bibinfo {volume} {12}},\ \bibinfo {pages} {121}
  (\bibinfo {year} {2017})},\ \Eprint {http://arxiv.org/abs/1709.09098}
  {arXiv:1709.09098 [hep-th]} \BibitemShut {NoStop}%
\bibitem [{\citenamefont {Knorr}\ and\ \citenamefont
  {Lippoldt}(2017)}]{Knorr:2017fus}%
  \BibitemOpen
  \bibfield  {author} {\bibinfo {author} {\bibfnamefont {B.}~\bibnamefont
  {Knorr}}\ and\ \bibinfo {author} {\bibfnamefont {S.}~\bibnamefont
  {Lippoldt}},\ }\href {\doibase 10.1103/PhysRevD.96.065020} {\bibfield
  {journal} {\bibinfo  {journal} {Phys. Rev. D}\ }\textbf {\bibinfo {volume}
  {96}},\ \bibinfo {pages} {065020} (\bibinfo {year} {2017})},\ \Eprint
  {http://arxiv.org/abs/1707.01397} {arXiv:1707.01397 [hep-th]} \BibitemShut
  {NoStop}%
\bibitem [{\citenamefont {Knorr}(2018)}]{Knorr:2017mhu}%
  \BibitemOpen
  \bibfield  {author} {\bibinfo {author} {\bibfnamefont {B.}~\bibnamefont
  {Knorr}},\ }\href {\doibase 10.1088/1361-6382/aabaa0} {\bibfield  {journal}
  {\bibinfo  {journal} {Class. Quant. Grav.}\ }\textbf {\bibinfo {volume}
  {35}},\ \bibinfo {pages} {115005} (\bibinfo {year} {2018})},\ \Eprint
  {http://arxiv.org/abs/1710.07055} {arXiv:1710.07055 [hep-th]} \BibitemShut
  {NoStop}%
\bibitem [{\citenamefont {De~Brito}\ \emph {et~al.}(2018)\citenamefont
  {De~Brito}, \citenamefont {Ohta}, \citenamefont {Pereira}, \citenamefont
  {Tomaz},\ and\ \citenamefont {Yamada}}]{DeBrito:2018hur}%
  \BibitemOpen
  \bibfield  {author} {\bibinfo {author} {\bibfnamefont {G.~P.}\ \bibnamefont
  {De~Brito}}, \bibinfo {author} {\bibfnamefont {N.}~\bibnamefont {Ohta}},
  \bibinfo {author} {\bibfnamefont {A.~D.}\ \bibnamefont {Pereira}}, \bibinfo
  {author} {\bibfnamefont {A.~A.}\ \bibnamefont {Tomaz}}, \ and\ \bibinfo
  {author} {\bibfnamefont {M.}~\bibnamefont {Yamada}},\ }\href {\doibase
  10.1103/PhysRevD.98.026027} {\bibfield  {journal} {\bibinfo  {journal} {Phys.
  Rev. D}\ }\textbf {\bibinfo {volume} {98}},\ \bibinfo {pages} {026027}
  (\bibinfo {year} {2018})},\ \Eprint {http://arxiv.org/abs/1805.09656}
  {arXiv:1805.09656 [hep-th]} \BibitemShut {NoStop}%
\bibitem [{\citenamefont {Eichhorn}\ \emph
  {et~al.}(2019{\natexlab{a}})\citenamefont {Eichhorn}, \citenamefont
  {Lippoldt}, \citenamefont {Pawlowski}, \citenamefont {Reichert},\ and\
  \citenamefont {Schiffer}}]{Eichhorn:2018ydy}%
  \BibitemOpen
  \bibfield  {author} {\bibinfo {author} {\bibfnamefont {A.}~\bibnamefont
  {Eichhorn}}, \bibinfo {author} {\bibfnamefont {S.}~\bibnamefont {Lippoldt}},
  \bibinfo {author} {\bibfnamefont {J.~M.}\ \bibnamefont {Pawlowski}}, \bibinfo
  {author} {\bibfnamefont {M.}~\bibnamefont {Reichert}}, \ and\ \bibinfo
  {author} {\bibfnamefont {M.}~\bibnamefont {Schiffer}},\ }\href {\doibase
  10.1016/j.physletb.2019.01.071} {\bibfield  {journal} {\bibinfo  {journal}
  {Phys. Lett. B}\ }\textbf {\bibinfo {volume} {792}},\ \bibinfo {pages} {310}
  (\bibinfo {year} {2019}{\natexlab{a}})},\ \Eprint
  {http://arxiv.org/abs/1810.02828} {arXiv:1810.02828 [hep-th]} \BibitemShut
  {NoStop}%
\bibitem [{\citenamefont {Falls}\ \emph {et~al.}(2019)\citenamefont {Falls},
  \citenamefont {Litim},\ and\ \citenamefont {Schr{\"o}der}}]{Falls:2018ylp}%
  \BibitemOpen
  \bibfield  {author} {\bibinfo {author} {\bibfnamefont {K.~G.}\ \bibnamefont
  {Falls}}, \bibinfo {author} {\bibfnamefont {D.~F.}\ \bibnamefont {Litim}}, \
  and\ \bibinfo {author} {\bibfnamefont {J.}~\bibnamefont {Schr{\"o}der}},\
  }\href {\doibase 10.1103/PhysRevD.99.126015} {\bibfield  {journal} {\bibinfo
  {journal} {Phys. Rev. D}\ }\textbf {\bibinfo {volume} {99}},\ \bibinfo
  {pages} {126015} (\bibinfo {year} {2019})},\ \Eprint
  {http://arxiv.org/abs/1810.08550} {arXiv:1810.08550 [gr-qc]} \BibitemShut
  {NoStop}%
\bibitem [{\citenamefont {Bosma}\ \emph {et~al.}(2019)\citenamefont {Bosma},
  \citenamefont {Knorr},\ and\ \citenamefont {Saueressig}}]{Bosma:2019aiu}%
  \BibitemOpen
  \bibfield  {author} {\bibinfo {author} {\bibfnamefont {L.}~\bibnamefont
  {Bosma}}, \bibinfo {author} {\bibfnamefont {B.}~\bibnamefont {Knorr}}, \ and\
  \bibinfo {author} {\bibfnamefont {F.}~\bibnamefont {Saueressig}},\ }\href
  {\doibase 10.1103/PhysRevLett.123.101301} {\bibfield  {journal} {\bibinfo
  {journal} {Phys. Rev. Lett.}\ }\textbf {\bibinfo {volume} {123}},\ \bibinfo
  {pages} {101301} (\bibinfo {year} {2019})},\ \Eprint
  {http://arxiv.org/abs/1904.04845} {arXiv:1904.04845 [hep-th]} \BibitemShut
  {NoStop}%
\bibitem [{\citenamefont {Knorr}\ \emph {et~al.}(2019)\citenamefont {Knorr},
  \citenamefont {Ripken},\ and\ \citenamefont {Saueressig}}]{Knorr:2019atm}%
  \BibitemOpen
  \bibfield  {author} {\bibinfo {author} {\bibfnamefont {B.}~\bibnamefont
  {Knorr}}, \bibinfo {author} {\bibfnamefont {C.}~\bibnamefont {Ripken}}, \
  and\ \bibinfo {author} {\bibfnamefont {F.}~\bibnamefont {Saueressig}},\
  }\href {\doibase 10.1088/1361-6382/ab4a53} {\bibfield  {journal} {\bibinfo
  {journal} {Class. Quant. Grav.}\ }\textbf {\bibinfo {volume} {36}},\ \bibinfo
  {pages} {234001} (\bibinfo {year} {2019})},\ \Eprint
  {http://arxiv.org/abs/1907.02903} {arXiv:1907.02903 [hep-th]} \BibitemShut
  {NoStop}%
\bibitem [{\citenamefont {Falls}\ \emph {et~al.}(2020)\citenamefont {Falls},
  \citenamefont {Ohta},\ and\ \citenamefont {Percacci}}]{Falls:2020qhj}%
  \BibitemOpen
  \bibfield  {author} {\bibinfo {author} {\bibfnamefont {K.}~\bibnamefont
  {Falls}}, \bibinfo {author} {\bibfnamefont {N.}~\bibnamefont {Ohta}}, \ and\
  \bibinfo {author} {\bibfnamefont {R.}~\bibnamefont {Percacci}},\ }\href
  {\doibase 10.1016/j.physletb.2020.135773} {\bibfield  {journal} {\bibinfo
  {journal} {Phys. Lett. B}\ }\textbf {\bibinfo {volume} {810}},\ \bibinfo
  {pages} {135773} (\bibinfo {year} {2020})},\ \Eprint
  {http://arxiv.org/abs/2004.04126} {arXiv:2004.04126 [hep-th]} \BibitemShut
  {NoStop}%
\bibitem [{\citenamefont {Kluth}\ and\ \citenamefont
  {Litim}(2023)}]{Kluth:2020bdv}%
  \BibitemOpen
  \bibfield  {author} {\bibinfo {author} {\bibfnamefont {Y.}~\bibnamefont
  {Kluth}}\ and\ \bibinfo {author} {\bibfnamefont {D.~F.}\ \bibnamefont
  {Litim}},\ }\href {\doibase 10.1103/PhysRevD.108.026005} {\bibfield
  {journal} {\bibinfo  {journal} {Phys. Rev. D}\ }\textbf {\bibinfo {volume}
  {108}},\ \bibinfo {pages} {026005} (\bibinfo {year} {2023})},\ \Eprint
  {http://arxiv.org/abs/2008.09181} {arXiv:2008.09181 [hep-th]} \BibitemShut
  {NoStop}%
\bibitem [{\citenamefont {Knorr}(2021)}]{Knorr:2021slg}%
  \BibitemOpen
  \bibfield  {author} {\bibinfo {author} {\bibfnamefont {B.}~\bibnamefont
  {Knorr}},\ }\href {\doibase 10.21468/SciPostPhysCore.4.3.020} {\bibfield
  {journal} {\bibinfo  {journal} {SciPost Phys. Core}\ }\textbf {\bibinfo
  {volume} {4}},\ \bibinfo {pages} {020} (\bibinfo {year} {2021})},\ \Eprint
  {http://arxiv.org/abs/2104.11336} {arXiv:2104.11336 [hep-th]} \BibitemShut
  {NoStop}%
\bibitem [{\citenamefont {Bonanno}\ \emph {et~al.}(2022)\citenamefont
  {Bonanno}, \citenamefont {Denz}, \citenamefont {Pawlowski},\ and\
  \citenamefont {Reichert}}]{Bonanno:2021squ}%
  \BibitemOpen
  \bibfield  {author} {\bibinfo {author} {\bibfnamefont {A.}~\bibnamefont
  {Bonanno}}, \bibinfo {author} {\bibfnamefont {T.}~\bibnamefont {Denz}},
  \bibinfo {author} {\bibfnamefont {J.~M.}\ \bibnamefont {Pawlowski}}, \ and\
  \bibinfo {author} {\bibfnamefont {M.}~\bibnamefont {Reichert}},\ }\href
  {\doibase 10.21468/SciPostPhys.12.1.001} {\bibfield  {journal} {\bibinfo
  {journal} {SciPost Phys.}\ }\textbf {\bibinfo {volume} {12}},\ \bibinfo
  {pages} {001} (\bibinfo {year} {2022})},\ \Eprint
  {http://arxiv.org/abs/2102.02217} {arXiv:2102.02217 [hep-th]} \BibitemShut
  {NoStop}%
\bibitem [{\citenamefont {Baldazzi}\ and\ \citenamefont
  {Falls}(2021)}]{Baldazzi:2021orb}%
  \BibitemOpen
  \bibfield  {author} {\bibinfo {author} {\bibfnamefont {A.}~\bibnamefont
  {Baldazzi}}\ and\ \bibinfo {author} {\bibfnamefont {K.}~\bibnamefont
  {Falls}},\ }\href {\doibase 10.3390/universe7080294} {\bibfield  {journal}
  {\bibinfo  {journal} {Universe}\ }\textbf {\bibinfo {volume} {7}},\ \bibinfo
  {pages} {294} (\bibinfo {year} {2021})},\ \Eprint
  {http://arxiv.org/abs/2107.00671} {arXiv:2107.00671 [hep-th]} \BibitemShut
  {NoStop}%
\bibitem [{\citenamefont {Sen}\ \emph {et~al.}(2022)\citenamefont {Sen},
  \citenamefont {Wetterich},\ and\ \citenamefont {Yamada}}]{Sen:2021ffc}%
  \BibitemOpen
  \bibfield  {author} {\bibinfo {author} {\bibfnamefont {S.}~\bibnamefont
  {Sen}}, \bibinfo {author} {\bibfnamefont {C.}~\bibnamefont {Wetterich}}, \
  and\ \bibinfo {author} {\bibfnamefont {M.}~\bibnamefont {Yamada}},\ }\href
  {\doibase 10.1007/JHEP03(2022)130} {\bibfield  {journal} {\bibinfo  {journal}
  {JHEP}\ }\textbf {\bibinfo {volume} {03}},\ \bibinfo {pages} {130} (\bibinfo
  {year} {2022})},\ \Eprint {http://arxiv.org/abs/2111.04696} {arXiv:2111.04696
  [hep-th]} \BibitemShut {NoStop}%
\bibitem [{\citenamefont {Mitchell}\ \emph {et~al.}(2022)\citenamefont
  {Mitchell}, \citenamefont {Morris},\ and\ \citenamefont
  {Stulga}}]{Mitchell:2021qjr}%
  \BibitemOpen
  \bibfield  {author} {\bibinfo {author} {\bibfnamefont {A.}~\bibnamefont
  {Mitchell}}, \bibinfo {author} {\bibfnamefont {T.~R.}\ \bibnamefont
  {Morris}}, \ and\ \bibinfo {author} {\bibfnamefont {D.}~\bibnamefont
  {Stulga}},\ }\href {\doibase 10.1007/JHEP01(2022)041} {\bibfield  {journal}
  {\bibinfo  {journal} {JHEP}\ }\textbf {\bibinfo {volume} {01}},\ \bibinfo
  {pages} {041} (\bibinfo {year} {2022})},\ \Eprint
  {http://arxiv.org/abs/2111.05067} {arXiv:2111.05067 [hep-th]} \BibitemShut
  {NoStop}%
\bibitem [{\citenamefont {Knorr}\ \emph {et~al.}(2022)\citenamefont {Knorr},
  \citenamefont {Ripken},\ and\ \citenamefont {Saueressig}}]{Knorr:2021iwv}%
  \BibitemOpen
  \bibfield  {author} {\bibinfo {author} {\bibfnamefont {B.}~\bibnamefont
  {Knorr}}, \bibinfo {author} {\bibfnamefont {C.}~\bibnamefont {Ripken}}, \
  and\ \bibinfo {author} {\bibfnamefont {F.}~\bibnamefont {Saueressig}},\
  }\href {\doibase 10.1393/ncc/i2022-22028-5} {\bibfield  {journal} {\bibinfo
  {journal} {Nuovo Cim. C}\ }\textbf {\bibinfo {volume} {45}},\ \bibinfo
  {pages} {28} (\bibinfo {year} {2022})},\ \Eprint
  {http://arxiv.org/abs/2111.12365} {arXiv:2111.12365 [hep-th]} \BibitemShut
  {NoStop}%
\bibitem [{\citenamefont {Baldazzi}\ \emph {et~al.}(2022)\citenamefont
  {Baldazzi}, \citenamefont {Falls},\ and\ \citenamefont
  {Ferrero}}]{Baldazzi:2021fye}%
  \BibitemOpen
  \bibfield  {author} {\bibinfo {author} {\bibfnamefont {A.}~\bibnamefont
  {Baldazzi}}, \bibinfo {author} {\bibfnamefont {K.}~\bibnamefont {Falls}}, \
  and\ \bibinfo {author} {\bibfnamefont {R.}~\bibnamefont {Ferrero}},\ }\href
  {\doibase 10.1016/j.aop.2022.168822} {\bibfield  {journal} {\bibinfo
  {journal} {Annals Phys.}\ }\textbf {\bibinfo {volume} {440}},\ \bibinfo
  {pages} {168822} (\bibinfo {year} {2022})},\ \Eprint
  {http://arxiv.org/abs/2112.02118} {arXiv:2112.02118 [hep-th]} \BibitemShut
  {NoStop}%
\bibitem [{\citenamefont {Fehre}\ \emph {et~al.}(2023)\citenamefont {Fehre},
  \citenamefont {Litim}, \citenamefont {Pawlowski},\ and\ \citenamefont
  {Reichert}}]{Fehre:2021eob}%
  \BibitemOpen
  \bibfield  {author} {\bibinfo {author} {\bibfnamefont {J.}~\bibnamefont
  {Fehre}}, \bibinfo {author} {\bibfnamefont {D.~F.}\ \bibnamefont {Litim}},
  \bibinfo {author} {\bibfnamefont {J.~M.}\ \bibnamefont {Pawlowski}}, \ and\
  \bibinfo {author} {\bibfnamefont {M.}~\bibnamefont {Reichert}},\ }\href
  {\doibase 10.1103/PhysRevLett.130.081501} {\bibfield  {journal} {\bibinfo
  {journal} {Phys. Rev. Lett.}\ }\textbf {\bibinfo {volume} {130}},\ \bibinfo
  {pages} {081501} (\bibinfo {year} {2023})},\ \Eprint
  {http://arxiv.org/abs/2111.13232} {arXiv:2111.13232 [hep-th]} \BibitemShut
  {NoStop}%
\bibitem [{\citenamefont {Baldazzi}\ \emph {et~al.}(2023)\citenamefont
  {Baldazzi}, \citenamefont {Falls}, \citenamefont {Kluth},\ and\ \citenamefont
  {Knorr}}]{Baldazzi:2023pep}%
  \BibitemOpen
  \bibfield  {author} {\bibinfo {author} {\bibfnamefont {A.}~\bibnamefont
  {Baldazzi}}, \bibinfo {author} {\bibfnamefont {K.}~\bibnamefont {Falls}},
  \bibinfo {author} {\bibfnamefont {Y.}~\bibnamefont {Kluth}}, \ and\ \bibinfo
  {author} {\bibfnamefont {B.}~\bibnamefont {Knorr}},\ }\href@noop {} {\
  (\bibinfo {year} {2023})},\ \Eprint {http://arxiv.org/abs/2312.03831}
  {arXiv:2312.03831 [hep-th]} \BibitemShut {NoStop}%
\bibitem [{\citenamefont {Saueressig}\ and\ \citenamefont
  {Wang}(2023)}]{Saueressig:2023tfy}%
  \BibitemOpen
  \bibfield  {author} {\bibinfo {author} {\bibfnamefont {F.}~\bibnamefont
  {Saueressig}}\ and\ \bibinfo {author} {\bibfnamefont {J.}~\bibnamefont
  {Wang}},\ }\href {\doibase 10.1007/JHEP09(2023)064} {\bibfield  {journal}
  {\bibinfo  {journal} {JHEP}\ }\textbf {\bibinfo {volume} {09}},\ \bibinfo
  {pages} {064} (\bibinfo {year} {2023})},\ \Eprint
  {http://arxiv.org/abs/2306.10408} {arXiv:2306.10408 [hep-th]} \BibitemShut
  {NoStop}%
\bibitem [{\citenamefont {Kawai}\ and\ \citenamefont
  {Ohta}(2023)}]{Kawai:2023rgy}%
  \BibitemOpen
  \bibfield  {author} {\bibinfo {author} {\bibfnamefont {H.}~\bibnamefont
  {Kawai}}\ and\ \bibinfo {author} {\bibfnamefont {N.}~\bibnamefont {Ohta}},\
  }\href {\doibase 10.1103/PhysRevD.107.126025} {\bibfield  {journal} {\bibinfo
   {journal} {Phys. Rev. D}\ }\textbf {\bibinfo {volume} {107}},\ \bibinfo
  {pages} {126025} (\bibinfo {year} {2023})},\ \Eprint
  {http://arxiv.org/abs/2305.10591} {arXiv:2305.10591 [hep-th]} \BibitemShut
  {NoStop}%
\bibitem [{\citenamefont {Eichhorn}(2018)}]{Eichhorn:2017egq}%
  \BibitemOpen
  \bibfield  {author} {\bibinfo {author} {\bibfnamefont {A.}~\bibnamefont
  {Eichhorn}},\ }\href {\doibase 10.1007/s10701-018-0196-6} {\bibfield
  {journal} {\bibinfo  {journal} {Found. Phys.}\ }\textbf {\bibinfo {volume}
  {48}},\ \bibinfo {pages} {1407} (\bibinfo {year} {2018})},\ \Eprint
  {http://arxiv.org/abs/1709.03696} {arXiv:1709.03696 [gr-qc]} \BibitemShut
  {NoStop}%
\bibitem [{\citenamefont {Percacci}(2017)}]{Percacci:2017fkn}%
  \BibitemOpen
  \bibfield  {author} {\bibinfo {author} {\bibfnamefont {R.}~\bibnamefont
  {Percacci}},\ }\href {\doibase 10.1142/10369} {\emph {\bibinfo {title} {{An
  Introduction to Covariant Quantum Gravity and Asymptotic Safety}}}},\
  \bibinfo {series} {100 Years of General Relativity}, Vol.~\bibinfo {volume}
  {3}\ (\bibinfo  {publisher} {World Scientific},\ \bibinfo {year}
  {2017})\BibitemShut {NoStop}%
\bibitem [{\citenamefont {Reuter}\ and\ \citenamefont
  {Saueressig}(2019)}]{Reuter:2019byg}%
  \BibitemOpen
  \bibfield  {author} {\bibinfo {author} {\bibfnamefont {M.}~\bibnamefont
  {Reuter}}\ and\ \bibinfo {author} {\bibfnamefont {F.}~\bibnamefont
  {Saueressig}},\ }\href@noop {} {\emph {\bibinfo {title} {{Quantum Gravity and
  the Functional Renormalization Group}: {The Road towards Asymptotic
  Safety}}}}\ (\bibinfo  {publisher} {Cambridge University Press},\ \bibinfo
  {year} {2019})\BibitemShut {NoStop}%
\bibitem [{\citenamefont {Pereira}(2019)}]{Pereira:2019dbn}%
  \BibitemOpen
  \bibfield  {author} {\bibinfo {author} {\bibfnamefont {A.~D.}\ \bibnamefont
  {Pereira}},\ }in\ \href@noop {} {\emph {\bibinfo {booktitle} {{Progress and
  Visions in Quantum Theory in View of Gravity}: {Bridging foundations of
  physics and mathematics}}}}\ (\bibinfo {year} {2019})\ \Eprint
  {http://arxiv.org/abs/1904.07042} {arXiv:1904.07042 [gr-qc]} \BibitemShut
  {NoStop}%
\bibitem [{\citenamefont {Reichert}(2020)}]{Reichert:2020mja}%
  \BibitemOpen
  \bibfield  {author} {\bibinfo {author} {\bibfnamefont {M.}~\bibnamefont
  {Reichert}},\ }\href {\doibase 10.22323/1.384.0005} {\bibfield  {journal}
  {\bibinfo  {journal} {PoS}\ }\textbf {\bibinfo {volume} {384}},\ \bibinfo
  {pages} {005} (\bibinfo {year} {2020})}\BibitemShut {NoStop}%
\bibitem [{\citenamefont {Pawlowski}\ and\ \citenamefont
  {Reichert}(2021)}]{Pawlowski:2020qer}%
  \BibitemOpen
  \bibfield  {author} {\bibinfo {author} {\bibfnamefont {J.~M.}\ \bibnamefont
  {Pawlowski}}\ and\ \bibinfo {author} {\bibfnamefont {M.}~\bibnamefont
  {Reichert}},\ }\href {\doibase 10.3389/fphy.2020.551848} {\bibfield
  {journal} {\bibinfo  {journal} {Front. in Phys.}\ }\textbf {\bibinfo {volume}
  {8}},\ \bibinfo {pages} {551848} (\bibinfo {year} {2021})},\ \Eprint
  {http://arxiv.org/abs/2007.10353} {arXiv:2007.10353 [hep-th]} \BibitemShut
  {NoStop}%
\bibitem [{\citenamefont {Eichhorn}(2022)}]{Eichhorn:2022jqj}%
  \BibitemOpen
  \bibfield  {author} {\bibinfo {author} {\bibfnamefont {A.}~\bibnamefont
  {Eichhorn}},\ }\href {\doibase 10.1393/ncc/i2022-22029-4} {\bibfield
  {journal} {\bibinfo  {journal} {Nuovo Cim. C}\ }\textbf {\bibinfo {volume}
  {45}},\ \bibinfo {pages} {29} (\bibinfo {year} {2022})},\ \Eprint
  {http://arxiv.org/abs/2201.11543} {arXiv:2201.11543 [gr-qc]} \BibitemShut
  {NoStop}%
\bibitem [{\citenamefont {Eichhorn}\ and\ \citenamefont
  {Schiffer}(2022)}]{Eichhorn:2022gku}%
  \BibitemOpen
  \bibfield  {author} {\bibinfo {author} {\bibfnamefont {A.}~\bibnamefont
  {Eichhorn}}\ and\ \bibinfo {author} {\bibfnamefont {M.}~\bibnamefont
  {Schiffer}},\ }\href@noop {} {\  (\bibinfo {year} {2022})},\ \Eprint
  {http://arxiv.org/abs/2212.07456} {arXiv:2212.07456 [hep-th]} \BibitemShut
  {NoStop}%
\bibitem [{\citenamefont {Saueressig}(2023)}]{Saueressig:2023irs}%
  \BibitemOpen
  \bibfield  {author} {\bibinfo {author} {\bibfnamefont {F.}~\bibnamefont
  {Saueressig}},\ }\enquote {\bibinfo {title} {{The Functional Renormalization
  Group in Quantum Gravity}},}\ \ (\bibinfo {year} {2023})\ \Eprint
  {http://arxiv.org/abs/2302.14152} {arXiv:2302.14152 [hep-th]} \BibitemShut
  {NoStop}%
\bibitem [{\citenamefont {Pawlowski}\ and\ \citenamefont
  {Reichert}(2024)}]{Pawlowski:2023gym}%
  \BibitemOpen
  \bibfield  {author} {\bibinfo {author} {\bibfnamefont {J.~M.}\ \bibnamefont
  {Pawlowski}}\ and\ \bibinfo {author} {\bibfnamefont {M.}~\bibnamefont
  {Reichert}},\ }\enquote {\bibinfo {title} {{Quantum Gravity from Dynamical
  Metric Fluctuations}},}\ \ (\bibinfo {year} {2024})\ \Eprint
  {http://arxiv.org/abs/2309.10785} {arXiv:2309.10785 [hep-th]} \BibitemShut
  {NoStop}%
\bibitem [{\citenamefont {D'Angelo}(2024)}]{DAngelo:2023wje}%
  \BibitemOpen
  \bibfield  {author} {\bibinfo {author} {\bibfnamefont {E.}~\bibnamefont
  {D'Angelo}},\ }\href {\doibase 10.1103/PhysRevD.109.066012} {\bibfield
  {journal} {\bibinfo  {journal} {Phys. Rev. D}\ }\textbf {\bibinfo {volume}
  {109}},\ \bibinfo {pages} {066012} (\bibinfo {year} {2024})},\ \Eprint
  {http://arxiv.org/abs/2310.20603} {arXiv:2310.20603 [hep-th]} \BibitemShut
  {NoStop}%
\bibitem [{\citenamefont {Saueressig}\ and\ \citenamefont
  {Wang}(2025)}]{Saueressig:2025ypi}%
  \BibitemOpen
  \bibfield  {author} {\bibinfo {author} {\bibfnamefont {F.}~\bibnamefont
  {Saueressig}}\ and\ \bibinfo {author} {\bibfnamefont {J.}~\bibnamefont
  {Wang}},\ }\href {\doibase 10.1103/PhysRevD.111.106007} {\bibfield  {journal}
  {\bibinfo  {journal} {Phys. Rev. D}\ }\textbf {\bibinfo {volume} {111}},\
  \bibinfo {pages} {106007} (\bibinfo {year} {2025})},\ \Eprint
  {http://arxiv.org/abs/2501.03752} {arXiv:2501.03752 [hep-th]} \BibitemShut
  {NoStop}%
\bibitem [{\citenamefont {Platania}\ and\ \citenamefont
  {Wetterich}(2020)}]{Platania:2020knd}%
  \BibitemOpen
  \bibfield  {author} {\bibinfo {author} {\bibfnamefont {A.}~\bibnamefont
  {Platania}}\ and\ \bibinfo {author} {\bibfnamefont {C.}~\bibnamefont
  {Wetterich}},\ }\href {\doibase 10.1016/j.physletb.2020.135911} {\bibfield
  {journal} {\bibinfo  {journal} {Phys. Lett. B}\ }\textbf {\bibinfo {volume}
  {811}},\ \bibinfo {pages} {135911} (\bibinfo {year} {2020})},\ \Eprint
  {http://arxiv.org/abs/2009.06637} {arXiv:2009.06637 [hep-th]} \BibitemShut
  {NoStop}%
\bibitem [{\citenamefont {Knorr}\ and\ \citenamefont
  {Schiffer}(2021)}]{Knorr:2021niv}%
  \BibitemOpen
  \bibfield  {author} {\bibinfo {author} {\bibfnamefont {B.}~\bibnamefont
  {Knorr}}\ and\ \bibinfo {author} {\bibfnamefont {M.}~\bibnamefont
  {Schiffer}},\ }\href {\doibase 10.3390/universe7070216} {\bibfield  {journal}
  {\bibinfo  {journal} {Universe}\ }\textbf {\bibinfo {volume} {7}},\ \bibinfo
  {pages} {216} (\bibinfo {year} {2021})},\ \Eprint
  {http://arxiv.org/abs/2105.04566} {arXiv:2105.04566 [hep-th]} \BibitemShut
  {NoStop}%
\bibitem [{\citenamefont {Platania}(2022)}]{Platania:2022gtt}%
  \BibitemOpen
  \bibfield  {author} {\bibinfo {author} {\bibfnamefont {A.}~\bibnamefont
  {Platania}},\ }\href {\doibase 10.1007/JHEP09(2022)167} {\bibfield  {journal}
  {\bibinfo  {journal} {JHEP}\ }\textbf {\bibinfo {volume} {09}},\ \bibinfo
  {pages} {167} (\bibinfo {year} {2022})},\ \Eprint
  {http://arxiv.org/abs/2206.04072} {arXiv:2206.04072 [hep-th]} \BibitemShut
  {NoStop}%
\bibitem [{\citenamefont {Pastor-Guti{\'e}rrez}\ \emph
  {et~al.}(2025)\citenamefont {Pastor-Guti{\'e}rrez}, \citenamefont
  {Pawlowski}, \citenamefont {Reichert},\ and\ \citenamefont
  {Ruisi}}]{Pastor-Gutierrez:2024sbt}%
  \BibitemOpen
  \bibfield  {author} {\bibinfo {author} {\bibfnamefont {{\'A}.}~\bibnamefont
  {Pastor-Guti{\'e}rrez}}, \bibinfo {author} {\bibfnamefont {J.~M.}\
  \bibnamefont {Pawlowski}}, \bibinfo {author} {\bibfnamefont {M.}~\bibnamefont
  {Reichert}}, \ and\ \bibinfo {author} {\bibfnamefont {G.}~\bibnamefont
  {Ruisi}},\ }\href {\doibase 10.1103/PhysRevD.111.106005} {\bibfield
  {journal} {\bibinfo  {journal} {Phys. Rev. D}\ }\textbf {\bibinfo {volume}
  {111}},\ \bibinfo {pages} {106005} (\bibinfo {year} {2025})},\ \Eprint
  {http://arxiv.org/abs/2412.13800} {arXiv:2412.13800 [hep-ph]} \BibitemShut
  {NoStop}%
\bibitem [{\citenamefont {Donoghue}(2020)}]{Donoghue:2019clr}%
  \BibitemOpen
  \bibfield  {author} {\bibinfo {author} {\bibfnamefont {J.~F.}\ \bibnamefont
  {Donoghue}},\ }\href {\doibase 10.3389/fphy.2020.00056} {\bibfield  {journal}
  {\bibinfo  {journal} {Front. in Phys.}\ }\textbf {\bibinfo {volume} {8}},\
  \bibinfo {pages} {56} (\bibinfo {year} {2020})},\ \Eprint
  {http://arxiv.org/abs/1911.02967} {arXiv:1911.02967 [hep-th]} \BibitemShut
  {NoStop}%
\bibitem [{\citenamefont {Bonanno}\ \emph {et~al.}(2020)\citenamefont
  {Bonanno}, \citenamefont {Eichhorn}, \citenamefont {Gies}, \citenamefont
  {Pawlowski}, \citenamefont {Percacci}, \citenamefont {Reuter}, \citenamefont
  {Saueressig},\ and\ \citenamefont {Vacca}}]{Bonanno:2020bil}%
  \BibitemOpen
  \bibfield  {author} {\bibinfo {author} {\bibfnamefont {A.}~\bibnamefont
  {Bonanno}}, \bibinfo {author} {\bibfnamefont {A.}~\bibnamefont {Eichhorn}},
  \bibinfo {author} {\bibfnamefont {H.}~\bibnamefont {Gies}}, \bibinfo {author}
  {\bibfnamefont {J.~M.}\ \bibnamefont {Pawlowski}}, \bibinfo {author}
  {\bibfnamefont {R.}~\bibnamefont {Percacci}}, \bibinfo {author}
  {\bibfnamefont {M.}~\bibnamefont {Reuter}}, \bibinfo {author} {\bibfnamefont
  {F.}~\bibnamefont {Saueressig}}, \ and\ \bibinfo {author} {\bibfnamefont
  {G.~P.}\ \bibnamefont {Vacca}},\ }\href {\doibase 10.3389/fphy.2020.00269}
  {\bibfield  {journal} {\bibinfo  {journal} {Front. in Phys.}\ }\textbf
  {\bibinfo {volume} {8}},\ \bibinfo {pages} {269} (\bibinfo {year} {2020})},\
  \Eprint {http://arxiv.org/abs/2004.06810} {arXiv:2004.06810 [gr-qc]}
  \BibitemShut {NoStop}%
\bibitem [{\citenamefont {Don{\`a}}\ \emph {et~al.}(2014)\citenamefont
  {Don{\`a}}, \citenamefont {Eichhorn},\ and\ \citenamefont
  {Percacci}}]{Dona:2013qba}%
  \BibitemOpen
  \bibfield  {author} {\bibinfo {author} {\bibfnamefont {P.}~\bibnamefont
  {Don{\`a}}}, \bibinfo {author} {\bibfnamefont {A.}~\bibnamefont {Eichhorn}},
  \ and\ \bibinfo {author} {\bibfnamefont {R.}~\bibnamefont {Percacci}},\
  }\href {\doibase 10.1103/PhysRevD.89.084035} {\bibfield  {journal} {\bibinfo
  {journal} {Phys. Rev. D}\ }\textbf {\bibinfo {volume} {89}},\ \bibinfo
  {pages} {084035} (\bibinfo {year} {2014})},\ \Eprint
  {http://arxiv.org/abs/1311.2898} {arXiv:1311.2898 [hep-th]} \BibitemShut
  {NoStop}%
\bibitem [{\citenamefont {Meibohm}\ \emph {et~al.}(2016)\citenamefont
  {Meibohm}, \citenamefont {Pawlowski},\ and\ \citenamefont
  {Reichert}}]{Meibohm:2015twa}%
  \BibitemOpen
  \bibfield  {author} {\bibinfo {author} {\bibfnamefont {J.}~\bibnamefont
  {Meibohm}}, \bibinfo {author} {\bibfnamefont {J.~M.}\ \bibnamefont
  {Pawlowski}}, \ and\ \bibinfo {author} {\bibfnamefont {M.}~\bibnamefont
  {Reichert}},\ }\href {\doibase 10.1103/PhysRevD.93.084035} {\bibfield
  {journal} {\bibinfo  {journal} {Phys. Rev. D}\ }\textbf {\bibinfo {volume}
  {93}},\ \bibinfo {pages} {084035} (\bibinfo {year} {2016})},\ \Eprint
  {http://arxiv.org/abs/1510.07018} {arXiv:1510.07018 [hep-th]} \BibitemShut
  {NoStop}%
\bibitem [{\citenamefont {Biemans}\ \emph
  {et~al.}(2017{\natexlab{b}})\citenamefont {Biemans}, \citenamefont
  {Platania},\ and\ \citenamefont {Saueressig}}]{Biemans:2017zca}%
  \BibitemOpen
  \bibfield  {author} {\bibinfo {author} {\bibfnamefont {J.}~\bibnamefont
  {Biemans}}, \bibinfo {author} {\bibfnamefont {A.}~\bibnamefont {Platania}}, \
  and\ \bibinfo {author} {\bibfnamefont {F.}~\bibnamefont {Saueressig}},\
  }\href {\doibase 10.1007/JHEP05(2017)093} {\bibfield  {journal} {\bibinfo
  {journal} {JHEP}\ }\textbf {\bibinfo {volume} {05}},\ \bibinfo {pages} {093}
  (\bibinfo {year} {2017}{\natexlab{b}})},\ \Eprint
  {http://arxiv.org/abs/1702.06539} {arXiv:1702.06539 [hep-th]} \BibitemShut
  {NoStop}%
\bibitem [{\citenamefont {Alkofer}\ and\ \citenamefont
  {Saueressig}(2018)}]{Alkofer:2018fxj}%
  \BibitemOpen
  \bibfield  {author} {\bibinfo {author} {\bibfnamefont {N.}~\bibnamefont
  {Alkofer}}\ and\ \bibinfo {author} {\bibfnamefont {F.}~\bibnamefont
  {Saueressig}},\ }\href {\doibase 10.1016/j.aop.2018.07.017} {\bibfield
  {journal} {\bibinfo  {journal} {Annals Phys.}\ }\textbf {\bibinfo {volume}
  {396}},\ \bibinfo {pages} {173} (\bibinfo {year} {2018})},\ \Eprint
  {http://arxiv.org/abs/1802.00498} {arXiv:1802.00498 [hep-th]} \BibitemShut
  {NoStop}%
\bibitem [{\citenamefont {Wetterich}\ and\ \citenamefont
  {Yamada}(2019)}]{Wetterich:2019zdo}%
  \BibitemOpen
  \bibfield  {author} {\bibinfo {author} {\bibfnamefont {C.}~\bibnamefont
  {Wetterich}}\ and\ \bibinfo {author} {\bibfnamefont {M.}~\bibnamefont
  {Yamada}},\ }\href {\doibase 10.1103/PhysRevD.100.066017} {\bibfield
  {journal} {\bibinfo  {journal} {Phys. Rev. D}\ }\textbf {\bibinfo {volume}
  {100}},\ \bibinfo {pages} {066017} (\bibinfo {year} {2019})},\ \Eprint
  {http://arxiv.org/abs/1906.01721} {arXiv:1906.01721 [hep-th]} \BibitemShut
  {NoStop}%
\bibitem [{\citenamefont {Korver}\ \emph {et~al.}(2024)\citenamefont {Korver},
  \citenamefont {Saueressig},\ and\ \citenamefont {Wang}}]{Korver:2024sam}%
  \BibitemOpen
  \bibfield  {author} {\bibinfo {author} {\bibfnamefont {G.}~\bibnamefont
  {Korver}}, \bibinfo {author} {\bibfnamefont {F.}~\bibnamefont {Saueressig}},
  \ and\ \bibinfo {author} {\bibfnamefont {J.}~\bibnamefont {Wang}},\ }\href
  {\doibase 10.1016/j.physletb.2024.138789} {\bibfield  {journal} {\bibinfo
  {journal} {Phys. Lett. B}\ }\textbf {\bibinfo {volume} {855}},\ \bibinfo
  {pages} {138789} (\bibinfo {year} {2024})},\ \Eprint
  {http://arxiv.org/abs/2402.01260} {arXiv:2402.01260 [hep-th]} \BibitemShut
  {NoStop}%
\bibitem [{\citenamefont {Eichhorn}\ \emph {et~al.}(2018)\citenamefont
  {Eichhorn}, \citenamefont {Labus}, \citenamefont {Pawlowski},\ and\
  \citenamefont {Reichert}}]{Eichhorn:2018akn}%
  \BibitemOpen
  \bibfield  {author} {\bibinfo {author} {\bibfnamefont {A.}~\bibnamefont
  {Eichhorn}}, \bibinfo {author} {\bibfnamefont {P.}~\bibnamefont {Labus}},
  \bibinfo {author} {\bibfnamefont {J.~M.}\ \bibnamefont {Pawlowski}}, \ and\
  \bibinfo {author} {\bibfnamefont {M.}~\bibnamefont {Reichert}},\ }\href
  {\doibase 10.21468/SciPostPhys.5.4.031} {\bibfield  {journal} {\bibinfo
  {journal} {SciPost Phys.}\ }\textbf {\bibinfo {volume} {5}},\ \bibinfo
  {pages} {031} (\bibinfo {year} {2018})},\ \Eprint
  {http://arxiv.org/abs/1804.00012} {arXiv:1804.00012 [hep-th]} \BibitemShut
  {NoStop}%
\bibitem [{\citenamefont {Eichhorn}\ \emph
  {et~al.}(2019{\natexlab{b}})\citenamefont {Eichhorn}, \citenamefont
  {Lippoldt},\ and\ \citenamefont {Schiffer}}]{Eichhorn:2018nda}%
  \BibitemOpen
  \bibfield  {author} {\bibinfo {author} {\bibfnamefont {A.}~\bibnamefont
  {Eichhorn}}, \bibinfo {author} {\bibfnamefont {S.}~\bibnamefont {Lippoldt}},
  \ and\ \bibinfo {author} {\bibfnamefont {M.}~\bibnamefont {Schiffer}},\
  }\href {\doibase 10.1103/PhysRevD.99.086002} {\bibfield  {journal} {\bibinfo
  {journal} {Phys. Rev. D}\ }\textbf {\bibinfo {volume} {99}},\ \bibinfo
  {pages} {086002} (\bibinfo {year} {2019}{\natexlab{b}})},\ \Eprint
  {http://arxiv.org/abs/1812.08782} {arXiv:1812.08782 [hep-th]} \BibitemShut
  {NoStop}%
\bibitem [{\citenamefont {Shaposhnikov}\ and\ \citenamefont
  {Wetterich}(2010)}]{Shaposhnikov:2009pv}%
  \BibitemOpen
  \bibfield  {author} {\bibinfo {author} {\bibfnamefont {M.}~\bibnamefont
  {Shaposhnikov}}\ and\ \bibinfo {author} {\bibfnamefont {C.}~\bibnamefont
  {Wetterich}},\ }\href {\doibase 10.1016/j.physletb.2009.12.022} {\bibfield
  {journal} {\bibinfo  {journal} {Phys. Lett. B}\ }\textbf {\bibinfo {volume}
  {683}},\ \bibinfo {pages} {196} (\bibinfo {year} {2010})},\ \Eprint
  {http://arxiv.org/abs/0912.0208} {arXiv:0912.0208 [hep-th]} \BibitemShut
  {NoStop}%
\bibitem [{\citenamefont {Eichhorn}\ and\ \citenamefont
  {Held}(2018{\natexlab{a}})}]{Eichhorn:2017ylw}%
  \BibitemOpen
  \bibfield  {author} {\bibinfo {author} {\bibfnamefont {A.}~\bibnamefont
  {Eichhorn}}\ and\ \bibinfo {author} {\bibfnamefont {A.}~\bibnamefont
  {Held}},\ }\href {\doibase 10.1016/j.physletb.2017.12.040} {\bibfield
  {journal} {\bibinfo  {journal} {Phys. Lett. B}\ }\textbf {\bibinfo {volume}
  {777}},\ \bibinfo {pages} {217} (\bibinfo {year} {2018}{\natexlab{a}})},\
  \Eprint {http://arxiv.org/abs/1707.01107} {arXiv:1707.01107 [hep-th]}
  \BibitemShut {NoStop}%
\bibitem [{\citenamefont {Eichhorn}\ and\ \citenamefont
  {Held}(2018{\natexlab{b}})}]{Eichhorn:2018whv}%
  \BibitemOpen
  \bibfield  {author} {\bibinfo {author} {\bibfnamefont {A.}~\bibnamefont
  {Eichhorn}}\ and\ \bibinfo {author} {\bibfnamefont {A.}~\bibnamefont
  {Held}},\ }\href {\doibase 10.1103/PhysRevLett.121.151302} {\bibfield
  {journal} {\bibinfo  {journal} {Phys. Rev. Lett.}\ }\textbf {\bibinfo
  {volume} {121}},\ \bibinfo {pages} {151302} (\bibinfo {year}
  {2018}{\natexlab{b}})},\ \Eprint {http://arxiv.org/abs/1803.04027}
  {arXiv:1803.04027 [hep-th]} \BibitemShut {NoStop}%
\bibitem [{\citenamefont {Alkofer}\ \emph {et~al.}(2020)\citenamefont
  {Alkofer}, \citenamefont {Eichhorn}, \citenamefont {Held}, \citenamefont
  {Nieto}, \citenamefont {Percacci},\ and\ \citenamefont
  {Schr{\"o}fl}}]{Alkofer:2020vtb}%
  \BibitemOpen
  \bibfield  {author} {\bibinfo {author} {\bibfnamefont {R.}~\bibnamefont
  {Alkofer}}, \bibinfo {author} {\bibfnamefont {A.}~\bibnamefont {Eichhorn}},
  \bibinfo {author} {\bibfnamefont {A.}~\bibnamefont {Held}}, \bibinfo {author}
  {\bibfnamefont {C.~M.}\ \bibnamefont {Nieto}}, \bibinfo {author}
  {\bibfnamefont {R.}~\bibnamefont {Percacci}}, \ and\ \bibinfo {author}
  {\bibfnamefont {M.}~\bibnamefont {Schr{\"o}fl}},\ }\href {\doibase
  10.1016/j.aop.2020.168282} {\bibfield  {journal} {\bibinfo  {journal} {Annals
  Phys.}\ }\textbf {\bibinfo {volume} {421}},\ \bibinfo {pages} {168282}
  (\bibinfo {year} {2020})},\ \Eprint {http://arxiv.org/abs/2003.08401}
  {arXiv:2003.08401 [hep-ph]} \BibitemShut {NoStop}%
\bibitem [{\citenamefont {Kowalska}\ \emph {et~al.}(2022)\citenamefont
  {Kowalska}, \citenamefont {Pramanick},\ and\ \citenamefont
  {Sessolo}}]{Kowalska:2022ypk}%
  \BibitemOpen
  \bibfield  {author} {\bibinfo {author} {\bibfnamefont {K.}~\bibnamefont
  {Kowalska}}, \bibinfo {author} {\bibfnamefont {S.}~\bibnamefont {Pramanick}},
  \ and\ \bibinfo {author} {\bibfnamefont {E.~M.}\ \bibnamefont {Sessolo}},\
  }\href {\doibase 10.1007/JHEP08(2022)262} {\bibfield  {journal} {\bibinfo
  {journal} {JHEP}\ }\textbf {\bibinfo {volume} {08}},\ \bibinfo {pages} {262}
  (\bibinfo {year} {2022})},\ \Eprint {http://arxiv.org/abs/2204.00866}
  {arXiv:2204.00866 [hep-ph]} \BibitemShut {NoStop}%
\bibitem [{\citenamefont {Pastor-Guti{\'e}rrez}\ \emph
  {et~al.}(2023)\citenamefont {Pastor-Guti{\'e}rrez}, \citenamefont
  {Pawlowski},\ and\ \citenamefont {Reichert}}]{Pastor-Gutierrez:2022nki}%
  \BibitemOpen
  \bibfield  {author} {\bibinfo {author} {\bibfnamefont {{\'A}.}~\bibnamefont
  {Pastor-Guti{\'e}rrez}}, \bibinfo {author} {\bibfnamefont {J.~M.}\
  \bibnamefont {Pawlowski}}, \ and\ \bibinfo {author} {\bibfnamefont
  {M.}~\bibnamefont {Reichert}},\ }\href {\doibase
  10.21468/SciPostPhys.15.3.105} {\bibfield  {journal} {\bibinfo  {journal}
  {SciPost Phys.}\ }\textbf {\bibinfo {volume} {15}},\ \bibinfo {pages} {105}
  (\bibinfo {year} {2023})},\ \Eprint {http://arxiv.org/abs/2207.09817}
  {arXiv:2207.09817 [hep-th]} \BibitemShut {NoStop}%
\bibitem [{\citenamefont {Canet}\ \emph
  {et~al.}(2003{\natexlab{a}})\citenamefont {Canet}, \citenamefont {Delamotte},
  \citenamefont {Mouhanna},\ and\ \citenamefont {Vidal}}]{Canet:2002gs}%
  \BibitemOpen
  \bibfield  {author} {\bibinfo {author} {\bibfnamefont {L.}~\bibnamefont
  {Canet}}, \bibinfo {author} {\bibfnamefont {B.}~\bibnamefont {Delamotte}},
  \bibinfo {author} {\bibfnamefont {D.}~\bibnamefont {Mouhanna}}, \ and\
  \bibinfo {author} {\bibfnamefont {J.}~\bibnamefont {Vidal}},\ }\href
  {\doibase 10.1103/PhysRevD.67.065004} {\bibfield  {journal} {\bibinfo
  {journal} {Phys. Rev. D}\ }\textbf {\bibinfo {volume} {67}},\ \bibinfo
  {pages} {065004} (\bibinfo {year} {2003}{\natexlab{a}})},\ \Eprint
  {http://arxiv.org/abs/hep-th/0211055} {arXiv:hep-th/0211055} \BibitemShut
  {NoStop}%
\bibitem [{\citenamefont {Canet}\ \emph
  {et~al.}(2003{\natexlab{b}})\citenamefont {Canet}, \citenamefont {Delamotte},
  \citenamefont {Mouhanna},\ and\ \citenamefont {Vidal}}]{Canet:2003qd}%
  \BibitemOpen
  \bibfield  {author} {\bibinfo {author} {\bibfnamefont {L.}~\bibnamefont
  {Canet}}, \bibinfo {author} {\bibfnamefont {B.}~\bibnamefont {Delamotte}},
  \bibinfo {author} {\bibfnamefont {D.}~\bibnamefont {Mouhanna}}, \ and\
  \bibinfo {author} {\bibfnamefont {J.}~\bibnamefont {Vidal}},\ }\href
  {\doibase 10.1103/PhysRevB.68.064421} {\bibfield  {journal} {\bibinfo
  {journal} {Phys. Rev. B}\ }\textbf {\bibinfo {volume} {68}},\ \bibinfo
  {pages} {064421} (\bibinfo {year} {2003}{\natexlab{b}})},\ \Eprint
  {http://arxiv.org/abs/hep-th/0302227} {arXiv:hep-th/0302227} \BibitemShut
  {NoStop}%
\bibitem [{\citenamefont {Duclut}\ and\ \citenamefont
  {Delamotte}(2017)}]{Duclut:2016jct}%
  \BibitemOpen
  \bibfield  {author} {\bibinfo {author} {\bibfnamefont {C.}~\bibnamefont
  {Duclut}}\ and\ \bibinfo {author} {\bibfnamefont {B.}~\bibnamefont
  {Delamotte}},\ }\href {\doibase 10.1103/PhysRevE.95.012107} {\bibfield
  {journal} {\bibinfo  {journal} {Phys. Rev. E}\ }\textbf {\bibinfo {volume}
  {95}},\ \bibinfo {pages} {012107} (\bibinfo {year} {2017})},\ \Eprint
  {http://arxiv.org/abs/1611.07301} {arXiv:1611.07301 [cond-mat.stat-mech]}
  \BibitemShut {NoStop}%
\bibitem [{\citenamefont {Balog}\ \emph {et~al.}(2019)\citenamefont {Balog},
  \citenamefont {Chat{\'e}}, \citenamefont {Delamotte}, \citenamefont
  {Marohnic},\ and\ \citenamefont {Wschebor}}]{Balog:2019rrg}%
  \BibitemOpen
  \bibfield  {author} {\bibinfo {author} {\bibfnamefont {I.}~\bibnamefont
  {Balog}}, \bibinfo {author} {\bibfnamefont {H.}~\bibnamefont {Chat{\'e}}},
  \bibinfo {author} {\bibfnamefont {B.}~\bibnamefont {Delamotte}}, \bibinfo
  {author} {\bibfnamefont {M.}~\bibnamefont {Marohnic}}, \ and\ \bibinfo
  {author} {\bibfnamefont {N.}~\bibnamefont {Wschebor}},\ }\href {\doibase
  10.1103/PhysRevLett.123.240604} {\bibfield  {journal} {\bibinfo  {journal}
  {Phys. Rev. Lett.}\ }\textbf {\bibinfo {volume} {123}},\ \bibinfo {pages}
  {240604} (\bibinfo {year} {2019})},\ \Eprint
  {http://arxiv.org/abs/1907.01829} {arXiv:1907.01829 [cond-mat.stat-mech]}
  \BibitemShut {NoStop}%
\bibitem [{\citenamefont {De~Polsi}\ \emph {et~al.}(2020)\citenamefont
  {De~Polsi}, \citenamefont {Balog}, \citenamefont {Tissier},\ and\
  \citenamefont {Wschebor}}]{DePolsi:2020pjk}%
  \BibitemOpen
  \bibfield  {author} {\bibinfo {author} {\bibfnamefont {G.}~\bibnamefont
  {De~Polsi}}, \bibinfo {author} {\bibfnamefont {I.}~\bibnamefont {Balog}},
  \bibinfo {author} {\bibfnamefont {M.}~\bibnamefont {Tissier}}, \ and\
  \bibinfo {author} {\bibfnamefont {N.}~\bibnamefont {Wschebor}},\ }\href
  {\doibase 10.1103/PhysRevE.101.042113} {\bibfield  {journal} {\bibinfo
  {journal} {Phys. Rev. E}\ }\textbf {\bibinfo {volume} {101}},\ \bibinfo
  {pages} {042113} (\bibinfo {year} {2020})},\ \Eprint
  {http://arxiv.org/abs/2001.07525} {arXiv:2001.07525 [cond-mat.stat-mech]}
  \BibitemShut {NoStop}%
\bibitem [{\citenamefont {De~Polsi}\ and\ \citenamefont
  {Wschebor}(2022)}]{DePolsi:2022wyb}%
  \BibitemOpen
  \bibfield  {author} {\bibinfo {author} {\bibfnamefont {G.}~\bibnamefont
  {De~Polsi}}\ and\ \bibinfo {author} {\bibfnamefont {N.}~\bibnamefont
  {Wschebor}},\ }\href {\doibase 10.1103/PhysRevE.106.024111} {\bibfield
  {journal} {\bibinfo  {journal} {Phys. Rev. E}\ }\textbf {\bibinfo {volume}
  {106}},\ \bibinfo {pages} {024111} (\bibinfo {year} {2022})},\ \Eprint
  {http://arxiv.org/abs/2204.09170} {arXiv:2204.09170 [cond-mat.stat-mech]}
  \BibitemShut {NoStop}%
\bibitem [{\citenamefont {Stevenson}(2022)}]{Stevenson:2022gcv}%
  \BibitemOpen
  \bibfield  {author} {\bibinfo {author} {\bibfnamefont {P.~M.}\ \bibnamefont
  {Stevenson}},\ }\href {\doibase 10.1142/12817} {\emph {\bibinfo {title}
  {{Renormalized Perturbation Theory and its Optimization by the Principle of
  Minimal Sensitivity}}}}\ (\bibinfo  {publisher} {World Scientific},\ \bibinfo
  {year} {2022})\BibitemShut {NoStop}%
\bibitem [{\citenamefont {Litim}(2000)}]{Litim:2000ci}%
  \BibitemOpen
  \bibfield  {author} {\bibinfo {author} {\bibfnamefont {D.~F.}\ \bibnamefont
  {Litim}},\ }\href {\doibase 10.1016/S0370-2693(00)00748-6} {\bibfield
  {journal} {\bibinfo  {journal} {Phys. Lett. B}\ }\textbf {\bibinfo {volume}
  {486}},\ \bibinfo {pages} {92} (\bibinfo {year} {2000})},\ \Eprint
  {http://arxiv.org/abs/hep-th/0005245} {arXiv:hep-th/0005245} \BibitemShut
  {NoStop}%
\bibitem [{\citenamefont {Litim}(2001)}]{Litim:2001up}%
  \BibitemOpen
  \bibfield  {author} {\bibinfo {author} {\bibfnamefont {D.~F.}\ \bibnamefont
  {Litim}},\ }\href {\doibase 10.1103/PhysRevD.64.105007} {\bibfield  {journal}
  {\bibinfo  {journal} {Phys. Rev. D}\ }\textbf {\bibinfo {volume} {64}},\
  \bibinfo {pages} {105007} (\bibinfo {year} {2001})},\ \Eprint
  {http://arxiv.org/abs/hep-th/0103195} {arXiv:hep-th/0103195} \BibitemShut
  {NoStop}%
\bibitem [{\citenamefont {Pawlowski}\ \emph {et~al.}(2004)\citenamefont
  {Pawlowski}, \citenamefont {Litim}, \citenamefont {Nedelko},\ and\
  \citenamefont {von Smekal}}]{Pawlowski:2003hq}%
  \BibitemOpen
  \bibfield  {author} {\bibinfo {author} {\bibfnamefont {J.~M.}\ \bibnamefont
  {Pawlowski}}, \bibinfo {author} {\bibfnamefont {D.~F.}\ \bibnamefont
  {Litim}}, \bibinfo {author} {\bibfnamefont {S.}~\bibnamefont {Nedelko}}, \
  and\ \bibinfo {author} {\bibfnamefont {L.}~\bibnamefont {von Smekal}},\
  }\href {\doibase 10.1103/PhysRevLett.93.152002} {\bibfield  {journal}
  {\bibinfo  {journal} {Phys. Rev. Lett.}\ }\textbf {\bibinfo {volume} {93}},\
  \bibinfo {pages} {152002} (\bibinfo {year} {2004})},\ \Eprint
  {http://arxiv.org/abs/hep-th/0312324} {arXiv:hep-th/0312324} \BibitemShut
  {NoStop}%
\bibitem [{\citenamefont {de~Brito}\ \emph {et~al.}()\citenamefont {de~Brito},
  \citenamefont {Reichert},\ and\ \citenamefont {Schiffer}}]{YukToApp}%
  \BibitemOpen
  \bibfield  {author} {\bibinfo {author} {\bibfnamefont {G.}~\bibnamefont
  {de~Brito}}, \bibinfo {author} {\bibfnamefont {M.}~\bibnamefont {Reichert}},
  \ and\ \bibinfo {author} {\bibfnamefont {M.}~\bibnamefont {Schiffer}},\
  }\href@noop {} {\ }\Eprint {http://arxiv.org/abs/To appear} {To appear}
  \BibitemShut {NoStop}%
\bibitem [{\citenamefont {Robinson}\ and\ \citenamefont
  {Wilczek}(2006)}]{Robinson:2005fj}%
  \BibitemOpen
  \bibfield  {author} {\bibinfo {author} {\bibfnamefont {S.~P.}\ \bibnamefont
  {Robinson}}\ and\ \bibinfo {author} {\bibfnamefont {F.}~\bibnamefont
  {Wilczek}},\ }\href {\doibase 10.1103/PhysRevLett.96.231601} {\bibfield
  {journal} {\bibinfo  {journal} {Phys. Rev. Lett.}\ }\textbf {\bibinfo
  {volume} {96}},\ \bibinfo {pages} {231601} (\bibinfo {year} {2006})},\
  \Eprint {http://arxiv.org/abs/hep-th/0509050} {arXiv:hep-th/0509050}
  \BibitemShut {NoStop}%
\bibitem [{\citenamefont {Pietrykowski}(2007)}]{Pietrykowski:2006xy}%
  \BibitemOpen
  \bibfield  {author} {\bibinfo {author} {\bibfnamefont {A.~R.}\ \bibnamefont
  {Pietrykowski}},\ }\href {\doibase 10.1103/PhysRevLett.98.061801} {\bibfield
  {journal} {\bibinfo  {journal} {Phys. Rev. Lett.}\ }\textbf {\bibinfo
  {volume} {98}},\ \bibinfo {pages} {061801} (\bibinfo {year} {2007})},\
  \Eprint {http://arxiv.org/abs/hep-th/0606208} {arXiv:hep-th/0606208}
  \BibitemShut {NoStop}%
\bibitem [{\citenamefont {Toms}(2007)}]{Toms:2007sk}%
  \BibitemOpen
  \bibfield  {author} {\bibinfo {author} {\bibfnamefont {D.~J.}\ \bibnamefont
  {Toms}},\ }\href {\doibase 10.1103/PhysRevD.76.045015} {\bibfield  {journal}
  {\bibinfo  {journal} {Phys. Rev. D}\ }\textbf {\bibinfo {volume} {76}},\
  \bibinfo {pages} {045015} (\bibinfo {year} {2007})},\ \Eprint
  {http://arxiv.org/abs/0708.2990} {arXiv:0708.2990 [hep-th]} \BibitemShut
  {NoStop}%
\bibitem [{\citenamefont {Ebert}\ \emph {et~al.}(2008)\citenamefont {Ebert},
  \citenamefont {Plefka},\ and\ \citenamefont {Rodigast}}]{Ebert:2007gf}%
  \BibitemOpen
  \bibfield  {author} {\bibinfo {author} {\bibfnamefont {D.}~\bibnamefont
  {Ebert}}, \bibinfo {author} {\bibfnamefont {J.}~\bibnamefont {Plefka}}, \
  and\ \bibinfo {author} {\bibfnamefont {A.}~\bibnamefont {Rodigast}},\ }\href
  {\doibase 10.1016/j.physletb.2008.01.037} {\bibfield  {journal} {\bibinfo
  {journal} {Phys. Lett. B}\ }\textbf {\bibinfo {volume} {660}},\ \bibinfo
  {pages} {579} (\bibinfo {year} {2008})},\ \Eprint
  {http://arxiv.org/abs/0710.1002} {arXiv:0710.1002 [hep-th]} \BibitemShut
  {NoStop}%
\bibitem [{\citenamefont {Toms}(2010)}]{Toms:2010vy}%
  \BibitemOpen
  \bibfield  {author} {\bibinfo {author} {\bibfnamefont {D.~J.}\ \bibnamefont
  {Toms}},\ }\href {\doibase 10.1038/nature09506} {\bibfield  {journal}
  {\bibinfo  {journal} {Nature}\ }\textbf {\bibinfo {volume} {468}},\ \bibinfo
  {pages} {56} (\bibinfo {year} {2010})},\ \Eprint
  {http://arxiv.org/abs/1010.0793} {arXiv:1010.0793 [hep-th]} \BibitemShut
  {NoStop}%
\bibitem [{\citenamefont {Anber}\ \emph {et~al.}(2011)\citenamefont {Anber},
  \citenamefont {Donoghue},\ and\ \citenamefont {El-Houssieny}}]{Anber:2010uj}%
  \BibitemOpen
  \bibfield  {author} {\bibinfo {author} {\bibfnamefont {M.~M.}\ \bibnamefont
  {Anber}}, \bibinfo {author} {\bibfnamefont {J.~F.}\ \bibnamefont {Donoghue}},
  \ and\ \bibinfo {author} {\bibfnamefont {M.}~\bibnamefont {El-Houssieny}},\
  }\href {\doibase 10.1103/PhysRevD.83.124003} {\bibfield  {journal} {\bibinfo
  {journal} {Phys. Rev. D}\ }\textbf {\bibinfo {volume} {83}},\ \bibinfo
  {pages} {124003} (\bibinfo {year} {2011})},\ \Eprint
  {http://arxiv.org/abs/1011.3229} {arXiv:1011.3229 [hep-th]} \BibitemShut
  {NoStop}%
\bibitem [{\citenamefont {de~Brito}\ and\ \citenamefont
  {Eichhorn}(2023)}]{deBrito:2022vbr}%
  \BibitemOpen
  \bibfield  {author} {\bibinfo {author} {\bibfnamefont {G.~P.}\ \bibnamefont
  {de~Brito}}\ and\ \bibinfo {author} {\bibfnamefont {A.}~\bibnamefont
  {Eichhorn}},\ }\href {\doibase 10.1140/epjc/s10052-023-11172-z} {\bibfield
  {journal} {\bibinfo  {journal} {Eur. Phys. J. C}\ }\textbf {\bibinfo {volume}
  {83}},\ \bibinfo {pages} {161} (\bibinfo {year} {2023})},\ \Eprint
  {http://arxiv.org/abs/2201.11402} {arXiv:2201.11402 [hep-th]} \BibitemShut
  {NoStop}%
\bibitem [{\citenamefont {Wetterich}(1993)}]{Wetterich:1992yh}%
  \BibitemOpen
  \bibfield  {author} {\bibinfo {author} {\bibfnamefont {C.}~\bibnamefont
  {Wetterich}},\ }\href {\doibase 10.1016/0370-2693(93)90726-X} {\bibfield
  {journal} {\bibinfo  {journal} {Phys. Lett. B}\ }\textbf {\bibinfo {volume}
  {301}},\ \bibinfo {pages} {90} (\bibinfo {year} {1993})},\ \Eprint
  {http://arxiv.org/abs/1710.05815} {arXiv:1710.05815 [hep-th]} \BibitemShut
  {NoStop}%
\bibitem [{\citenamefont {Manrique}\ and\ \citenamefont
  {Reuter}(2011)}]{Manrique:2009tj}%
  \BibitemOpen
  \bibfield  {author} {\bibinfo {author} {\bibfnamefont {E.}~\bibnamefont
  {Manrique}}\ and\ \bibinfo {author} {\bibfnamefont {M.}~\bibnamefont
  {Reuter}},\ }\href {\doibase 10.22323/1.079.0001} {\bibfield  {journal}
  {\bibinfo  {journal} {PoS}\ }\textbf {\bibinfo {volume} {CLAQG08}},\ \bibinfo
  {pages} {001} (\bibinfo {year} {2011})},\ \Eprint
  {http://arxiv.org/abs/0905.4220} {arXiv:0905.4220 [hep-th]} \BibitemShut
  {NoStop}%
\bibitem [{\citenamefont {Morris}\ and\ \citenamefont
  {Slade}(2015)}]{Morris:2015oca}%
  \BibitemOpen
  \bibfield  {author} {\bibinfo {author} {\bibfnamefont {T.~R.}\ \bibnamefont
  {Morris}}\ and\ \bibinfo {author} {\bibfnamefont {Z.~H.}\ \bibnamefont
  {Slade}},\ }\href {\doibase 10.1007/JHEP11(2015)094} {\bibfield  {journal}
  {\bibinfo  {journal} {JHEP}\ }\textbf {\bibinfo {volume} {11}},\ \bibinfo
  {pages} {094} (\bibinfo {year} {2015})},\ \Eprint
  {http://arxiv.org/abs/1507.08657} {arXiv:1507.08657 [hep-th]} \BibitemShut
  {NoStop}%
\bibitem [{\citenamefont {Fraaije}\ \emph {et~al.}(2022)\citenamefont
  {Fraaije}, \citenamefont {Platania},\ and\ \citenamefont
  {Saueressig}}]{Fraaije:2022uhg}%
  \BibitemOpen
  \bibfield  {author} {\bibinfo {author} {\bibfnamefont {M.}~\bibnamefont
  {Fraaije}}, \bibinfo {author} {\bibfnamefont {A.}~\bibnamefont {Platania}}, \
  and\ \bibinfo {author} {\bibfnamefont {F.}~\bibnamefont {Saueressig}},\
  }\href {\doibase 10.1016/j.physletb.2022.137399} {\bibfield  {journal}
  {\bibinfo  {journal} {Phys. Lett. B}\ }\textbf {\bibinfo {volume} {834}},\
  \bibinfo {pages} {137399} (\bibinfo {year} {2022})},\ \Eprint
  {http://arxiv.org/abs/2206.10626} {arXiv:2206.10626 [hep-th]} \BibitemShut
  {NoStop}%
\bibitem [{\citenamefont {Morris}(1994)}]{Morris:1993qb}%
  \BibitemOpen
  \bibfield  {author} {\bibinfo {author} {\bibfnamefont {T.~R.}\ \bibnamefont
  {Morris}},\ }\href {\doibase 10.1142/S0217751X94000972} {\bibfield  {journal}
  {\bibinfo  {journal} {Int. J. Mod. Phys. A}\ }\textbf {\bibinfo {volume}
  {9}},\ \bibinfo {pages} {2411} (\bibinfo {year} {1994})},\ \Eprint
  {http://arxiv.org/abs/hep-ph/9308265} {arXiv:hep-ph/9308265} \BibitemShut
  {NoStop}%
\bibitem [{\citenamefont {Ellwanger}(1994)}]{Ellwanger:1993mw}%
  \BibitemOpen
  \bibfield  {author} {\bibinfo {author} {\bibfnamefont {U.}~\bibnamefont
  {Ellwanger}},\ }\href {\doibase 10.1007/BF01555911} {\bibfield  {journal}
  {\bibinfo  {journal} {Z. Phys. C}\ }\textbf {\bibinfo {volume} {62}},\
  \bibinfo {pages} {503} (\bibinfo {year} {1994})},\ \Eprint
  {http://arxiv.org/abs/hep-ph/9308260} {arXiv:hep-ph/9308260} \BibitemShut
  {NoStop}%
\bibitem [{\citenamefont {Dupuis}\ \emph {et~al.}(2021)\citenamefont {Dupuis},
  \citenamefont {Canet}, \citenamefont {Eichhorn}, \citenamefont {Metzner},
  \citenamefont {Pawlowski}, \citenamefont {Tissier},\ and\ \citenamefont
  {Wschebor}}]{Dupuis:2020fhh}%
  \BibitemOpen
  \bibfield  {author} {\bibinfo {author} {\bibfnamefont {N.}~\bibnamefont
  {Dupuis}}, \bibinfo {author} {\bibfnamefont {L.}~\bibnamefont {Canet}},
  \bibinfo {author} {\bibfnamefont {A.}~\bibnamefont {Eichhorn}}, \bibinfo
  {author} {\bibfnamefont {W.}~\bibnamefont {Metzner}}, \bibinfo {author}
  {\bibfnamefont {J.~M.}\ \bibnamefont {Pawlowski}}, \bibinfo {author}
  {\bibfnamefont {M.}~\bibnamefont {Tissier}}, \ and\ \bibinfo {author}
  {\bibfnamefont {N.}~\bibnamefont {Wschebor}},\ }\href {\doibase
  10.1016/j.physrep.2021.01.001} {\bibfield  {journal} {\bibinfo  {journal}
  {Phys. Rept.}\ }\textbf {\bibinfo {volume} {910}},\ \bibinfo {pages} {1}
  (\bibinfo {year} {2021})},\ \Eprint {http://arxiv.org/abs/2006.04853}
  {arXiv:2006.04853 [cond-mat.stat-mech]} \BibitemShut {NoStop}%
\bibitem [{\citenamefont {Litim}\ and\ \citenamefont
  {Pawlowski}(2002)}]{Litim:2002ce}%
  \BibitemOpen
  \bibfield  {author} {\bibinfo {author} {\bibfnamefont {D.~F.}\ \bibnamefont
  {Litim}}\ and\ \bibinfo {author} {\bibfnamefont {J.~M.}\ \bibnamefont
  {Pawlowski}},\ }\href {\doibase 10.1088/1126-6708/2002/09/049} {\bibfield
  {journal} {\bibinfo  {journal} {JHEP}\ }\textbf {\bibinfo {volume} {09}},\
  \bibinfo {pages} {049} (\bibinfo {year} {2002})},\ \Eprint
  {http://arxiv.org/abs/hep-th/0203005} {arXiv:hep-th/0203005} \BibitemShut
  {NoStop}%
\bibitem [{\citenamefont {Ward}(1950)}]{Ward:1950xp}%
  \BibitemOpen
  \bibfield  {author} {\bibinfo {author} {\bibfnamefont {J.~C.}\ \bibnamefont
  {Ward}},\ }\href {\doibase 10.1103/PhysRev.78.182} {\bibfield  {journal}
  {\bibinfo  {journal} {Phys. Rev.}\ }\textbf {\bibinfo {volume} {78}},\
  \bibinfo {pages} {182} (\bibinfo {year} {1950})}\BibitemShut {NoStop}%
\bibitem [{\citenamefont {Takahashi}(1957)}]{Takahashi:1957xn}%
  \BibitemOpen
  \bibfield  {author} {\bibinfo {author} {\bibfnamefont {Y.}~\bibnamefont
  {Takahashi}},\ }\href {\doibase 10.1007/BF02832514} {\bibfield  {journal}
  {\bibinfo  {journal} {Nuovo Cim.}\ }\textbf {\bibinfo {volume} {6}},\
  \bibinfo {pages} {371} (\bibinfo {year} {1957})}\BibitemShut {NoStop}%
\bibitem [{\citenamefont {de~Brito}\ and\ \citenamefont
  {Schiffer}()}]{GustavoInPrep}%
  \BibitemOpen
  \bibfield  {author} {\bibinfo {author} {\bibfnamefont {G.}~\bibnamefont
  {de~Brito}}\ and\ \bibinfo {author} {\bibfnamefont {M.}~\bibnamefont
  {Schiffer}},\ }\href@noop {} {\ }\Eprint {http://arxiv.org/abs/To appear} {To
  appear} \BibitemShut {NoStop}%
\end{thebibliography}%
\end{document}